\documentclass[a4paper,11pt]{article}
\pdfoutput=1
\usepackage{jheppub}
\usepackage{multirow}
\usepackage{graphicx}
\usepackage{amssymb}
\usepackage{amsmath}
\usepackage{bbold,amsfonts}
\usepackage[utf8]{inputenc}
\usepackage{bm}
\usepackage{xcolor}
\usepackage{soul}
\usepackage{float}
\usepackage{slashed}
\usepackage{subcaption}
\usepackage{mathtools}

\DeclarePairedDelimiter\floor{\lfloor}{\rfloor}

\newcommand{\ii}{\mathrm{i}}
\newcommand{\rme}{\mathrm{e}}

\newcommand{\vev}[1]{{\left\langle #1 \right\rangle}}

\DeclareMathOperator{\tr}{tr}

\makeatletter
\newcommand*{\letterdef@}{}
\newcommand*{\letterdef}[3]{%
	\def\letterdef@##1{\expandafter\newcommand\csname #1\endcsname{#2{##1}}}%
	\@tfor\@tempa :=#3\do{\expandafter\letterdef@\expandafter{\@tempa}}}
\makeatother
\letterdef{c#1} {\mathcal}{ABCDEFGHIJKLMNOPQRSTUVWXYZ} 
\letterdef{rm#1}{\mathrm} {dDmM} 

\newdimen\tableauside\tableauside=1.0ex
\newdimen\tableaurule\tableaurule=0.4pt
\newdimen\tableaustep
\def\phantomhrule#1{\hbox{\vbox to0pt{\hrule height\tableaurule
			width#1\vss}}}
\def\phantomvrule#1{\vbox{\hbox to0pt{\vrule width\tableaurule
			height#1\hss}}}
\def\sqr{\vbox{%
		\phantomhrule\tableaustep
		\hbox{\phantomvrule\tableaustep\kern\tableaustep\phantomvrule\tableaustep}%
		\hbox{\vbox{\phantomhrule\tableauside}\kern-\tableaurule}}}
\def\squares#1{\hbox{\count0=#1\noindent\loop\sqr
		\advance\count0 by-1 \ifnum\count0>0\repeat}}
\def\tableau#1{\vcenter{\offinterlineskip
		\tableaustep=\tableauside\advance\tableaustep by-\tableaurule
		\kern\normallineskip\hbox
		{\kern\normallineskip\vbox
			{\gettableau#1 0 }%
			\kern\normallineskip\kern\tableaurule}%
		\kern\normallineskip\kern\tableaurule}}
\def\gettableau#1 {\ifnum#1=0\let\next=\null\else
	\squares{#1}\let\next=\gettableau\fi\next}
\newcommand{\parenth}[1]{\left( #1 \right)}

\renewcommand{\a}{\alpha}

\title{\boldmath  Wilson loop correlators in
	$\cN=2$ superconformal quivers}

\author[a]{Francesco Galvagno,}
\affiliation[a]{Institut f\"ur Theoretische Physik, ETH Z\"urich
\\
	Wolfgang-Pauli-Strasse 27, 8093 Z\"urich, Switzerland
\\}
\emailAdd{fgalvagno@phys.ethz.ch}

\author[b]{Michelangelo Preti}
\affiliation[b]{Mathematics Department, King’s College London
\\
The Strand, London WC2R 2LS, UK
\\}
\emailAdd{michelangelo.preti@gmail.com}

\abstract{We complete the program of \cite{Galvagno:2020cgq} about perturbative approaches for $\mathcal{N}=2$ superconformal quiver theories in four dimensions. We consider several classes of observables in presence of Wilson loops, and we evaluate them with the help of supersymmetric localization. We compute Wilson loop vacuum expectation values, correlators of multiple coincident Wilson loops and one-point functions of chiral operators in presence of them acting as superconformal defects. We extend this analysis to the most general case considering chiral operators and multiple Wilson loops scattered in all the possible ways among the vector multiplets of the quiver. Finally, we identify twisted and untwisted observables which probe the orbifold of $AdS_5\times S^5$ with the aim of testing possible holographic perspectives of quiver theories in $\cN=2$.}


\begin{document}
	\maketitle
	\flushbottom

\section{Introduction}
\label{sec:intro}
Gauge theories with extended supersymmetry played and still play a primary role in the context of $AdS/CFT$ correspondence. The most studied example is given by $\cN=4$ Super-Yang-Mills (SYM) theory. This is the maximal supersymmetric theory in four dimensions, it is integrable in the planar limit and, given the large amount of symmetries, it allows for exact results both in the coupling and in the gauge group rank $N$. Therefore, it is possible to interpolate between the weak and strong coupling regimes of the theory, allowing to probe $AdS/CFT$ correspondence. The most powerful technique that generates such exact results is supersymmetric localization. 

In this framework, one of the main observables that has been studied is the 1/2 BPS circular Wilson loop. In $\cN=4$ SYM, its expectation value can be localized to a Gaussian matrix model on a four sphere \cite{Erickson:2000af,Drukker:2000rr,Pestun:2007rz}. 
Supersymmetric localization allows also to compute more general correlation functions that include local operators \cite{Semenoff:2001xp,Pestun:2002mr,Drukker:2007qr,Pestun:2009nn,Giombi:2009ds,Giombi:2009ek,Giombi:2012ep,Bonini:2014vta,Bonini:2015fng}), the Bremsstrahlung function \cite{Correa:2012at,Lewkowycz:2013laa} or multiple insertions of Wilson loops and Wilson loops in higher dimensional representations \cite{Okuyama:2018aij,Correa:2018lyl,Correa:2018pfn, CanazasGaray:2019mgq, Beccaria:2020ykg}.
Interesting attempts to extend these exact results to different frameworks have been achieved in the three-dimensional analogue of $\cN=4$, $ABJM$ theory, where localization was still successfully implemented \cite{Kapustin:2009kz,Marino:2009jd,Drukker:2010nc,Bianchi:2018bke,Griguolo:2021rke} and several observables have been computed exactly \cite{Bianchi:2014laa,Bianchi:2017svd,Bianchi:2017ozk,Bianchi:2018scb}, see \cite{Drukker:2019bev} for a recent review.

Sticking to four dimensions, the natural direction to extend $\cN=4$ features is the less supersymmetric $\cN=2$ case, where it is possible to directly follow some guidelines from $\cN=4$. The most studied $\cN=2$ theory is superconformal QCD (SCQCD), described by an $SU(N)$ gauge group with $2N$ hypermultiplets as a conformal matter content. The action of supersymmetric localization in $\cN=2$ is still powerful enough to reduce the Wilson loop vacuum expectation value (vev) to a matrix model, even though it is no longer Gaussian. 
Several quantities have been analysed for $\cN=2$ theories, such as the Wilson loop vev  \cite{Andree:2010na,Passerini:2011fe,Bourgine:2011ie,Billo:2019fbi,Beccaria:2021vuc} 
together with correlators of chiral operators and Wilson loops \cite{Rodriguez-Gomez:2016cem,Billo:2018oog,Beccaria:2020hgy} and the Bremsstrahlung function \cite{Fiol:2015mrp,Bianchi:2018zpb,Bianchi:2019dlw}.

Another class of $\cN=2$ Lagrangian theories which is central in the present paper is represented by a circular quiver with $q$ nodes and denoted as $A_{q-1}$. Each node corresponds to a vector multiplet with $SU(N)$ gauge group, the lines between the nodes stand for bifundamental hypermultiplets, which represent the conformal matter content of the theory. These special $\cN=2$ theories, under special conditions, are known to possess a holographic dual \cite{Kachru:1998ys,Gukov:1998kk} defined as a type IIB string theory on a background in presence of a $\mathbb{Z}_q$ orbifold, and have been studied in integrability contexts \cite{Gadde:2009dj,Gadde:2010zi,Pomoni:2011jj,Gadde:2012rv,Pomoni:2013poa,Mitev:2014yba,Mitev:2015oty,Pomoni:2019oib,Niarchos:2019onf,Niarchos:2020nxk}. For these reasons, $A_{q-1}$ quiver theories can be considered the most similar $\cN=2$ theories to $\cN=4$ SYM. Indeed, they are also often called interpolating theories placed in the middle between $\cN=4$ SYM and the conventional $\cN=2$ SCQCD\footnote{In \cite{Bourget:2018fhe,Billo:2019fbi,Beccaria:2020hgy,Beccaria:2021ksw} a special $\cN=2$ SCFT with symmetric and antisymmetric matter content has been considered, and such theory also enjoys nice holographic properties. It would be very interesting to fully understand the common properties shared with $A_{q-1}$ theories.}.
$A_{q-1}$ theories also admit a localization approach that localizes observables on a multi-matrix model. Many results have been recently obtained using such matrix model, in particular for Wilson loop vevs \cite{Fiol:2020ojn,Zarembo:2020tpf,Ouyang:2020hwd,Beccaria:2021ksw} and for chiral/antichiral correlators \cite{Pini:2017ouj,Galvagno:2020cgq}. In \cite{Galvagno:2020cgq} a complete analysis of $A_{q-1}$ matrix model has been developed and applied to the two-point correlators of chiral/anti-chiral operators, by solving the mixing problem in moving from the flat space to $S^4$ and finding general formulas for the two-point chiral correlators for any values of the dimension $n$ and the number of vector multiplets $q$. 

In the present paper we exploit the technical achievements of \cite{Galvagno:2020cgq} to be applied to a wider range of observables, in presence of Wilson loop insertions. In particular we consider the circular Wilson loop vev, correlation functions of multiple coincident Wilson loops and one-point functions of chiral operators in presence of a Wilson loop defects. In full generality, we can also produce results for the correlation function of a chiral operator in presence of any number of coincident Wilson loops. We discuss various cases, playing with the number of nodes $q$, the number of Wilson loops inserted in each node and the dimension $n$ of chiral operators. We write the perturbative results organized as expansions in transcendentality (labelled by odd Riemann zetas ${\color{red}\zeta}_3,{\color{red}\zeta}_5,\dots$), where each transcendentality term can be written in terms of derivatives of the $\cN=4$ exact results\footnote{See section \ref{sec:WLSCQCD} for a detailed explanation of these mothod and expansion.}
We also generate a database of perturbative results for generic values of the couplings ${\color{blue}\lambda}_1,\dots,{\color{blue}\lambda}_q$ and at finite-$N$. This was computed with the package \cite{Preti2021maybe} and collected in the Mathematica notebook \texttt{WLcorrelators.nb} attached to this manuscript. 

The main results of the present paper which are most relevant in the context of $\cN=2$ theories  can be summarized as follows:
\begin{itemize}
\item 
The Wilson loop vev in $A_{q-1}$ theories in the large-$N$ limit and at the orbifold point (\textit{i.e.} the limit where all the coupling are equivalent) is the same as the $\cN=4$, see equation \eqref{wqvanishes}. The same happens for correlators of multiple coincident Wilson loops, which coincide with the $\cN=4$ analogue in the planar limit, see equation \eqref{wvecIqvanishes}. These facts confirm at weak coupling a well known holographic result \cite{Rey:2010ry,Zarembo:2020tpf,Ouyang:2020hwd,Beccaria:2021ksw}, since the minimal surface described by the Wilson loop is unaffected by the orbifold $\mathbb Z_q$. See also section \ref{sec:WuWt} for additional comments about Wilson loop vevs for $AdS/CFT$ perspectives.
\item 
Among all the possible one-point functions with the Wilson loop, we isolate the combinations corresponding to the twisted and untwisted sectors under the orbifold action. In particular untwisted operators reproduce the same exact result in ${\color{blue}\lambda}$ as the $\cN=4$, see equation \eqref{UntwlikeN4}, whereas twisted operators deviate from the corresponding $\cN=4$ results with a full perturbative expansion, see equations \eqref{TwistedOrbifold}, \eqref{TwistedGeneral} and \eqref{TwistedOrbifoldq=2}, and represent the ideal observables to probe $A_{q-1}$ theories at a holographic level.
\item
The vev of a circular Wilson loop in SCQCD has a full perturbative expansion, see equations \eqref{WL2N} and \eqref{WL2NlargeN}. Going to high transcendentality orders, we are able to find an exponentiation property of the ${\color{red}\zeta}_3$ transcendentality term, see equation \eqref{expresum1}.
\item
We consistently check the results from the matrix model approach with standard Feynman diagrams computations using the $\cN=1$ superspace formalism. In particular we prove how the ${\color{red}\zeta}_3$ part of Wilson loop vev in SCQCD enjoys exponentiation properties in terms of pure combinatorics of insertions of special two-loops diagrams (see equation \eqref{N2WLvev}) and the cancellations at the level of color factors for the $A_{q-1}$ Wilson loop vev at the orbifold point, see equation \eqref{Aqscalarprop}. Finally we identify special diagrams which act as building blocks for several computations of a wide number of observables, confirming a pattern already discussed in \cite{Galvagno:2020cgq}.
\end{itemize}
The structure of the manuscript is the following. In section \ref{Sec:2} we introduce the $A_{q-1}$ quiver theories and we define all the observables that will be studied across the text. In section \ref{sec:MM} all the technical set up is introduced: we build the multi-matrix model using localization and we define all the observables in the matrix model. Sections \ref{sec:<W>} and \ref{sec:5} are devoted to all the results of the paper using the multi-matrix model, while in section \ref{Sec:FieldTheory} such results are discussed in terms of Feynman diagrams. Additional results and some technical material are stored in the appendices.

\section{Wilson loops and chiral operators in superconformal quivers}\label{Sec:2}

\begin{figure}[!t]
\begin{center}
\includegraphics[scale=0.6]{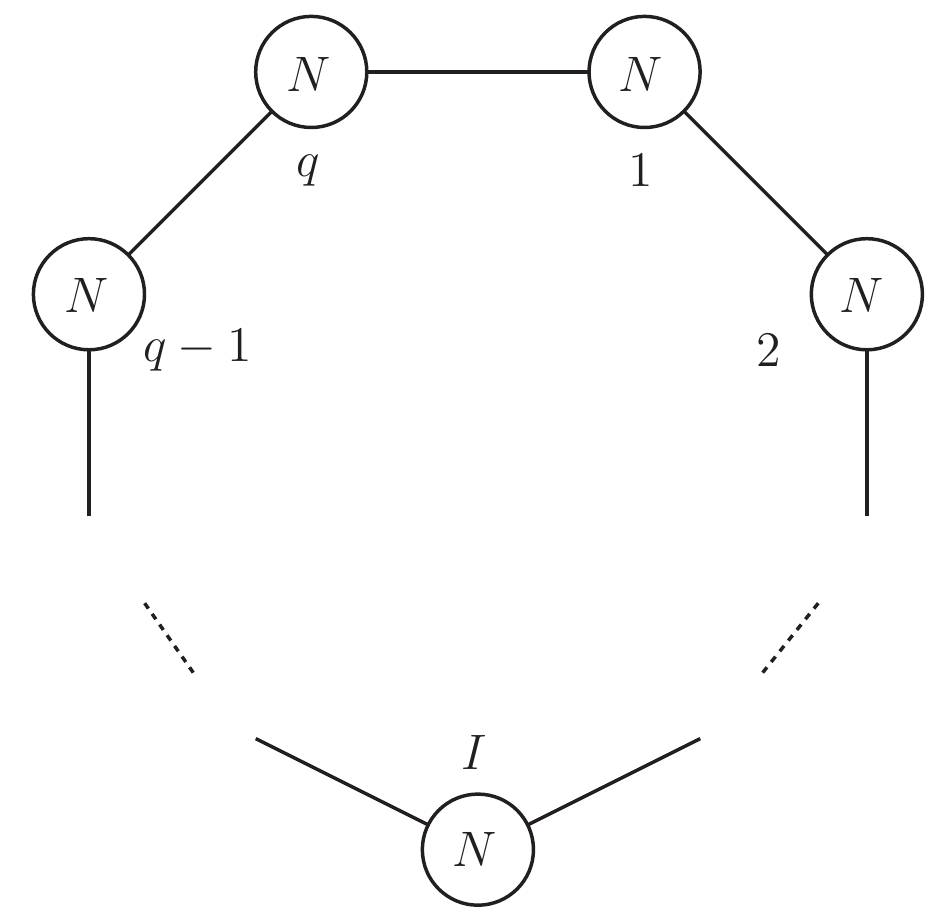}
\caption{$A_{q-1}$ theories as circular quivers with $q$ gauge nodes.}
\label{Fig:circquiver}
\end{center}
\end{figure}

Our setup is the same considered in \cite{Galvagno:2020cgq}, namely the family of $\cN=2$ superconformal theories, denoted as $A_{q-1}$. 
The gauge structure of those theories can be represented by a circular quiver with $q$ nodes, see Figure \ref{Fig:circquiver}. 
Each node $I$ stands for a $SU(N)_I$ vector multiplet, whereas lines between nodes represent the conformal matter content, \textit{i.e.} a hypermultiplet in the bifundamental of $SU(N)_I \times SU(N)_{I+1}$. Finally, any node $I$ is associated to a non-running coupling constant $g_I$, which can be recast in the usual 't Hooft combination defined as
\begin{align}\label{tHooftcouplings}
{\color{blue}\lambda}_I = g_I^2 N~.
\end{align}

\noindent
As anticipated in the introduction,  $A_{q-1}$ theories enjoy the nice property of interpolating between two interesting gauge theories. Switching off all the gauge couplings except one (${\color{blue}\lambda}_{I\neq 1} =0$) reproduces $\cN=2$ SCQCD, a theory with a single gauge group and $2N$ hypermultiplets as conformal matter content, while if we tune ${\color{blue}\lambda}_{I} = {\color{blue}\lambda},~\forall I$ we obtain the $\mathbb Z_q$ orbifold point of $\cN=4$. In this limit $A_{q-1}$ theories admit a AdS$_5 \times (S^5/\mathbb Z_q)$ dual geometry and hence they are interesting for holographic perspectives.\\
$A_{q-1}$ theories admit a Lagrangian description in terms of $\cN=1$ superfields, which is shown in appendix \ref{App:FieldTheory}. The complete formalism in Euclidean space can be found in \cite{Galvagno:2020cgq}. We also refer to that paper for all the conventions about normalization of the fields and the complete list of Feynman rules. 
In the following, we introduce all the observables we are going to analyse throughout the paper.

The leading role in this paper is played by the 1/2 BPS circular Wilson loop \cite{Rey:2010ry,Andree:2010na,Passerini:2011fe}. Such non-local operator measures the holonomy of the gauge connection around a circular path $C$ and it represents a conformal defect in the theory. Considering the presence of multiple vector multiplets in the necklace quiver, the theories $A_{q-1}$ admit $q$ copies of the Wilson loop defined as
\begin{equation}
	\label{WLdef}
		W_I=\frac{1}{N}\tr \mathcal{P}
		\exp \bigg\{g_{I} \oint_C d\tau \Big[\ii \,A^{I}_{\mu}(x)\,\dot{x}^{\mu}(\tau)
		+\frac{R}{\sqrt{2}}\big(\varphi_{I}(x) +\bar \varphi_{I}(x)\big)\Big]\bigg\}\,,
\end{equation}
where $``\tr"$ is the trace over the fundamental representation of $SU(N)$, $\mathcal{P}$ denotes the path-ordering, $R$ is the radius of the circle $C$ parametrized by $x^\mu(\tau)$ and $\{A^{I}_{\mu},\varphi_{I}\}$ are the gauge and the complex scalar field belonging to the $I$-th vector multiplet.

In this paper we focus on observables involving the Wilson operator \eqref{WLdef} captured by 
localization.  
This technique, relying on supersymmetry, yields exact results for a specific set of observables invariant with respect to a subset of the supersymmetry charges or a combination of them. In the following we identify those observables and, in section \ref{sec:MM}, we present the localization based method to compute them.\\
The simplest one is certainly the vacuum expectation
value of the Wilson loop itself
\begin{equation}
\label{Wvev}
\vev{W_I}_q \equiv w_I^{(q)}({\color{blue}\lambda}_1,...,{\color{blue}\lambda}_q,N)~,
\end{equation}
where $\vev{\;\;}_q$ indicates that the average is computed in a theory with $q$ vector multiplets. Due to the circular symmetry of the quiver, under a suitable rotation of the couplings ${\color{blue}\lambda}_I$, the functions $w_I^{(q)}$ are all equivalent for $I=1,...,q$. Then, for simplicity, we consider the Wilson loop always belonging to the first node of the quiver and we drop the index $I$ such that $w^{(q)}_1=w^{(q)}$. However, the supersymmetry of the theory allows us to consider also more involved objects such as correlation functions of multiple Wilson loops.
These observables have been widely explored in $\cN=4$ SYM in many different set ups \cite{Drukker:2000rr, Okuyama:2018aij,Correa:2018lyl,Correa:2018pfn, CanazasGaray:2019mgq, Muck:2019hnz,Beccaria:2020ykg}. In the present paper we study correlators of multiple coincident circular Wilson loop\footnote{We need them to be coincident in order to preserve enough supersymmetry to allow a localization approach.} defined as
\begin{equation}
\label{multWvev}
\vev{W_{\vec{I}}}_q\equiv\vev{W_{I_1}W_{I_2}...W_{I_n}}_q \equiv w_{\vec{I}}^{(q)}({\color{blue}\lambda}_1,...,{\color{blue}\lambda}_q,N)~,
\end{equation}
where $\vec{I}=[I_1,I_2,...,I_n]$. The loop operators appearing in the left-hand side of \eqref{multWvev} have all the same radius and they belong to any vector multiplet of the $A_{q-1}$ theory. It is understood that the $n=1$ case coincides with the vev of the single Wilson loop \eqref{Wvev}.

In superconformal theory, Wilson loops that preserve a subgroup of the superconformal symmetry can be interpreted as conformal defect \cite{Billo:2016cpy,Billo:2018oog}. Correlators of Wilson loop defect with bulk operators are constrained by the residual symmetry, and special classes of such local operators can be studied using supersymmetric localization.
We introduce the following scalar multi-trace local operator\footnote{We choose a different normalization for the operator \eqref{Ondefinition} with respect to \cite{Galvagno:2020cgq} in order to simplify the overall factors in the results of section \ref{sec:5}. Similarly we choose a different normalization for operators \eqref{ut}}
\begin{align}
	\label{Ondefinition}
		O^{(I)}_{\vec{n}}(x) \equiv 
		\parenth{\frac{{\color{blue}\lambda}_I}{N}}^{\frac{n}{2}-t}\;\tr \varphi_I^{n_1}(x)\, \tr \varphi_I^{n_2}(x) \ldots\,\tr \varphi_I^{n_t}(x)~,
\end{align}
defined as the combinations of $t$ traces of the scalar field $\varphi_I$ belonging to the $I$-th node of the quiver and labelled by the vector $\vec{n}=[n_1,n_2,...,n_t]$. 
Since they are annihilated by half of the supercharges, the operators \eqref{Ondefinition} are chiral and also known as 1/2 BPS. Their total R-charge is given by $n = \sum_{i=1}^t n_i$ and they are normal-ordered by construction.
The choice of $SU(N)$ as gauge group of the vector multiplets restricts the powers $n_i$ to be $n_i\geq 2$ since $\tr \varphi_I=0$.

We also define some special combinations of the operators \eqref{Ondefinition} and \eqref{WLdef} originally introduced in \cite{Gukov:1998kk,Lee:1998bxa}. They correspond to twisted and untwisted sectors of the $A_{q-1}$ quiver theories for local operators\footnote{In the case $q=2$ there is only one independent twisted operators since $T^{(2)}_{\vec n}=-T^{(1)}_{\vec n}$.}:
\begin{align}\label{ut}
U_{\vec n} = \sum_{I=1}^q O_{\vec n}^{(I)}~, \hspace{1cm} T^{(I)}_{\vec n} = \left(O_{\vec n}^{(I)}-O_{\vec n}^{(I+1)}\right)~.
\end{align}
and for Wilson loops
\begin{equation}\label{WuWt}
    W^u=\frac{1}{q}\sum_{I=1}^q W_{I}\,,
    \qquad
    W^t_I=\sum_{J=1}^{q}\omega^{I(J-1)}W_J~,
\end{equation}
where $\omega$ are the $q$ roots of unity $\omega=e^{2\pi i/q}$.
For a given quiver with $q$ nodes, one can define a single untwisted operator and $q$ twisted ones that are even and odd respectively under gauge group exchange.
They enjoy good transformation properties under the orbifold action of $\mathbb{Z}_q$, and therefore they represent the ideal variables for holographic perspectives in $\cN=2$ context. The two-point functions of chiral/antichiral operators $O$, $U$ and $T$ have been discussed in \cite{Pini:2017ouj,Galvagno:2020cgq}.
In this paper we extend that analysis to one-point functions of local operators \eqref{Ondefinition} and \eqref{ut} in presence of a Wilson loop defect for the $A_{q-1}$ theories.
We also briefly explore correlators involving the twisted an untwisted Wilson loops defined in \eqref{WuWt}.\\
Considering the operator \eqref{Ondefinition}, we define the following quantity
\begin{equation}
\label{OnWdef}
\vev{W_{\vec{I}}\;\,O^{(J)}_{\vec{n}}(0)}_q = \frac{\mathcal{A}_{\vec{n}}^{(\vec{I},J)}({\color{blue}\lambda}_1,...,{\color{blue}\lambda}_q,N)}{(2\pi R)^n}~,
\end{equation}
and correspondingly 
\begin{equation}
\begin{split}
\label{TUdef}
\vev{W_{\vec{I}}\;\,U_{\vec{n}}^{\phantom{(I)}}\!\!\!(0)}_q &= \frac{\cU_{\vec n}^{(\vec{I}\,)}({\color{blue}\lambda}_1,...,{\color{blue}\lambda}_q,N)}{(2\pi R)^n}~,\\
\vev{W_{\vec{I}}\;\,T^{(J)}_{\vec{n}}(0)}_q &= \frac{\cT_{\vec n}^{(\vec{I},J)}({\color{blue}\lambda}_1,...,{\color{blue}\lambda}_q,N)}{(2\pi R)^n}~.
\end{split}
\end{equation}
for the untwisted and twisted cases respectively. The quantities \eqref{OnWdef} and \eqref{TUdef} are completely fixed by conformal invariance (see \cite{Billo:2016cpy,Billo:2018oog} for further details). Similarly to the case of the Wilson loop alone, when the vector $\vec{I}$ has a single element $I$, the functions  $\mathcal{A}_{\vec{n}}^{(I,J)}$, $\mathcal{U}_{\vec{n}}^{(I)}$ and $\mathcal{T}_{\vec{n}}^{(I,J)}$ take the same value for any $I$ if one rotates the 
quiver couplings in a suitable way. 
Given the definition \eqref{ut}, it is understood that the quantities $\mathcal{U}$ and $\mathcal{T}$ can be written as combinations of the functions $\mathcal{A}$
\begin{equation}\label{UTinA}
   \cU_{\vec n}^{(\vec{I}\,)}=  \sum_{J=1}^q\mathcal{A}_{\vec{n}}^{(\vec{I},J)}\,,\quad\qquad
   \cT_{\vec n}^{(\vec{I},J)}= \left(\mathcal{A}_{\vec{n}}^{(\vec{I},J)}-\mathcal{A}_{\vec{n}}^{(\vec{I},J+1)}\right)\,.
\end{equation}
The same conclusions can be made for correlators involving the Wilson loops \eqref{WuWt}. 
In the following section, we generalize the localization techniques to the $A_{q-1}$ theories case, establishing a connection between the gauge theory correlators above and some corresponding correlation functions in a multi-matrix model.

\section{Wilson loop correlation functions in the multi-matrix model}\label{sec:MM}

Any $\cN=2$ Lagrangian theory can be localized to a finite dimensional integral on a four sphere \cite{Pestun:2007rz}. In particular, $A_{q-1}$ theories are reduced to a multi-matrix model. 
In this section we review its construction and we describe a method to compute correlation function in this framework.
This method, based on the full Lie Algebra, has many advantages with respect to the eigenvalue distribution method (which consists in going to the Cartan subalgebra of the gauge group), allowing us to implement it in an efficient algorithmic way. 
The procedure for the $A_{q-1}$ theories is described in full details in \cite{Galvagno:2020cgq}, here we just outline the main steps.

\subsection{From the localized partition function to correlators}

Placing $A_{q-1}$ theories on the sphere, the partition function is reduced to the following multi-matrix model
\begin{equation}
	\label{Zmulti}
		\cZ = \int \prod_{I=1}^q da_I ~ e^{-\tr a_I^2} \,\big| Z_{\mathrm{inst}}\big|^2\, \big| Z_{\mathrm{1-loop}} (a_I)\big|^2\,~,
\end{equation}
where each matrix $a_I$ can be decomposed over the generators $T_a$ of $\mathfrak{su}(N)$
\begin{equation}
	\label{aont}
	a_I = a_I^b \,T_b~,\hspace{0.5cm} b = 1,\ldots, N^2-1~,
\end{equation}
normalized as $\tr \,T_b \,T_c=\delta_{bc}/2$. Each node in \eqref{Zmulti} contributes with a Gaussian term and, in order to normalize it, we consider a flat integration measure for each matrix defined as follows
\begin{align}
	\label{defMeasure}
		da_I = \prod_{b=1}^{N^2-1} \frac{da_I^b}{\sqrt{2\pi}}~.
\end{align}
Then $Z_{\mathrm{inst}}$ corresponds to instanton contribution. In this paper we will only consider the zero-instanton sector of the theory, hence we can neglect it setting $Z_{\mathrm{inst}}=1$. Moreover, we will mainly study observables in the large $N$ limit where instantons are exponentially suppressed. The 1-loop contribution $Z_{\mathrm{1-loop}}$ introduces interaction terms and it can be recast in the following exponential form
\begin{equation}
	\label{ZtoSint}
	\big|Z_{\mathrm{1-loop}}\big|^2\, = \prod_{I=1}^q \mathrm{e}^{-\mathcal{S}_{\mathrm{int}}(a_I,a_{I+1})}~,
\end{equation}
where the interacting action is given by
\begin{equation}\label{S_int}
\mathcal{S}_{\mathrm{int}} =  \!\!\sum_{m=2}^\infty \sum_{\ell=0}^{2m} \frac{(-1)^{m+\ell}}{(8\pi^2)^m m}\frac{{\color{blue}\lambda}_I^m}{ N^{2m}}\binom{2m}{\ell}  {{\color{red}\zeta}}_{2m-1} \!
\left[\tr a_I^{2m-\ell}\tr a_{I}^\ell-\frac{{\color{blue}\lambda}_{I+1}^{\ell/2}}{{\color{blue}\lambda}_I^{\ell/2}}\tr a_I^{2m-\ell}\tr a_{I+1}^\ell\right]~.
\end{equation} 
Note that the product over the nodes in \eqref{ZtoSint} is meant to take into account the circularity of the quiver. Therefore node $q+1$ is identified with node $1$. Finally the partition function \eqref{Zmulti} can be written in a compact way as
\begin{equation}\label{partfun}
\cZ=\int \prod_{I=1}^q \parenth{da_I~\rme^{-\tr\, a_I^2 - \cS_{\mathrm{int}}(a_I,a_{I+1})}} = \prod_{I=1}^q\vev{\rme^{- \cS_{\mathrm{int}}(a_I,a_{I+1})}}_0~,
\end{equation}
where the subscript $0$ indicates that the vev is computed in the Gaussian matrix model. A theory without interacting action ($\mathcal{S}_{\mathrm{int}}=0$) is interpreted as $q$ copies of a pure Gaussian matrix model corresponding to the $\cN=4$ SYM. Turning on the interacting action \eqref{S_int}, one can 
systematically expand it in perturbation theory with respect to the couplings ${\color{blue}\lambda}_I$ and then treat the resulting terms as correlation function in $q$ multiple copies of the free Gaussian matrix model describing the $\cN=4$ theory.

Analogously, any gauge invariant observable generically represented by a function $f(a_J^k)$ is evaluated in the multi-matrix model using the following definition
\begin{align}
	\label{vevf}
		\vev{ f(a_J^k) }_q\, 
		\!=\! \frac{1}
		{\cZ} \int\! \prod_{I=1}^q da_I~\rme^{-\mathrm{tr}\, a_I^2 - \cS_{\mathrm{int}}(a_I,a_{I+1})}\,f( a_J^k)\,	\!=\! \frac{1}
		{\cZ} \,\prod_{I=1}^q\vev{
		\rme^{- \cS_{\mathrm{int}}(a_I,a_{I+1})}\,f( a_J^k)}_0~.
\end{align} 
where we have the same factorised structure of \eqref{partfun}. In other words, we reduced the computation of vevs in the interacting matrix model to vevs in a Gaussian model. Then we focus on the latter.

Considering the formulas \eqref{partfun} and \eqref{vevf}, 
it is clear that the basic elements to compute in the $A_{q-1}$ multi-matrix model
are the expectation values of the multi-trace operators $\vev{\tr a_I^{n_1}\tr a_I^{n_2}\dots}_0$ in the Gaussian theory.
So we conveniently introduce the following notation:
\begin{equation}
	\label{rectn}
		t^{(I)}_{n_1,n_2,\dots} = t^{(I)}_{\vec{n}}= \vev{\tr a_I^{n_1}\tr a_I^{n_2}\dots}_0~.
\end{equation}
Considering the following initial conditions
\begin{align}
	\label{rectnodd}
		t^{(I)}_0=N~,~~~~~~~t^{(I)}_n = 0~~~\text{for $n$ odd}~,
\end{align}
one can evaluate the general expression \eqref{rectn} for $t^{(I)}_{\vec{n}}$ using the following recursion relations originally derived in \cite{Billo:2017glv}:
\begin{equation}\begin{split}\label{recursion}
t^{(I)}_{[n_1,n_2,\dots,n_t]} =   &\frac{1}{2} \sum_{m=0}^{n_1-2}  \Big( t^{(I)}_{[m,n_1-m-2,n_2,\dots,n_t]}
		-\frac{1}{N}\,   t^{(I)}_{[n_1-2,n_2,\dots,n_t]}  \Big) \\ 
		&+ \sum_{k=2}^{t}\frac{n_k}{2} \,\Big(  t^{(I)}_{[n_1+n_k-2,n_2,\dots,\slashed{n_k},\dots,n_t]} -\frac{1}{N} \,t^{(I)}_{[n_1-1,n_2,\dots,n_k-1,\dots,n_t]} \Big)~,
\end{split}\end{equation}
where ${[n_1,\dots,\slashed{n_k},\dots,n_t]}$ stands for the vector of ${[n_1,\dots,n_t]}$ indices without the $k$-th one. Since all the vector multiplets in the superconformal quiver we are considering have the same gauge group, for a given vector $\vec{n}$, the correlators $t^{(I)}_{\vec{n}}$ are all the same for any $I$. Then for simplicity we drop the index $I$ in the rest of the paper.

\subsection{Wilson loops and chiral operators in the multi-matrix model}\label{sec:3.1}

In the matrix model, the Wilson loop \eqref{WLdef} defined on a circle of radius $R=1$ is given by the following operator \cite{Pestun:2007rz}
\begin{equation}
\label{Wlmm}
\mathcal{W}_I = \frac{1}{N}\tr\, \exp \left[\sqrt{\frac{{\color{blue}\lambda}_I}{2N}}\,a_I\right] = \frac{1}{N} \sum_{\ell = 0}^{\infty}\frac{1}{\ell!} \parenth{\frac{{\color{blue}\lambda}_I}{2 N}}^{\frac{\ell}{2}}\tr a_I^\ell~,
\end{equation}
Its expectation value is related to \eqref{Wvev} as follows
\begin{equation}\label{vevWmm}
  w_I^{(q)}({\color{blue}\lambda}_1,...,{\color{blue}\lambda}_q,N)
  =\vev{\mathcal{W}_I}_q~,
\end{equation}
where the right-hand side can be written in terms of the functions \eqref{rectn} using \eqref{vevf} and then it can be computed using the recursion relation \eqref{recursion}.

Similarly to the gauge theory case, we can define the operator corresponding to multiple coincident circular Wilson loops of radius $R=1$
\begin{equation}
\label{Wlmultiplomm}
\mathcal{W}_{\vec{I}} \equiv
\mathcal{W}_{I_1}\mathcal{W}_{I_2}...\mathcal{W}_{I_n}=
\frac{1}{N^n} \sum_{\ell_1 ,\ell_2,...,\ell_n }\left[\prod_{i=1}^n\frac{1}{\ell_i!} \parenth{\frac{{\color{blue}\lambda}_{I_i}}{2 N}}^{\frac{\ell_i}{2}}\right]\tr a_{I_1}^{\ell_1}\tr a_{I_2}^{\ell_2}...\tr a_{I_n}^{\ell_n}~,
\end{equation}
whose expectation value is related to \eqref{multWvev} as follows
\begin{equation}\label{WvecImm}
  w_{\vec{I}}^{(q)}({\color{blue}\lambda}_1,...,{\color{blue}\lambda}_q,N)
  =\vev{\mathcal{W}_{\vec{I}}}_q~.
\end{equation}
In this case, the right-hand side can be written in terms of one or more functions \eqref{rectn} depending on the choices of the indices $I_i$.

Let's now consider the matrix model version of the multi-trace chiral operator defined in \eqref{Ondefinition}.
It would seem natural to associate it to a multi-matrix model operator $\mathcal{O}_{\vec{n}}^{(I)}$ with precisely the same expression but with the scalar fields $\varphi_I(x)$ replaced by the matrix $a_I$, namely
\begin{align}
	\label{Onmmdefinition}
		\mathcal{O}^{(I)}_{\vec{n}} \equiv 
		\parenth{\frac{{\color{blue}\lambda}_I}{N}}^{\frac{n}{2}-t}\;\tr a_I^{n_1}\, \tr a_I^{n_2} \ldots\,\tr a_I^{n_t}~.
\end{align}
However, identifying the correct chiral operators in the matrix model for $\cN=2$ theories is a non-trivial task. Indeed, since the propagator in gauge theory connect scalar field with their complex conjugates, operators \eqref{Ondefinition} have no self-contraction by construction whereas $\mathcal{O}^{(I)}_{\vec{n}}$ defined in \eqref{Onmmdefinition} do not share this property. Therefore, one needs to subtract all the self-contraction from \eqref{Onmmdefinition} by making it normal-ordered as described in \cite{Gerchkovitz:2016gxx,Rodriguez-Gomez:2016cem,Rodriguez-Gomez:2016ijh,Billo:2017glv,Billo:2018oog}. This is equivalent to impose orthogonality  conditions to $\mathcal{O}^{(I)}_{\vec{n}}$ and all the lower-dimensional operators.
The case of $A_{q-1}$ theories has been fully described in \cite{Galvagno:2020cgq}, we refer to that paper for all the details, here we only recap the main steps of such procedure.

Given the operator \eqref{Onmmdefinition} with scaling dimension $n$, its normal-ordered version is defined by the following linear combination
\begin{equation}\label{basis}
:\mathcal{O}^{(I)}_{\vec n}:=\mathcal{O}^{(I)}_{\vec n}+\sum_{\substack{\vec p=\text{partitions} \\ \text{of dim. $\{p\}$}}}
\;\sum_{\substack{J=\text{nodes of} \\ \text{quiver $A_{q-1}$}}}
\alpha^{(I,J)}_{\vec n,\vec p}\;\mathcal{O}^{(J)}_{\vec p}~,
\end{equation}
where the coefficients $\alpha$ are functions of the couplings ${\color{blue}\lambda}_1,...,{\color{blue}\lambda}_q$ and $N$. The basis of operators $\{\mathcal{O}^{(J)}_{\vec p}\}$ contains operators with scaling dimensions $\{p\}=\{n-2,n-4,...\}$. For instance if $n=6$, the basis contains operators of dimensions $\{p\}=\{4,2,0\}$ and, considering all the possible partitions, is given by $\{\mathcal{O}^{(J)}_{\vec p}\}=\{\mathcal{O}^{(J)}_{[4]},\mathcal{O}^{(J)}_{[2,2]},\mathcal{O}^{(J)}_{[2]},\mathbb{1}\}$. In this example, since we choose $n$ to be even, the dimension of the last operator is zero an it corresponds to the identity operator\footnote{In \eqref{basis}, when $\mathcal{O}^{(J)}_{\vec p}$ is the identity $\mathbb{1}$, our convention is to set $J=I$ since identity operator appears only once.}. However, this is not always the case. Indeed, if $n$ is odd, the lowest possible dimension is three, corresponding to the operator $\mathcal{O}^{(I)}_{[3]}$\footnote{Operators with dimension 1 do not appear in the mixing since in $SU(N)$ we have $\tr a_I=0$.}.

Mixing coefficients $\alpha^{(I,J)}_{\vec n,\vec p}$ can be determined through the Gram-Schmidt procedure as follows
\begin{equation}\label{coeffGS}
\alpha^{(I,J)}_{\vec n,\vec p}=-\sum_{\substack{\text{nodes $K$} \\ \text{partitions $\vec s$ of $\{s\}$}}}\langle \mathcal{O}^{(I)}_{\vec n} \;\mathcal{O}^{(K)}_{\vec s} \rangle_q\;\left(M_{\vec s,\vec p}^{(K,J)}\right)^{-1}~,
\end{equation}
where $M^{-1}$ is the inverse of the matrix
\begin{equation}\label{matrixM}
M_{\vec s,\vec p}^{(K,J)}=\langle \mathcal{O}^{(K)}_{\vec s} \;\mathcal{O}^{(J)}_{\vec p} \rangle_q\,,
\end{equation}
Here $\{\mathcal{O}^{(J)}_{\vec p}\}$ and $\{\mathcal{O}^{(K)}_{\vec s}\}$ are two copies of the basis associated to $\mathcal{O}^{(I)}_{\vec n}$ with $K,J=1,2,...,q$ and $\vec s$ and $\vec p$ all the possible partitions of the dimensions $\{s\}$ and $\{p\}$ (see \cite{Galvagno:2020cgq} for several examples).
The Gram-Schmidt coefficient of the identity operator can be easily computed considering the one-point functions of all the operators involved in the normal ordering with their corresponding coefficients
\begin{equation}\label{alpha0}
\alpha^{(I,I)}_{\vec n,[0]}=-\,\langle\,\mathcal{O}^{(I)}_{\vec n}\,\rangle_q-\sum_{\substack{\vec p=\text{partitions} \\ \text{of dim. $\{p\}\setminus 0$}}}
\;\sum_{\substack{J=\text{nodes of} \\ \text{quiver $A_{q-1}$}}}
\alpha^{(I,J)}_{\vec n,\vec p}\;\langle\,\mathcal{O}^{(J)}_{\vec p}\,\rangle_q~,
\end{equation}
where the set $\{\mathcal{O}^{(J)}_{\vec p}\}$ in this case does not include the identity.

Finally the normal-ordered operator \eqref{basis} with coefficients given by \eqref{coeffGS} is now orthogonal by construction to all the lower-dimensional operators and then it's perfectly equivalent to its field theory companion. 
Indeed, accordingly with the gauge theory, its one-point function vanishes
\begin{equation}\label{oneptvanish}
\langle \;:\mathcal{O}^{(I)}_{\vec n}: \;\rangle_q=0~,
\end{equation}
while it is non-trivial when computed in presence of a Wilson loop defect. In particular, now we can relate the observable defined in \eqref{OnWdef} with its matrix model relative as follows
\begin{equation}\label{Adef}
    \mathcal{A}_{\vec{n}}^{(\vec{I},J)}({\color{blue}\lambda}_1,...,{\color{blue}\lambda}_q,N)=\vev{\mathcal{W}_{\vec{I}}\;\,:\mathcal{O}^{(J)}_{\vec{n}}:}_q~,
\end{equation}
and consequently also the observables \eqref{TUdef} using \eqref{UTinA}.

\section{Wilson loops correlators}\label{sec:<W>}

In this section we use the recursion relation \eqref{recursion} to compute several observables only involving Wilson loops, hence Wilson loop vevs and correlators of multiple coincident Wilson loops. Since this procedure lends itself to being treated algorithmically, it is possible to implement it in a Mathematica package \cite{Preti2021maybe}. The data presented in this section and in the attached notebook \texttt{WLcorrelators.nb} are generated by that package.

\subsection{Correlators in the pure Gaussian model}\label{sec:WN4}

Let's start considering the expectation value of a single Wilson loop in the pure Gaussian model. This corresponds to set $\mathcal{S}_{\text{int}}=0$ in the definition \eqref{vevf}. As a consequence, given the factorization of the quiver theory in $q$ copies of the pure Gaussian model, the expectation value $\vev{\mathcal{W}_I}_q$ is simply replaced by $\vev{\mathcal{W}_I}_0$ where any information about the quiver structure of the theory is lost. Then, using the definition \eqref{Wlmm} and \eqref{rectn} we have
\begin{equation}\begin{split}\label{wgauss}
    \vev{\mathcal{W}_I}_0 &\equiv w_I({\color{blue}\lambda}_I,N)=
 \frac{1}{N} \sum_{\ell = 0}^{\infty}\frac{1}{\ell!} \parenth{\frac{{\color{blue}\lambda}_I}{2 N}}^{\frac{\ell}{2}}\,t_{[\ell]}\\
 &=1+\frac{(N^2\!-\!1){\color{blue}\lambda}_I}{8N^2}+\frac{(2N^4\!-\!5N^2\!+\!3){\color{blue}\lambda}_I^2}{384N^4}+\frac{(N^6\!-\!4N^4\!+\!6N^2\!-\!3){\color{blue}\lambda}_I^3}{9216N^6}+O({\color{blue}\lambda}_I^4)~,
\end{split}\end{equation}
where the $t$-functions are computed with the recursion \eqref{recursion}. Summing up all the terms, one obtains the exact formula for the circular Wilson loop in $\cN=4$ SYM in terms of Laguerre polynomials \cite{Drukker:2000rr,Pestun:2007rz}
\begin{equation}\label{exactWL}
w_I({\color{blue}\lambda}_I,N)= \frac{1}{N}\,L_{N-1}^{1}\Big(-\frac{{\color{blue}\lambda}_I}{4 N}\Big)\,\exp\left[\frac{{\color{blue}\lambda}_I}{8N}\Big(1-\frac{1}{N}\Big)\right]~,
\end{equation}
that, in the large $N$ limit, reduces to the well-known result in terms of the Bessel function \cite{Erickson:2000af}
\begin{equation}\label{exactWLlargeN}
w_I({\color{blue}\lambda}_I,N)\xrightarrow{N\rightarrow \infty} 
\frac{2}{\sqrt{{\color{blue}\lambda}_I}}\,I_{1}\left(\sqrt{{\color{blue}\lambda}_I}\right)~.
\end{equation}

A similar procedure can be repeated for the expectation value of multiple coincident Wilson loops \eqref{WvecImm}. However, in this case the computation is more subtle since we have to distinguish the occasions in which the loop operators belong to the same vector multiplet or not. Let's proceed considering the simplest example, \textit{i.e.} the expectation value of $\mathcal{W}_{[I,J]}$ defined in \eqref{Wlmultiplomm}. If the operators belong to two different nodes of the quiver, namely when $I\neq J$,  we have
\begin{equation}\begin{split}\label{factN4IneqJ}
    \vev{\mathcal{W}_{[I,J]}}_0 \equiv w_{[I,J]}({\color{blue}\lambda}_I,{\color{blue}\lambda}_J,N)=\!\!
 \left[ \frac{1}{N}\!\sum_{\ell_1 = 0}^{\infty}\frac{1}{\ell_1!} \parenth{\frac{{\color{blue}\lambda}_I}{2 N}}^{\frac{\ell_1}{2}}\!t_{[\ell_1]}\right]\!\!
 \left[ \frac{1}{N}\!\sum_{\ell_2 = 0}^{\infty}\frac{1}{\ell_2!} \parenth{\frac{{\color{blue}\lambda}_J}{2 N}}^{\frac{\ell_2}{2}}\!t_{[\ell_2]}\right]
\end{split}\end{equation}
 Then it's easy to conclude that the correlator factorizes in the following way
\begin{equation}
    w_{[I,J]}({\color{blue}\lambda}_I,{\color{blue}\lambda}_J,N)=w_{I}({\color{blue}\lambda}_I,N)w_{J}({\color{blue}\lambda}_J,N)~,\qquad\quad I\neq J~,
\end{equation}
in terms of the $\cN=4$ Wilson loop given by \eqref{exactWL} or equivalently by \eqref{exactWLlargeN} at large $N$.
On the other hand, when the two Wilson loops belong to the same vector multiplet labelled by $I$, we obtain
\begin{equation}\begin{split}\label{wIIexpansionN=4}
    \vev{\mathcal{W}_{[I,I]}}_0 \equiv& w_{[I,I]}({\color{blue}\lambda}_I,N)=\!\!
 \frac{1}{N^2}\sum_{\ell_1 = 0}^{\infty}\sum_{\ell_2 = 0}^{\infty}\frac{1}{\ell_1!\ell_2!} \parenth{\frac{{\color{blue}\lambda}_I}{2 N}}^{\frac{\ell_1+\ell_2}{2}}\!t_{[\ell_1,\ell_2]}\\
 =&1+\frac{(N^2\!-\!1){\color{blue}\lambda}_I}{4N^2}+\frac{5(N^2\!-\!1){\color{blue}\lambda}_I^2}{192N^2}+\frac{(7N^6\!+\!5N^4\!-\!60N^2\!+\!48){\color{blue}\lambda}_I^3}{4608N^6}+O({\color{blue}\lambda}_I^4)\,,
\end{split}\end{equation}
where the last line can be computed using the recursion \eqref{recursion}. This perturbative series is the weak coupling expansion of the $SU(N)$ version of the exact result found in \cite{Drukker:2000rr,Kawamoto:2008gp,Okuyama:2018aij} for two coincident circular Wilson loops in $\mathcal{N}=4$ SYM
\begin{equation}\begin{split}\label{wIIexactN=4}
w_{[I,I]}&({\color{blue}\lambda}_I,N)= \frac{1}{N^2}\;e^{\frac{{\color{blue}\lambda}_I}{2N}\;\left(1-\frac{1}{N}\right)}L_{N-1}^{1}\Big(-\frac{{\color{blue}\lambda}_I}{N}\Big)\,+\\
&\!\!+\frac{2}{N^2}e^{\frac{{\color{blue}\lambda}_I}{4N}\left(1-\frac{2}{N}\right)}\sum_{i=0}^{N-1}\sum_{j=0}^{i-1}\!\left[
L_{i}\Big(\!\!-\frac{{\color{blue}\lambda}_I}{4N}\Big)
L_{j}\Big(\!\!-\frac{{\color{blue}\lambda}_I}{4N}\Big)\!-
\frac{j!}{i!}\Big(\!\!-\frac{{\color{blue}\lambda}_I}{4N}\Big)^{i-j}
L_{j}^{i-j}\Big(\!\!-\frac{{\color{blue}\lambda}_I}{4N}\Big)^2\right]
\end{split}\end{equation}
where $L_i=L_i^0$. Notice that this result drastically simplifies in the large $N$ limit, showing a factorization property:
\begin{equation}\begin{split}
w_{[I,I]}({\color{blue}\lambda}_I,N)\xrightarrow{N\rightarrow \infty} 
w_I({\color{blue}\lambda}_I)^2=\frac{4}{{\color{blue}\lambda}_I}\,I_{1}\left(\sqrt{{\color{blue}\lambda}_I}\right)^2~.
\end{split}\end{equation}

Finally we can generalize this example for any length of the vector $\vec{I}$
\begin{equation}\label{factorizedwvecIN4}
w_{\vec{I}}=\prod_{i=\;\substack{\text{repeated} \\
\text{nodes in } \vec{I}}} w_{[\underbrace{I_i,...,I_i}_{\substack{\text{\texttt{\#} of times it}\\\text{appears in $\vec{I}$}}}]}({\color{blue}\lambda}_{I_i},N)
\prod_{j=\;\substack{\text{non-repeated} \\
\text{nodes in } \vec{I}}} w_{I_j}({\color{blue}\lambda}_{I_j},N)~,
\end{equation}
where all the $w$ functions appearing in the right-hand side of the formula are the $\cN=4$ SYM results. For instance $w_{[I_1,I_2,I_2,I_3]}=w_{I_1}w_{[I_2,I_2]}w_{I_3}$. In the large $N$ limit the result is even simpler since all the $w$'s factorize\footnote{This fact could be easily explained by implementing recursion relations directly in the large $N$ regime, see \cite{Beccaria:2020hgy} for further details.}, then
\begin{equation}\begin{split}\label{factorizationwlargeN}
w_{\vec{I}}\xrightarrow{N\rightarrow \infty} 
\prod_{i=1}^n w_{I_i}({\color{blue}\lambda}_{I_i})~,
\end{split}\end{equation}
where $\vec{I}=[I_1,...,I_n]$. In the following sections we will see how those quantities behave when we turn on the interactions action $\mathcal{S}_{\text{int}}$.

\subsection{Correlators in SCQCD}\label{sec:WLSCQCD}

The simplest quiver theory we can study is SCQCD, which has a single gauge node with $2N$ fundamental hypermultiplets as matter content. 
In this framework we need to study Wilson loop vevs in presence of a non-trivial interaction action \eqref{S_int}, setting to zero all the couplings except for ${\color{blue}\lambda}_1$.

\paragraph{Wilson loop vev.} The first observable we compute is the expectation value of the circular Wilson loop \eqref{vevWmm}. Since the theory is built on a quiver with only one node, the only possible value for $I$ is 1.  Then, using the definition \eqref{vevf} we have
\begin{equation}\begin{split}\label{w11SCQCD}
w_1^{(1)}({\color{blue}\lambda}_1,N)=
 \frac{1}{N} \sum_{\ell = 0}^{\infty}&\frac{1}{\ell!} \parenth{\frac{{\color{blue}\lambda}_1}{2 N}}^{\frac{\ell}{2}}\left[t_{[\ell]}
 +\frac{3{\color{blue}\lambda}_1^2{\color{red}\zeta}_3}{64\pi^4N^2}(t_{[\ell]}t_{[2,2]}-t_{[\ell,2,2]})\right.\\
 &\left.-\frac{5{\color{blue}\lambda}_1^3{\color{red}\zeta}_5}{768\pi^6N^3}\left(t_{[\ell]}(3t_{[2,4]}-2t_{[3,3]})-3t_{[\ell,2,4]}+2t_{[\ell,3,3]}\right)+O({\color{blue}\lambda}_1^4)\right]
\end{split}\end{equation}
where ${\color{red}\zeta}_n$ are the Riemann zetas.
The naive way to proceed is first to fix the value of $\ell$ in the sum and then to compute the $t$-functions using the recursion relation \eqref{recursion}. In this way it is possible to systematically expand $w_1^{(1)}$ in the coupling ${\color{blue}\lambda}_1$ setting a high enough cut off for the sum. 
Indeed, even if the perturbative expansion in the square brackets corresponds to the honest weak coupling expansion of the path-integral \eqref{vevf}, the presence of ${\color{blue}\lambda}_1^{\ell/2}$ and the sum in front of it
mixes up all the powers of ${\color{blue}\lambda}_1$. Then, the first few orders reads
\begin{equation}\footnotesize\begin{split}
&w_1^{(1)}({\color{blue}\lambda}_1,N)\!=\!
 1\!+\!\frac{(N^2\!-\!1){\color{blue}\lambda}_1}{8N^2}
 \!+\!\frac{(2N^4\!-\!5N^2\!+\!3){\color{blue}\lambda}_1^2}{384N^4}
\!+\!\frac{\!(N^2\!-\!1)(N^4\!-\!3N^2\!+\!3\!-\!\frac{54N^2}{\pi^4}(N^2\!+\!1){\color{red}\zeta}_3){\color{blue}\lambda}_1^3}{9216N^6}\\
&\!+\!\frac{(N^2\!-\!1)(2N^6\!\!-\!8N^4\!\!+\!15N^2\!\!\!-\!15\!-\!\frac{360N^2}{\pi^4}(2N^4\!\!+\!N^2\!\!-\!6){\color{red}\zeta}_3\!+\!
\frac{2700N^2}{\pi^6}(2N^4\!\!+\!N^2\!\!-\!1){\color{red}\zeta}_5){\color{blue}\lambda}_1^4}{1474560N^8}\!+\!O({\color{blue}\lambda}_1^5)\,.
\end{split}\end{equation}
However, it is convenient to use a different approach. Let's leave the sum in \eqref{w11SCQCD} aside for a moment and use the recursion relation \eqref{recursion} until the $t$-functions depending on $\ell$ are reduced to have a single index as follows
\begin{equation}\small\begin{split}\label{texamplesqcd}
   t_{[\ell,2,2]}&\!\!=\tfrac{1}{4}((\ell+N^2)^2-1)t_{[\ell]},\\
   t_{[\ell,2,4]}&\!\!=\!\tfrac{(\ell\!+\!N^2\!+\!3)}{16 N^2}\!\Big[\ell(\ell\!-\!1)
   (\ell\!+\!N^2\!\!+\!3)t_{[\ell-2]}\!+\!2N(\!(2\ell\!-\!5)N^2\!\!-\!(\ell\!-\!1)(\ell\!+\!3)\!+\!2N^4)t_{[\ell]}\!+\!2\ell N t_{[\ell+2]}\!\Big]\\
   t_{[\ell,3,3]}&\!\!=\tfrac{1}{16 N^2}\Big[\ell(\ell\!-\!1)
   (\ell\!+\!N^2\!-\!3)(\ell\!+\!N^2\!+\!4)t_{[\ell-2]}\!+\!2N(3(N^4\!-\!5N^2\!+\!4)\\
   &\qquad\qquad\qquad\qquad\qquad\qquad\qquad\qquad\quad-\!2\ell(\ell(\ell\!+\!N^2\!+\!3)\!+\!2))t_{[\ell]}\!+\!2\ell (\ell\!+\!4) N t_{[\ell+2]}\Big]~.
\end{split}\end{equation}
Let's now turn on the sum and see how it acts on the functions \eqref{texamplesqcd}. Shifting $\ell$ in a suitable way, it is possible to reduce any term to be proportional to $t_{[\ell]}$ multiplied to some polynomial in $\ell$. Then, recalling the definition \eqref{wgauss}, it is possible to solve the archetype of these terms using 
\begin{equation}\label{derdef}
\frac{1}{N} \sum_{\ell = 0}^{\infty}\frac{1}{\ell!} \parenth{\frac{{\color{blue}\lambda}_1}{2 N}}^{\frac{\ell}{2}}\;\ell^n\;t_{[\ell]}=2^n [{\color{blue}\lambda}_1\partial_{1}]^n w_1({\color{blue}\lambda}_1,N)~,
\end{equation}
where $\partial_{1}X=dX/d{\color{blue}\lambda}_1$ and $w_1$ is the exact vev of the $\mathcal{N}=4$ SYM circular Wilson loop \eqref{exactWL} (or \eqref{exactWLlargeN} in the large $N$ case). In \eqref{derdef} the power of the differential operator represents the number of nested applications of such operator to the function $w_1$. For instance, using \eqref{derdef} and the first line of \eqref{texamplesqcd}, the term proportional to $t_{[\ell,2,2]}$ in \eqref{w11SCQCD} becomes
\begin{equation}\label{tlexample}
\frac{1}{N} \sum_{\ell = 0}^{\infty}\frac{1}{\ell!} \parenth{\frac{{\color{blue}\lambda}_1}{2 N}}^{\frac{\ell}{2}}t_{[\ell,2,2]}=
{\color{blue}\lambda}_1^2\partial_{1}^2w_1+(N^2+1){\color{blue}\lambda}_1\partial_{1}w_1+\frac{N^4-1}{4}w_1~,
\end{equation}
where we already solved the chain rules for derivatives that now act directly on $w_1$. Repeating the same procedure for all the $t$-functions appearing in the expansion of $w_1^{(1)}$, we obtain 
\begin{footnotesize}\begin{align}\label{WL2N}
&w_1^{(1)}\!=w_1\!-\!\frac{3 {\color{red}\zeta}_3 {\color{blue}\lambda}_1^3}{64 \pi ^4 N^2} \bigg[{\color{blue}\lambda}_1 \partial_1^2w_1\!+\!\left(N^2\!+\!1\right) \partial_1w_1\bigg]+\frac{5 {\color{red}\zeta}_5 {\color{blue}\lambda}_1 ^4 }{96\pi ^6N^6}\bigg[
{\color{blue}\lambda}_1^2  N^4 \partial_1^4w_1\!+\! \frac{{\color{blue}\lambda}_1  N^2}{8}\!\left({\color{blue}\lambda}_1\!+\!12 N^4\!+\!32 N^2\right)\! \partial_1^3w_1\nonumber\\
&+\frac{1}{64}\left(\left({\color{blue}\lambda}_1 ^2+144 N^6+4 (7 {\color{blue}\lambda}_1 +36) N^4+12 {\color{blue}\lambda}_1  N^2\right)\partial_1^2w_1+\left(2 {\color{blue}\lambda}_1 +24 N^6+22 N^4+({\color{blue}\lambda}_1 -6) N^2\right) \partial_1w_1\right)\nonumber\\
&+\frac{1}{256}\left(N^4+2 N^2-3\right) w_1\bigg]+
\frac{9{\color{red}\zeta}_3^2 {\color{blue}\lambda}_1^5}{8192 \pi ^8 N^4} \bigg[
{\color{blue}\lambda}_1^2 \left({\color{blue}\lambda}_1 \partial_1^4w_1+2 \left(N^2+5\right) \partial_1^3w_1\right)+
{\color{blue}\lambda}_1 \left(N^4+12 N^2+23\right)\partial_1^2w_1\nonumber\\
&+4 \left(N^4+3 N^2+2\right) \partial_1w_1\bigg]
+ \dots
\end{align}\end{footnotesize}
showing only the first few orders.
The original expansion in \eqref{w11SCQCD} can now be interpreted as an expansion in transcendentality where dots represent higher transcendentality terms. Since $\mathcal{S}_{\text{int}}$ contains only odd zetas, for a given transcendentality $\tau$ the only combinations that can appear in $w_1^{(1)}$ are the integer partitions of $\tau$ in odd numbers. For instance for $\tau=7$ we only have ${\color{red}\zeta}_7$, $\tau=8\rightarrow {\color{red}\zeta}_3{\color{red}\zeta}_5$, $\tau=9\rightarrow \{{\color{red}\zeta}_3^2,{\color{red}\zeta}_9\}$, $\tau=10\rightarrow \{{\color{red}\zeta}_3{\color{red}\zeta}_7,{\color{red}\zeta}_5^2\}$ and so on.
The crucial point of this approach is that we end up with an expression for $w_1^{(1)}$ that is exact in the coupling for fixed transcendentality! 

We can drastically simplify the result considering the large $N$ limit as follows
\begin{equation}\footnotesize\begin{split}\label{WL2NlargeN}
&w_1^{(1)}\xrightarrow{N\rightarrow \infty}\;w_1
-\frac{3 {\color{red}\zeta}_3 {\color{blue}\lambda}_1^3}{(8 \pi ^2)^2}\partial_1w_1
+\frac{10 {\color{red}\zeta}_5 {\color{blue}\lambda}_1 ^4 }{(8\pi ^2)^3}\bigg[
4{\color{blue}\lambda}_1\partial_1^3w_1+6\partial_1^2w_1+\partial_1w_1\bigg]
+\frac{9 {\color{red}\zeta}_3^2 {\color{blue}\lambda}_1 ^5 }{2(8\pi^2) ^4}\bigg[
{\color{blue}\lambda}_1\partial_1^2w_1+4\partial_1w_1\bigg]\\
&-\frac{7 {\color{red}\zeta}_7 {\color{blue}\lambda}_1 ^5 }{4(8\pi^2) ^4}\bigg[
256{\color{blue}\lambda}_1({\color{blue}\lambda}_1\partial_1^5w_1+5\partial_1^4w_1)+16(7{\color{blue}\lambda}_1+60)\partial_1^3w_1+168\partial_1^2w_1+21\partial_1w_1\bigg]\\
&-\!\frac{15 {\color{red}\zeta}_3{\color{red}\zeta}_5 {\color{blue}\lambda}_1 ^6 }{(8\pi^2)^5}\!\bigg[
4{\color{blue}\lambda}_1(2{\color{blue}\lambda}_1\partial_1^4w_1\!+\!13\partial_1^3w_1)\!+\!2({\color{blue}\lambda}_1\!\!+\!24)\partial_1^2w_1\!+\!11\partial_1w_1\!\bigg]
\!-\!\frac{9 {\color{red}\zeta}_3^3 {\color{blue}\lambda}_1 ^7 }{2(8\pi^2) ^6}\!\bigg[
{\color{blue}\lambda}_1^2\partial_1^3w_1\!+\!12{\color{blue}\lambda}_1\partial_1^2w_1\!+\!30\partial_1w_1\!\bigg]
\!\!+\!...
\end{split}\raisetag{44pt}
\end{equation}\normalsize
with $w_1$ the exact vev of the $\cN=4$ SYM circular Wilson loop at large $N$ \eqref{exactWLlargeN}. Notice that the first term in both \eqref{WL2N} and \eqref{WL2NlargeN} is simply given by the $\cN=4$ vev that, indeed, it corresponds to the zero-th order of the expansion of the interaction action \eqref{S_int} as in the pure Gaussian model described in section \ref{sec:WN4}. \\
Finally, let's concentrate on the terms in \eqref{WL2NlargeN} proportional to pure powers of ${\color{red}\zeta}_3$. Analyzing the pattern appearing in the expansion, it is possible to resum those terms obtaining the following compact expression 
\begin{equation}\label{expresum1}
    w_1^{(1)}\bigg|_{{\color{red}\zeta}_3^m}=
    \exp\left[{-\frac{1+6\left(\tfrac{{\color{blue}\lambda}_1}{8\pi^2}\right)^2{\color{red}\zeta}_3-\sqrt{1+12\left(\tfrac{{\color{blue}\lambda}_1}{8\pi^2}\right)^2{\color{red}\zeta}_3}}{6\left(\tfrac{{\color{blue}\lambda}_1}{8\pi^2}\right)^2{\color{red}\zeta}_3}}\;\mathcal{D}_1\right]\;w_1~,
\end{equation}
where the differential operator $\mathcal{D}_1\equiv{\color{blue}\lambda}_1\partial_1$ and its coefficient comes from the resummation of $\sum_{m=1}^\infty C_m \left(-\frac{3{\color{blue}\lambda}_1^2{\color{red}\zeta}_3}{64\pi^4}\right)^m$ with $C_m$ being the Catalan numbers. In order to recover the expansion, one needs to expand \eqref{expresum1} for small ${\color{blue}\lambda}_1$ and then substitute the action of the differential operator on $w_1$ with $\mathcal{D}^m_1w_1={\color{blue}\lambda}^m_1\partial_1^mw_1$. Such exponentiation property enjoys a nice diagrammatic interpretation, discussed in section \ref{FT:SCQCD}.

\paragraph{Multiple coincident Wilson loops.} The other observable we want to compute is the correlation function of multiple Wilson loops defined in \eqref{WvecImm}. In SCQCD $\vec{I}$ is constrained to be a vector of only ones. Let's start with the simplest one $\vec{I}=[1,1]$ and then we generalize for any number of coincident loops. Expanding the matrix model we have
\begin{equation}\begin{split}\label{w1_11SCQCD}
w_{[1,1]}^{(1)}({\color{blue}\lambda}_1,N)=&
 \frac{1}{N^2} \!\sum_{\ell_1,\,\ell_2}\!\frac{1}{\ell_1!\ell_2!}\! \parenth{\frac{{\color{blue}\lambda}_1}{2 N}}^{\!\frac{\ell_1+\ell_2}{2}}\!\left[t_{[\ell_1,\ell_2]}
 +\frac{3{\color{blue}\lambda}_1^2{\color{red}\zeta}_3}{64\pi^4N^2}(t_{[\ell_1,\ell_2]}t_{[2,2]}-t_{[\ell_1,\ell_2,2,2]})\right.\\
 &\!\!\!\!\!\!\!\left.-\frac{5{\color{blue}\lambda}_1^3{\color{red}\zeta}_5}{768\pi^6N^3}\left(t_{[\ell_1,\ell_2]}(3t_{[2,4]}-2t_{[3,3]})-3t_{[\ell_1,\ell_2,2,4]}+2t_{[\ell_1,\ell_2,3,3]}\right)+O({\color{blue}\lambda}_1^4)\right]
\end{split}\end{equation}
This case is way more tricky than the one studied above. Indeed, repeating the procedure introduced for $w_1^{(1)}$, one can find that the double sum of some $t$-functions appearing in \eqref{w1_11SCQCD} cannot be written in terms of derivatives of the $\mathcal{N}=4$ SYM results.
This is due to the fact that the coupling appears with the power $(\ell_1+\ell_2)$ but the recursion relation \eqref{recursion} doesn't always produce a result proportional to $(\ell_1+\ell_2)^k t_{[\ell_1,\ell_2]}$ that can be traded with $k$ derivatives respect to the coupling. This is the case for instance of the sum of $t_{[\ell_1,\ell_2,2,4]}$ and $t_{[\ell_1,\ell_2,3,3]}$.

However, there are some interesting exceptions. The first term in \eqref{w1_11SCQCD} can be immediately recognized to be the $\cN=4$ result given by \eqref{wIIexpansionN=4} and \eqref{wIIexactN=4}. The term proportional to $t_{[\ell_1,\ell_2,2,2]}$ can also be computed exactly since
\begin{equation}\label{tlexample2indices}
\!\sum_{\ell_1,\,\ell_2}\!\frac{1}{\ell_1!\ell_2!}\! \parenth{\frac{{\color{blue}\lambda}_1}{2 N}}^{\!\frac{\ell_1+\ell_2}{2}}t_{[\ell_1,\ell_2,2,2]}
=
{\color{blue}\lambda}_1^2\partial_{1}^2w_{[1,1]}+(N^2+1){\color{blue}\lambda}_1\partial_{1}w_{[1,1]}+\frac{N^4-1}{4}w_{[1,1]}~,
\end{equation}
where $w_{[1,1]}$ is again given by \eqref{wIIexactN=4}.
But there is more. Indeed, all the coefficients of pure powers of ${\color{red}\zeta}_3$ depend on the following $t$-functions
\begin{equation}\label{t222222}
t_{[\ell_1,\ell_2,...,\ell_n,\underbrace{2,2,...,2}_{\text{$m$-times}}]}=\frac{1}{2^m}\prod_{k=1}^m(\sum_{i=0}^n\ell_i+N^2+2k-3)\;t_{[\ell_1,\ell_2,...,\ell_n]}~,
\end{equation}
that, for instance in the case $n=2$, can be systematically written in terms of derivatives of $w_{[1,1]}$ with respect to the coupling. Moreover, as long as we substitute $w_1$ with $w_{[1,1]}$, the resulting expansion is the same as the one appearing in the vev of the single Wilson loop and at large $N$ it can be resummed as in \eqref{expresum1}.
Finally, even if we cannot compute them exactly, the remaining terms that are not proportional to ${\color{red}\zeta}_3$ can still be studied perturbatively cutting the sums at a high enough value and evaluating the resulting $t$-functions with \eqref{recursion}. For instance, the transcendentality ${\color{red}\zeta}_5$ term reads
\begin{equation}\footnotesize\begin{split}
    &w_{[1,1]}^{(1)}\bigg|_{{\color{red}\zeta}_5}=
    \frac{5(N^2-1){\color{red}\zeta}_5{\color{blue}\lambda}_1^3}{768\pi^6N^3}
    \bigg[\frac{9(N^2+1)(2N^2-1){\color{blue}\lambda}_1}{16N^3}+
    \frac{(31N^2(N^2+2)-9){\color{blue}\lambda}_1^2}{128N^3}\\
    &+\!\frac{\!\!(224N^8\!\!+\!1245N^6\!\!-\!1049N^4\!-\!3540N^2\!+\!4240){\color{blue}\lambda}_1^3}{10240N^7}
    \!+\!\frac{\!\!(23N^{10}\!\!+\!278N^8\!\!-\!433N^6\!\!-\!1163N^4\!-\!4360N^2\!+\!4640){\color{blue}\lambda}_1^4}{20480N^9}\!+\!...\!\bigg],
\end{split}\raisetag{46pt}\normalsize\end{equation}
as well as the transcendentality ${\color{red}\zeta}_7$
\begin{equation}\footnotesize\begin{split}
    w_{[1,1]}^{(1)}\bigg|_{{\color{red}\zeta}_7}=
    &-\frac{7(N^2-1){\color{red}\zeta}_7{\color{blue}\lambda}_1^4}{8192\pi^8N^4}
    \bigg[\frac{5(8N^6+4N^4-3N^2+3){\color{blue}\lambda}_1}{8N^4}+
    \frac{(422N^6+1105N^4+153N^2+60){\color{blue}\lambda}_1^2}{384N^4}\\
    &\qquad\qquad\qquad-\frac{(311N^{10}+2206N^8-1032N^6-3837N^4+5580N^2-6960){\color{blue}\lambda}_1^3}{3072N^8}
    +...\bigg]~,
\end{split}\raisetag{46pt}\normalsize\end{equation}
and so on. 

The generalization to any number $n$ of coincident Wilson loops is pretty straightforward. Using the definition \eqref{Wlmultiplomm}, one can expand the vev as in \eqref{w1_11SCQCD} but this time with $n$ sums. All the terms proportional to powers of ${\color{red}\zeta}_3$ contains combinations of the $t$-functions \eqref{tlexample2indices}.
Then, using the technique presented above, one can compute them exactly finding the same structures appearing in \eqref{WL2N} but with the substitution $w_1\rightarrow w_{[1,1,...,1]}$. At large $N$, we can resum them obtaining
\begin{equation}\label{expresumgeneral}
    w_{[\underbrace{1,1,...,1}_{\text{$n$-times}}]}^{(1)}\bigg|_{{\color{red}\zeta}_3^m}=
    \exp\left[{-\frac{1+6\left(\tfrac{{\color{blue}\lambda}_1}{8\pi^2}\right)^2{\color{red}\zeta}_3-\sqrt{1+12\left(\tfrac{{\color{blue}\lambda}_1}{8\pi^2}\right)^2{\color{red}\zeta}_3}}{6\left(\tfrac{{\color{blue}\lambda}_1}{8\pi^2}\right)^2{\color{red}\zeta}_3}}\;\mathcal{D}_1\right]\;w_1^n~,
\end{equation}
with the same differential operator of \eqref{expresum1} acting on $w_{[1,1,...,1]}=w_1^n$ where we used the factorisation property given in \eqref{factorizationwlargeN}. Similarly to the example above, all the other terms in transcendentality can be analyzed perturbatively.

\subsection{Correlators in the $A_{q-1}$ theories}\label{Sec4.3}

In this section we generalize the procedure introduced in section \ref{sec:WLSCQCD} to the general quiver theories $A_{q-1}$ with number of nodes $q\geq 2$. 

\paragraph{Wilson loop vev.} Let's consider first the expectation value of the circular Wilson loop belonging to a vector multiplet labelled by $I$ as defined in \eqref{vevWmm}. Since the theories $A_{q-1}$ are invariant under cyclic reparametrisations of the node labels, one can always reduce the computation of $w_I^{(q)}$ to $w_1^{(q)}$ with a suitable change of the couplings. Then, using the definition \eqref{vevf} and the interaction action \eqref{S_int} we find:
\begin{equation}\small\begin{split}\label{w11generalq}
w_1^{(q)}\!\!=
 \!\frac{1}{N}\! \sum_{\ell = 0}^{\infty}&\frac{1}{\ell!}\! \parenth{\!\frac{{\color{blue}\lambda}_1}{2 N}\!}^{\!\!\frac{\ell}{2}}\!\!\left[t_{[\ell]}
 \!+\!\frac{3{\color{blue}\lambda}_1{\color{red}\zeta}_3}{64\pi^4N^2}\!\left[(t_{[\ell]}t_{[2,2]}-t_{[\ell,2,2]}){\color{blue}\lambda}_1\!-\!(t_{[\ell]}t_{[2]}-t_{[\ell,2]})t_{[2]}({\color{blue}\lambda}_2\!+\!{\color{blue}\lambda}_q)\right]\!+\!...\right]
\end{split}\end{equation}
where the dots stand for higher orders in all the couplings ${\color{blue}\lambda}_I$ with $I=1,...,q$. From the expansion \eqref{w11generalq}, it is clear that at the first transcendental order ${\color{red}\zeta}_3$ the operator in the node 1 only interacts with the neighboring nodes 2 and $q$ through the bi-fundamental hypers. Defining the distance between two nodes $I$ and $J$ as
\begin{equation}\label{distance}
    d(q,I,J)\equiv |||J-I|-q/2|-q/2|~,
\end{equation}
new nodes with increasing distance will contribute at higher orders in transcendentality.  Consequently, increasing the perturbative order, the results become more and more involved, then in this section we show only the first few orders of the most interesting cases.

Computing the $t$-functions of \eqref{w11generalq} using the recursion relation \eqref{recursion} and then performing the sum, we obtain
\begin{equation}\footnotesize\begin{split}\label{w1q}
&w_{1}^{(q)}=
    w_1
    -\frac{3{\color{red}\zeta}_3{\color{blue}\lambda}_1^2}{128\pi^4N^2}
    \bigg[2{\color{blue}\lambda}_1^2\partial_1^2w_1+(2(N^2+1){\color{blue}\lambda}_1-(N^2-1)({\color{blue}\lambda}_2+{\color{blue}\lambda}_q))\partial_1w_1\bigg]
    +\frac{5{\color{red}\zeta}_5{\color{blue}\lambda}_1^2}{49152\pi^6N^6}\bigg[
    512 N^4 {\color{blue}\lambda} _{1}^4 \partial_1^4w_1\\
    &+8 N^2 {\color{blue}\lambda} _{1}^2 \left(4 \!\left(3 N^2\!+\!8\right) N^2 {\color{blue}\lambda} _{1}\!-\!6 \left(N^2\!-\!1\right) \!N^2\! \left({\color{blue}\lambda} _{2}\!+\!{\color{blue}\lambda} _{q}\right)\!+\!{\color{blue}\lambda} _{1}^2\right) \partial^3_1w_1\!+\!(4 \left(7 N^2\!+\!3\right) \!N^2 {\color{blue}\lambda} _{1}^2\!-\!72 \left(N^2\!-\!1\right) \!N^4 \!\left({\color{blue}\lambda} _{2}\!+\!{\color{blue}\lambda} _{q}\right)\\
    &+6 N^2 {\color{blue}\lambda} _{1} \left(\left(N^2\!-\!1\right) ({\color{blue}\lambda} _{2}\!+\!{\color{blue}\lambda} _{q})\!+\!24 \left(N^4\!+\!N^2\right)\right)\!+\!{\color{blue}\lambda} _{1}^3) \partial_1^2w_1
    \!+\!2 (4 \left(N^2\!+\!2\right) {\color{blue}\lambda} _{1}^3\!-\!12 N^2\! \left(2 N^4\!-\!5 N^2\!+\!3\right)\! \left({\color{blue}\lambda} _{2}^2\!+\!{\color{blue}\lambda} _{q}^2\right)\\
    &-{\color{blue}\lambda} _{1}^2 (3 \left(N^2-1\right) {\color{blue}\lambda} _{2}+3 \!\left(N^2\!\!-\!1\right) {\color{blue}\lambda} _{q}\!-\!8\! \left(12 N^4\!+\!11 N^2\!-\!3\right) \!N^2)\!-\!24 N^2\! \left(N^4\!\!-\!3 N^2\!+\!2\right)\! \left({\color{blue}\lambda} _{2}\!+\!{\color{blue}\lambda} _{q}\right) {\color{blue}\lambda} _{1}\!) \partial_1w_1\\
    &+\left(N^2-1\right) {\color{blue}\lambda} _{1} \left(2 \left(N^2+3\right) {\color{blue}\lambda} _{1}-3 \left(N^2-1\right) \left({\color{blue}\lambda} _{2}+{\color{blue}\lambda} _{q}\right)\right) w_1
    \bigg]+...
\end{split}\raisetag{17pt}\normalsize\end{equation}
where $w_1({\color{blue}\lambda}_1,N)$ is the Wilson loop expectation value of $\mathcal{N}=4$ SYM given by \eqref{exactWL}.
At large $N$ the result drastically simplifies and reads
\begin{equation}\footnotesize\begin{split}\label{WqLargeN}
w_{1}^{(q)}\xrightarrow{N\rightarrow \infty}&
    \;w_1
    -\frac{3{\color{red}\zeta}_3{\color{blue}\lambda}_1^2(2{\color{blue}\lambda}_1-{\color{blue}\lambda}_{2}-{\color{blue}\lambda}_{q})\partial_1w_1}{128\pi^4}
    +\frac{5{\color{red}\zeta}_5{\color{blue}\lambda}_1^2}{1024\pi^6}\bigg[4{\color{blue}\lambda}_1(2{\color{blue}\lambda}_1-{\color{blue}\lambda}_{2}-{\color{blue}\lambda}_{q})(2{\color{blue}\lambda}_1\partial_1^3w_1+3\partial_1^2w_1)\\
    &+(4{\color{blue}\lambda}_1^2-({\color{blue}\lambda}_{2}+{\color{blue}\lambda}_{q}){\color{blue}\lambda}_1-{\color{blue}\lambda}_{2}^2-{\color{blue}\lambda}_{q}^2)\partial_1w_1\bigg]
    +\frac{9{\color{red}\zeta}_3^2{\color{blue}\lambda}_1^2}{32768\pi^8}\bigg[
    (2{\color{blue}\lambda}_1-{\color{blue}\lambda}_{2}-{\color{blue}\lambda}_{q})^2{\color{blue}\lambda}_1^2\partial_1^2w_1+2(8{\color{blue}\lambda}_1^3\\
    &-6({\color{blue}\lambda}_{2}+{\color{blue}\lambda}_{q}){\color{blue}\lambda}_{1}^2+2({\color{blue}\lambda}_{2}^2+{\color{blue}\lambda}_{2}{\color{blue}\lambda}_{q}+{\color{blue}\lambda}_{q}^2){\color{blue}\lambda}_{1}-2{\color{blue}\lambda}_{2}^3+({\color{blue}\lambda}_{q-1}-2{\color{blue}\lambda}_{q}){\color{blue}\lambda}_{q}^2+{\color{blue}\lambda}_{2}^2{\color{blue}\lambda}_{3})\partial_1w_1\bigg]+...
\end{split}\normalsize\end{equation}
Notice that at ${\color{red}\zeta}_3^2$ order the couplings ${\color{blue}\lambda}_3$ and ${\color{blue}\lambda}_{q-1}$ appear, confirming the interaction between the Wilson loop in node 1 and the nodes 3 and $q-1$. The SCQCD results \eqref{WL2N} and \eqref{WL2NlargeN} can be recovered from expansions \eqref{w1q} and \eqref{WqLargeN} by setting all the couplings but ${\color{blue}\lambda}_1$ to zero. 

Analyzing the Wilson loop vev in SCQCD and in the general $A_{q-1}$ theories, we see that both significantly deviates from the $\cN=4$ vev even in the large $N$ limit. However, something different happens for $A_{q-1}$ theories. Evaluating \eqref{WqLargeN} at the orbifold point, namely ${\color{blue}\lambda}_I={\color{blue}\lambda}$ for all $I=1,...,q$, all the terms in the transcendentality expansion vanish and only the very first one will survive such that
\begin{equation}\label{wqvanishes}
    w_{1}^{(q)}\bigg|_{{\color{blue}\lambda}_I = {\color{blue}\lambda}}\xrightarrow{N\rightarrow \infty} w_1({\color{blue}\lambda}) =\frac{2}{\sqrt{{\color{blue}\lambda}}}I_1(\sqrt{{\color{blue}\lambda}})~, \hspace{1cm} q\neq 1~.
\end{equation}
The Wilson loop vev in $A_{q-1}$ theories at the orbifold point does not deviate from the $\cN=4$ exact result \eqref{exactWLlargeN}. We specified $q\neq 1$ since such result does not hold for SCQCD (see equation \eqref{WL2NlargeN}).
We showed this fact starting from a weak coupling expansion, and in section \ref{FT:Aq} we shall see direct cancellations at the level of Feynman diagrams, proving \eqref{wqvanishes}. Besides, this result has been confirmed at strong coupling as well \cite{Rey:2010ry}.
Such disparity between the SCQCD \eqref{WL2NlargeN} and the quiver theories \eqref{WqLargeN} results  is crucial in the problem of extending the holographic duality to $\cN=2$ theories. Hence, in the context of $\cN=2$ theories, $A_{q-1}$ quivers represent the ideal bridge between $\cN=4$ SYM and the SCQCD. 
Besides, there exists a way to probe the difference between $\cN=4$ and $A_{q-1}$ theories using our techniques. In section \ref{sec:TwistUntw} we will explore this possibility studying some observables belonging to the twisted sector under the action of $\mathbb{Z}_q$. 

\paragraph{Multiple coincident Wilson loops.} Let's consider now the correlation functions of multiple Wilson loops defined in \eqref{WvecImm}. In the general theory with $q$ vector multiplets, the possible choices of the vector $\vec{I}$ are endless. In particular in this framework we can consider not only Wilson loops in the same node, but also Wilson loops belonging to multiple different nodes. Again, here we present only the most significant cases and we refer to the attached notebook \texttt{WLcorrelators.nb} for a more in-depth analysis.

\begin{figure}[!t]
\begin{minipage}[t]{.32\textwidth}
        \centering
        \includegraphics[width=\textwidth]{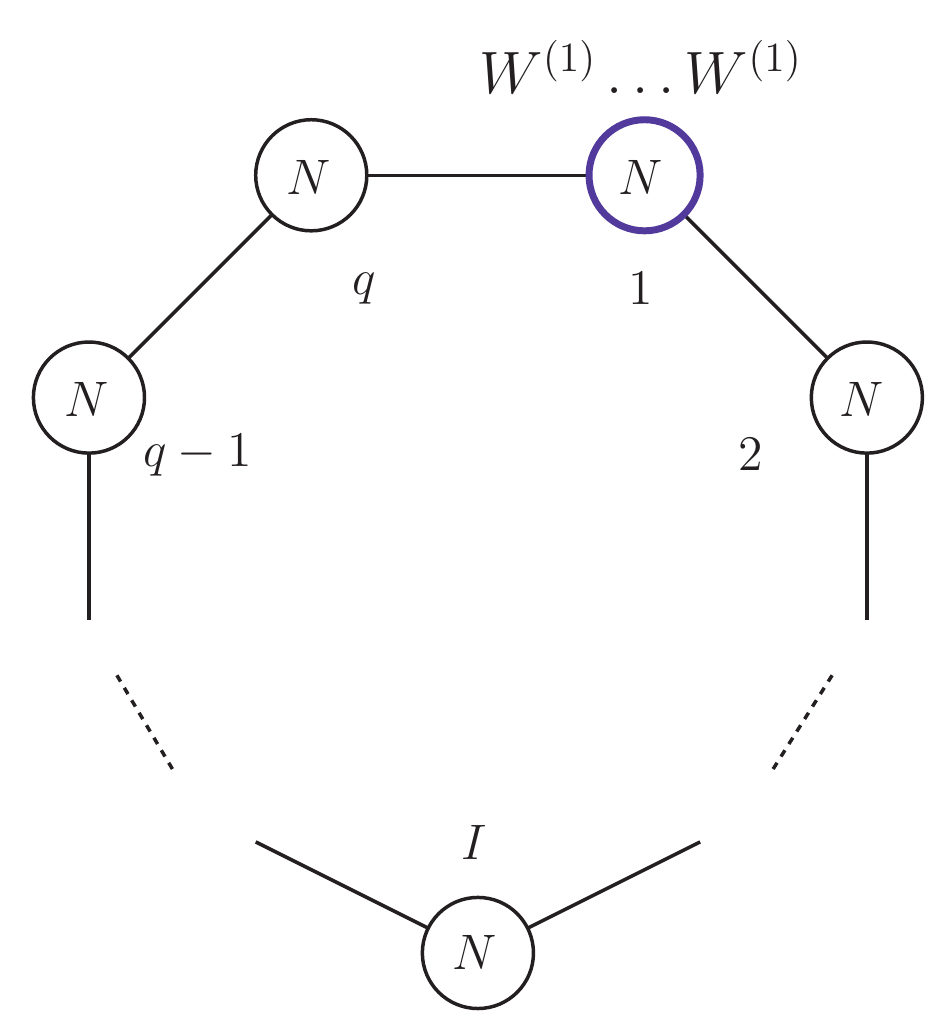}
        \subcaption{Observable  $w_{[1,1,...,1]}^{(q)}$}\label{Fig:quiverSameWL}
    \end{minipage}
    \hfill
    \begin{minipage}[t]{.32\textwidth}
        \centering
        \includegraphics[width=\textwidth]{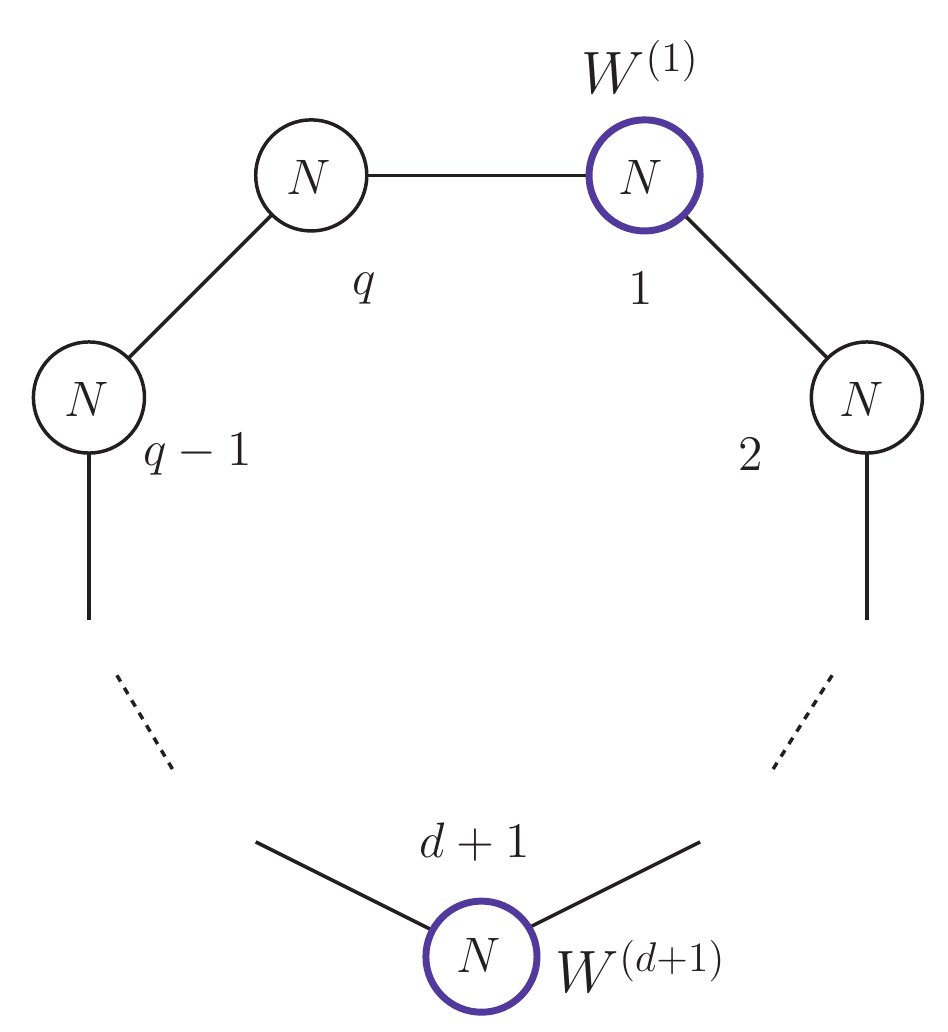}
        \subcaption{Observable $ w_{[1,d+1]}^{(q)}$}\label{fig:3a}
    \end{minipage}
    \hfill
    \begin{minipage}[t]{.32\textwidth}
        \centering
        \includegraphics[width=\textwidth]{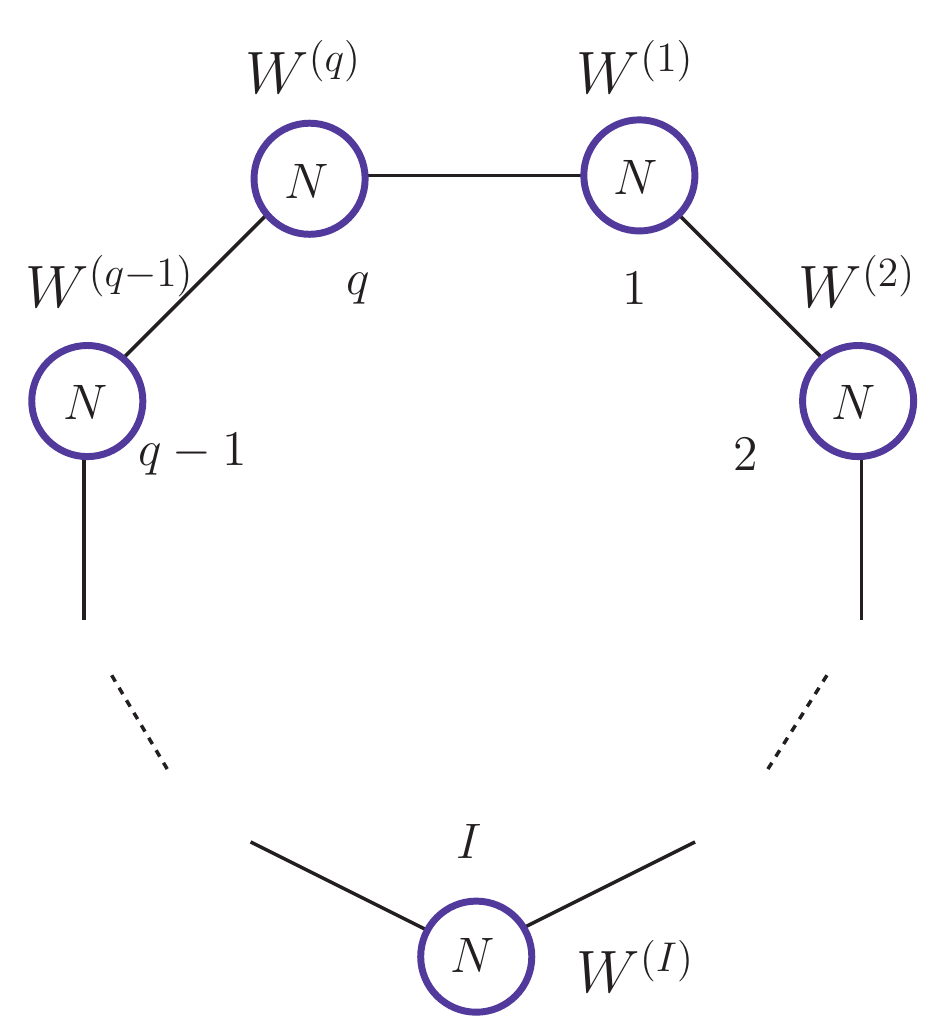}
        \subcaption{Observable $ w_{[1,2,\dots,q]}^{(q)}$}\label{fig:3b}
    \end{minipage}  
    \caption{Multiple insertions of coincident Wilson loops in different nodes. On the left multiple Wilson loops in the same node, at the center two Wilson loops in two nodes with a distance $d$, on the right $q$ Wilson loops, all belonging to different nodes.}
\end{figure}

As first, we study $n$ Wilson operators belonging to the same node 1. Such observable is graphically displayed in Figure \ref{Fig:quiverSameWL}. Writing the observable in terms of the $t$-functions, we end up with the same expansion of \eqref{w11generalq} with $n$ sums instead of one and with $\ell$ replaced by the set of indices $\ell_1,...,\ell_n$. Similarly to the SCQCD case, only the terms proportional to powers of ${\color{red}\zeta}_3$ can be solved exactly, obtaining
\begin{equation}\label{zeta3w1111q}
    w_{[1,1,...,1]}^{(q)}\bigg|_{{\color{red}\zeta}_3^m}=w_{1}^{(q)}\bigg|_{{\color{red}\zeta}_3^m}~,\qquad\text{with}\quad{w_1\rightarrow w_{[1,1,...,1]}}~.
\end{equation}
The remaining parts of the expansion with different transcendentality can be systematically computed in perturbation theory. As in \eqref{wqvanishes}, the large $N$ limit is special. First of all, in this limit $w_{[1,1,...,1]}$ in \eqref{zeta3w1111q} factorizes according to \eqref{factorizationwlargeN}, then the replacement becomes $w_1\rightarrow w_1^n$. Moreover, if in addition the theory is at the orbifold point, all the terms in the transcendentality expansion vanish and we have
\begin{equation}\label{wvecIqvanishes}
    w_{[\underbrace{1,1,...,1}_{\text{$n$-times}}]}^{(q)}\bigg|_{{\color{blue}\lambda}_I = {\color{blue}\lambda}} \xrightarrow{N\rightarrow \infty} w_1({\color{blue}\lambda})^n = \parenth{\frac{2}{\sqrt{{\color{blue}\lambda}}}I_1(\sqrt{{\color{blue}\lambda}})}^n~, \hspace{1cm} q\neq 1~.
\end{equation}

In general $A_{q-1}$ theories, we can study also correlation function of Wilson loops belonging to different vector multiplets. Let's start with the simplest example, namely a Wilson loop in the node 1 and another one in a node $J$ at distance $d(q,1,J)\geq 1$, where $d$ is defined in \eqref{distance}. The graphical representation of this observable can be found in Figure \ref{fig:3a}. As always, it is possible to recover all the other possible cases performing a rotation in the nodes  and couplings labels. It is convenient to represent this quantity as the sum of the disconnected part, namely the product of the two factorized Wilson loops, and the connected part as follows
\begin{equation}\small\begin{split}\label{w[12]}
    w_{[1,d+1]}^{(q)}\!=w_{1}^{(q)}w_{d+1}^{(q)}\!+\!
    \begin{cases}
      \!\frac{3^d(N^2-1)^{d-1}{\color{red}\zeta}_3^d}{2^{d-1}(8\pi^2 N)^{2d}}{\color{blue}\lambda}_1^2{\color{blue}\lambda}_{d+1}^2\!\!\left[{\displaystyle\prod_{i=2}^d{\color{blue}\lambda}^2_i}+\!\!\!{\displaystyle\prod_{j=d+2}^q\!\!{\color{blue}\lambda}^2_j}\right]\!\!\partial_1w_1\partial_{d+1}w_{d+1}, & \!\!\!\text{if}\ d=q/2 \\
      \!\frac{3^d(N^2-1)^{d-1}{\color{red}\zeta}_3^d}{2^{d-1}(8\pi^2 N)^{2d}}\!\left[{\displaystyle\prod_{i=1}^{d+1}{\color{blue}\lambda}^2_i}\right]\;\partial_1w_1\partial_{d+1}w_{d+1}, & \!\!\!\text{otherwise}
    \end{cases}\!\!+\!...
\end{split}\normalsize\end{equation}
where dots represents higher orders in transcendentality that in this case are starting from ${\color{red}\zeta}_3^{d-1}{\color{red}\zeta}_5$.
Accordingly with the definition of distance \eqref{distance}, for a fixed value of $q\geq 2$ in \eqref{w[12]}, $d$ can take value in the interval $[1,\floor{q/2}]$ where $\floor{x}$ is the integer part of $x$. For each value of this interval, the connected part of the two loops takes the form of the second line of \eqref{w[12]}. However there is an interesting exception. Indeed, when the $d$ saturates the bound $d=q/2$ and $q$ is even, the distance between the two nodes 1 and $d+1$ is equal if one goes across the quiver clockwise or anti-clockwise. Then the first connected term is the sum of the two equal contributions and indeed it contains all the couplings appearing in the theory. Results like \eqref{w[12]} can be explained also at a diagrammatical level, see section \ref{FT:WLcoincident}.

Counting the powers of $N$ in the connected part of \eqref{w[12]}, we can conclude that at large $N$ only the disconnected part survives
\begin{equation}\label{w1dLargeN}
      w_{[1,d+1]}^{(q)}\xrightarrow{N\rightarrow \infty}w_{1}^{(q)}w_{d+1}^{(q)}~,
\end{equation}
where $w_{1}^{(q)}$ and $w_{d+1}^{(q)}$ are given by \eqref{WqLargeN}. Moreover, at the orbifold point, given the result \eqref{wqvanishes}, the correlation function is not only factorized but it can be written in terms of $\cN=4$ SYM exact results
\begin{equation}\label{w1dLargeNOrbifold}
      w_{[1,d+1]}^{(q)}\bigg|_{{\color{blue}\lambda}_I = {\color{blue}\lambda}}\xrightarrow{N\rightarrow \infty}w_{1}({\color{blue}\lambda})^2=\frac{4}{{\color{blue}\lambda}}I_1(\sqrt{{\color{blue}\lambda}})^2~,
\end{equation}
where we used \eqref{exactWLlargeN}.

Another interesting example to consider is the correlation function of $q$ Wilson loops, where each of them belongs to a different node of the quiver, as displayed in Figure \ref{fig:3b}. Also in this case it is convenient to split the results in the disconnected part, given by the factorized product of the single Wilson loop vevs, and the connected part as follows
\begin{equation}\begin{split}\label{w[12...q]}
    w_{[1,2,...,q]}^{(q)}\!=\prod_{i=1}^qw_{i}^{(q)}+
    \frac{3{\color{red}\zeta}_3}{64\pi^4 N^2}
    \sum_{i=1}^q{\color{blue}\lambda}_i^2{\color{blue}\lambda}_{i+1}^2\partial_{i}\partial_{i+1}\prod_{j=1}^q w_{j}^{(q)}+...
\end{split}\end{equation}
where $q+1=1$ due to the cyclicity of the quiver labels.
The dots in \eqref{w[12...q]} stand for higher transcendentality terms that in this case starts with ${\color{red}\zeta}_5$ with all the possible combinations of 4 derivatives on the product of $w_{j}^{(q)}$. Similarly to the previous example, all the connected terms of the expansion are subleading in $N$, then in the large $N$ limit only the disconnected part remains. At the orbifold point, the simplification it's even more drastic. Indeed,
since the Wilson loop vevs reduce to their $\cN=4$ SYM relatives \eqref{wqvanishes}, we have
\begin{equation}\label{w[12...q]largeN}
      w_{[1,2,...,q]}^{(q)}\bigg|_{{\color{blue}\lambda}_I = {\color{blue}\lambda}}\xrightarrow{N\rightarrow \infty} w_{1}({\color{blue}\lambda})^q=\frac{2^q}{{\color{blue}\lambda}^{q/2}}I_1(\sqrt{{\color{blue}\lambda}})^q~,
\end{equation}
where $w_{1}$ is given by \eqref{exactWLlargeN}. Again, the results \eqref{w[12...q]} and \eqref{w[12...q]largeN} will be discussed at a diagrammatical level in section \ref{FT:WLcoincident}.

Finally, one can consider mixed correlators of Wilson loops appearing multiple times on different quiver nodes. This is the most general case, but we can draw some conclusions about the large $N$ limit where, likewise the pure Gaussian case \eqref{factorizedwvecIN4} and the examples above, the following factorization occurs
\begin{equation}\label{factorizedwvecIN4q}
w_{\vec{I}}^{(q)}\xrightarrow{N\rightarrow\infty}\prod_{i=\;\substack{\text{repeated} \\
\text{nodes in } \vec{I}}} w^{(q)}_{[\underbrace{I_i,...,I_i}_{\substack{\text{\texttt{\#} of times it}\\\text{appears in $\vec{I}$}}}]}
\;\,\prod_{j=\;\substack{\text{non-repeated} \\
\text{nodes in } \vec{I}}} w_{I_j}^{(q)}~.
\end{equation}
In addition, at the orbifold point, all the contributions appearing in the right-hand side of \eqref{factorizedwvecIN4q} reduces to copies of the $\mathcal{N}=4$ SYM Wilson loop vev as shown in \eqref{wqvanishes} and \eqref{wvecIqvanishes}. 
Then we obtain
\begin{equation}
w_{\vec{I}}^{(q)}\bigg|_{{\color{blue}\lambda}_I = {\color{blue}\lambda}}\xrightarrow{N\rightarrow\infty} w_{1}({\color{blue}\lambda})^n=\frac{2^n}{{\color{blue}\lambda}^{n/2}}I_1(\sqrt{{\color{blue}\lambda}})^n~,
\end{equation}
where $n$ is the length of the vector $\vec{I}$.

\subsection{Wilson loops in the twisted and untwisted sectors}\label{sec:WuWt}

Given the definition of \eqref{WuWt}, we can briefly discuss the Wilson loops in the twisted and untwisted sector of the orbifold theory using the results collected in the previous sections. The operators $W^u$ and $W^t$ can be considered the natural choice for a base of Wilson loops that enjoy good transformation properties under the orbifold action of $\mathbb{Z}_q$.
The Wilson loop belonging to the untwisted sector can be easily obtained summing $q$ copies of \eqref{w1q}, for instance for $q=2$
\begin{equation}\small\begin{split}
    \vev{W^u}_2\equiv w^u=
    \frac{w_1+w_2}{2}
    -\frac{3{\color{red}\zeta}_3}{128\pi^4N^2}
    \bigg[{\color{blue}\lambda}_1^4&\partial_1^2w_1+{\color{blue}\lambda}_2^4\partial_2^2w_2
    +(N^2+1)({\color{blue}\lambda}_1^3\partial_1w_1+{\color{blue}\lambda}_2^3\partial_2w_2)\\
    &-(N^2-1){\color{blue}\lambda}_2{\color{blue}\lambda}_1({\color{blue}\lambda}_1\partial_1w_1+{\color{blue}\lambda}_2\partial_2w_2) \bigg]+...
\end{split}\normalsize\end{equation}
where dots stand for higher transcendentality terms.
Taking the large $N$ limit drastically simplifies the output. Moreover, the theory at the orbifold point present an interesting feature, indeed given the reduction \eqref{wqvanishes}, we have
\begin{equation}\label{wuorb}
    w^{u}\bigg|_{{\color{blue}\lambda}_I = {\color{blue}\lambda}}\xrightarrow{N\rightarrow\infty} w_{1}({\color{blue}\lambda})=\frac{2}{\sqrt{{\color{blue}\lambda}}}I_1(\sqrt{{\color{blue}\lambda}})~,\qquad \forall\;q~.
\end{equation}

The twisted sector is given by $q-1$ independent Wilson loops.
Their expansions at finite $N$ are cumbersome and then we don't show them here. They can be evaluated taking the data from the notebook \texttt{WLcorrelators.nb} attached to this manuscript and inserting them in the definition \eqref{WuWt}.
However, we still want to mention one interesting property. Indeed, taking the large $N$ limit in the theory at the orbifold point, since all the Wilson loops are equal \eqref{wqvanishes}, we have
\begin{equation}\label{wtorb}
     w^{t}\bigg|_{{\color{blue}\lambda}_I = {\color{blue}\lambda}}\xrightarrow{N\rightarrow\infty} 0~,\qquad \forall\;q~,
\end{equation}
due to the fact that the sum of the roots of unity is zero \cite{Rey:2010ry}.

\section{One-point functions of chiral operators in presence of Wilson loops}\label{sec:5}
We now move to the second set of observables that includes local operators besides Wilson loops.
We mainly consider the case of a chiral operator inserted with a single Wilson loop, with the goal of building up the main ingredients in the pure Gaussian case and SCQCD, then moving to the $q$ nodes case, in order to finally discuss the twisted and untwisted sectors in section \ref{sec:TwistUntw}. 
However, our algorithm allows for the most general case, i.e. one point functions in presence of multiple coincident Wilson loops. Hence this analysis shall be completed with the results of appendix \ref{App:OmultipleW} as well as the data generated using the package \cite{Preti2021maybe}, which are stored in the attached notebook \texttt{WLcorrelators.nb}.

\subsection{Defect correlators in the pure Gaussian model}\label{sec:WOgauss}

Following the same structure of section \ref{sec:<W>}, we consider first correlators in the pure Gaussian model, corresponding to the case of $\mathcal{S}_{\text{int}}=0$ in \eqref{vevf}. As before, the quiver theory factorizes in $q$ copies of the pure Gaussian model and then the expectation value $\big\langle\mathcal{W}_{\vec{I}}:\mathcal{O}_{\vec{n}}^{(J)}:\big\rangle_q$ can be replaced by $\big\langle\mathcal{W}_{\vec{I}}:\mathcal{O}_{\vec{n}}^{(J)}:\big\rangle_0$. Moreover, since in this framework there are no matter fields connecting different nodes of the quiver, the normal ordered operator is forced to share the vector multiplet with at least one of the Wilson loops. All the other cases are trivial due to the fact that the normal-ordered operators have vanishing one-point function \eqref{oneptvanish}.

Let's start considering, for instance, the normal-ordered single trace operator of dimension $n=6$\footnote{For further details about the mixing of operators and explicit examples see \cite{Galvagno:2020cgq}} 
\begin{equation}\begin{split}\label{:Op6:}
:\mathcal{O}^{(I)}_{[6]}:\;\equiv\; :\tr a_I^6:\;=&
\mathcal{O}^{(I)}_{[6]}
+\alpha^{(I,I)}_{[6],[4]}\;\mathcal{O}^{(I)}_{[4]}
+\alpha^{(I,I)}_{[6],[2,2]}\;\mathcal{O}^{(I)}_{[2,2]}
+\alpha^{(I,I)}_{[6],[2]}\;\mathcal{O}^{(I)}_{[2]}
+\alpha^{(I,I)}_{[6],[0]}~,
\end{split}\end{equation}
where we used \eqref{coeffGS} and \eqref{matrixM}. 
Even if the pure Gaussian theory is defined in a quiver with $q$ nodes, since $\mathcal{S}_{\text{int}}=0$, only operators sharing the same vector multiplet index can mix. The functions $\alpha$'s are the Gram-Schmidt coefficients where, since the operator dimension is even, the last one is associated to the mixing with the identity operator $\mathbb{1}$ and computed according to \eqref{alpha0}. In the pure Gaussian model they read
\begin{equation}\begin{split}\label{GS6}
\alpha^{(I,I)}_{[6],[4]}&=\frac{15-6N^2}{2 N},\qquad\qquad\qquad
\alpha^{(I,I)}_{[6],[2,2]}=-\frac{3}{2},\\
\alpha^{(I,I)}_{[6],[2]}&=\frac{15 \left(N^4-3 N^2+3\right)}{4 N^2},\qquad\;\;
\alpha^{(I,I)}_{[6],[0]}=-\frac{5 \left(N^6-4 N^4+6 N^2-3\right)}{8 N^2}.
\end{split}\end{equation}
Hence, the one-point function of $:\mathcal{O}^{(J)}_{[6]}:$ in presence of Wilson loops can be written as a linear combination of correlators between $\mathcal{W}_{\vec{I}}$ and the operators appearing in the right-hand side of \eqref{:Op6:}. The latter, correspond to the $t$-functions defined in \eqref{rectn} that can be computed  with the recursion relation \eqref{recursion}. This procedure can systematically be applied to any value $\vec{n}$.

 The simplest possible example is to consider the one-point function of \eqref{:Op6:} in presence of only one circular Wilson loop, setting $\vec{I}=I$ and $J=I$
\begin{equation}\small\begin{split}
    \vev{\!\mathcal{W}_I\! :\!\mathcal{O}^{(I)}_{[6]}\!\!:\!}_0\!\!\!\equiv
     \!\mathcal{A}_{[6]}^I\!=\!
     \frac{1}{N}\!\! \sum_{\ell = 0}^{\infty}\frac{1}{\ell!}\!\! \parenth{\!\!\frac{{\color{blue}\lambda}_I}{2 N}\!\!}^{\!\!\frac{\ell}{2}}\!\!\bigg[t_{[\ell,6]}\!+\!
     \alpha^{(I,I)}_{[6],[4]}t_{[\ell,4]}
\!+\!\alpha^{(I,I)}_{[6],[2,2]}t_{[\ell,2,2]}
\!+\!\alpha^{(I,I)}_{[6],[2]}t_{[\ell,2]}
\!+\!\alpha^{(I,I)}_{[6],[0]}t_{[\ell]}
     \!\bigg]
\end{split}\normalsize\end{equation}
Using the technique presented in section \ref{sec:WLSCQCD} and the Gram-Schmidt coefficients \eqref{GS6}, we can compute this observable exactly, obtaining
\begin{equation}\footnotesize\begin{split}\label{A6Igauss}
&\mathcal{A}_{[6]}^I=
\frac{{\color{blue}\lambda} _{I}^3}{128 N^8}\bigg[
64 \left(8 N^6 {\color{blue}\lambda} _{I} \left(8 N^2 {\color{blue}\lambda} _{I} \partial_I^5w_I\!+\!\left(40 N^2\!-\!{\color{blue}\lambda} _{I}\right) \partial_I^4w_I\right)\!-\!N^4 \left(16 N^4 \left({\color{blue}\lambda} _{I}\!-\!15\right)\!-\!16 N^2 {\color{blue}\lambda} _{I}\!-\!{\color{blue}\lambda} _{I}^2\right) \partial_I^3w_I\right)\\
&-\!8 N^6\! \left(12 \tfrac{ {\color{blue}\lambda} _{I}}{N^2}\!+\!\tfrac{{\color{blue}\lambda} _{I}^2}{N^4}\!+\!48 \left(4 N^2\!-\!9\right) \!\right) \!\partial_I^2w_I
\!+\!2N^4\! \left(8 \tfrac{ {\color{blue}\lambda} _{I}}{N^2}\!+\!\tfrac{{\color{blue}\lambda} _{I}^2}{N^4}\!+\!48\! \left(N^4\!-\!5 N^2\!+\!3\right) \!\right) \!\partial_Iw_I
\!+\!\left(N^2\!\!-\!1\right)\! \left({\color{blue}\lambda} _{I}\!+\!12 N^2\right) \!w_I\!\bigg],
\end{split}\normalsize\end{equation}
where $w_I$ is the exact expectation value of the Wilson loop in $\mathcal{N}=4$ SYM \eqref{exactWL}.

Following the same procedure we can compute $\mathcal{A}_{\vec{n}}^I$ for any value of $\vec{n}$, for example
\begin{equation}\footnotesize\begin{split}\label{examplegaussWO}
\mathcal{A}_{[2]}^I&={\color{blue}\lambda}_I\partial_Iw_I~,\\
\mathcal{A}_{[3]}^I&=
\frac{{\color{blue}\lambda} _{I}}{4 \sqrt{2} N^2} \bigg[\left(8 N^2-2 {\color{blue}\lambda} _{I}\right)\partial_Iw_I +16 N^2 {\color{blue}\lambda} _{I} \partial_I^2w_I-\left(N^2-1\right) w_I\bigg]~,\\
\mathcal{A}_{[4]}^I&=
\frac{{\color{blue}\lambda} _{I}^2}{16 N^4} \bigg[\left(N^2\!-\!1\right) w_I\!+\!2 \left(64 N^4 {\color{blue}\lambda} _{I} \partial_I^3w_I\!+\!\left({\color{blue}\lambda} _{I}\!-\!8 N^4\!+\!8 N^2\right) \partial_Iw_I\!+\!\left(96 N^4\!-\!8 N^2 {\color{blue}\lambda} _{I}\right) \partial_I^2w_I\right)\bigg]~,\\
\mathcal{A}_{[2,2]}^I&={\color{blue}\lambda}_I^2\partial_I^2w_I~,\\
\mathcal{A}_{[5]}^I&=
\frac{{\color{blue}\lambda} _{I}^2}{32 \sqrt{2} N^6} \bigg[\left(N^2\!-\!1\right) \left(4 N^4\!-12 N^2\!-{\color{blue}\lambda} _{I}\right) w_I\!-2 \left(8 N^2 {\color{blue}\lambda} _{I}\!+{\color{blue}\lambda} _{I}^2\!+48 N^6\!-4 N^4 \left({\color{blue}\lambda} _{I}\!+24\right)\right) \partial_Iw_I\\
&\quad+\!16 \left(8 N^4 {\color{blue}\lambda} _{I} \left(8 N^2 {\color{blue}\lambda} _{I} \partial_I^4w_I\!+\!\left(24 N^2\!-\!{\color{blue}\lambda} _{I}\right) \partial_I^3w_I\right)\!+\!\left(12 N^4 {\color{blue}\lambda} _{I}\!+\!N^2 {\color{blue}\lambda} _{I}^2\!-\!12 N^6 \left({\color{blue}\lambda} _{I}\!-\!4\right)\right) \partial_I^2w_I\right)\bigg]~,\\
\mathcal{A}_{[3,2]}^I&=
\frac{{\color{blue}\lambda} _{I}}{4 \sqrt{2} N^2} \bigg[\left({\color{blue}\lambda} _{I}\!-\!N^2 \left({\color{blue}\lambda} _{I}\!+\!8\right)\right) \partial_Iw_I\!+\!2 {\color{blue}\lambda} _{I} \left(\left(4 N^2\!-\!{\color{blue}\lambda} _{I}\right) \partial_I^2w_I\!+\!8 N^2 {\color{blue}\lambda} _{I} \partial_I^3w_I\right)\!+\!\left(N^2\!-\!1\right) w_I\bigg]~.
\end{split}\normalsize\end{equation}
As expected, all the previous results are the finite $N$ versions of those obtained in $\mathcal{N}=4$ SYM \cite{Semenoff:2001xp,Pestun:2002mr,Rodriguez-Gomez:2016cem}. Moreover, taking the large $N$ limit and then substituting $w_I$ with \eqref{exactWLlargeN}, we immediately recover the general formula for the correlation function of a Wilson loop and a single trace operator originally derived in \cite{Semenoff:2001xp} and also studied in \cite{Billo:2018oog,Beccaria:2020hgy}
\begin{align}\label{WLexactLambda}
\mathcal{A}_{[n]}^I({\color{blue}\lambda}_I) = \frac{n}{2^{n/2}}{\color{blue}\lambda}_I^{\frac{n}{2}-1} I_n(\sqrt{{\color{blue}\lambda}_I})~.
\end{align}

In the following sections we study the same observables presented here including the insertion of the interacting action $\mathcal{S}_{\text{int}}$.

\subsection{Defect correlators in SCQCD}\label{sec:WOSCQCD}
 
SCQCD is defined on a single vector multiplet conventionally chosen with label $I=1$. The simplest one-point coefficient one can study is the following correlation function 
\begin{equation}\label{WO2SCQCD}
   \mathcal{A}_{[2]}^{(1,1)}
   =\vev{\mathcal{W}_1 \,\mathcal{O}_{[2]}^{(1)}}_1-\alpha_{[2],[0]}^{(1,1)}\; w_1^{(1)}~,
\end{equation}
where $\alpha_{[2],[0]}^{(1,1)}$ is given by \eqref{alpha0} and $w_1^{(1)}$ is the SCQCD vev of the Wilson loop computed in \eqref{WL2N}. The vev in \eqref{WO2SCQCD} can be computed through the recursion relation \eqref{recursion} and it corresponds to the expansion \eqref{w11SCQCD} with an additional 2 in the $t$-functions indices. Then, we obtain
\begin{equation}\footnotesize\begin{split}\label{WO2SCQCDfiniteN}
   &\mathcal{A}_{[2]}^{(1,1)}
   =
   {\color{blue}\lambda} _{1} \partial_1w_1-\frac{3 {\color{red}\zeta} _{3} {\color{blue}\lambda} _{1}^3}{64 \pi ^4 N^2} \!\bigg[3 \left(N^2\!+\!1\right)\! \partial_1w_1\!+\!{\color{blue}\lambda} _{1} \left(\left(N^2\!+\!5\right) \!\partial_1^2w_1\!+\!{\color{blue}\lambda} _{1} \partial_1^3w_1\right)\!\!\bigg]\!
   \!+\!\frac{5 {\color{red}\zeta} _{5} {\color{blue}\lambda} _{1}^4}{6144 \pi ^6 N^6} \bigg[\!\!\left(N^4\!+\!2 N^2\!\!-\!3\right)\!w_1\\
   &+\!\tfrac{N^4}{4}\!\!\left({\color{blue}\lambda} _{1}\!+\!352\!+\!\tfrac{ 22 {\color{blue}\lambda} _{1}\!-\!96}{N^2}\!+\!\tfrac{37 {\color{blue}\lambda} _{1}}{N^4}\!+\!384 N^2\right) \!\partial_1w_1
   \!+\!\left(24 N^6 \!\left({\color{blue}\lambda} _{1}\!+\!24\right)\!+\!18 N^4\! \left(9 {\color{blue}\lambda} _{1}\!+\!32\right)\!+\!N^2 {\color{blue}\lambda} _{1} \!\left({\color{blue}\lambda} _{1}\!+\!54\right)\!+\!8 {\color{blue}\lambda} _{1}^2\right)\! \partial_1^2w_1\\
   &+{\color{blue}\lambda} _{1} \!\left(8 N^2 {\color{blue}\lambda} _{1} \left(8 N^2 {\color{blue}\lambda} _{1} \partial_1^5w_1\!+\!\left({\color{blue}\lambda} _{1}\!+\!12 N^4+80 N^2\right) \partial_1^4w_1\right)
   \!+\!\left(4 N^4 \left(7 {\color{blue}\lambda} _{1}\!+\!356\right)\!+\!60 N^2 {\color{blue}\lambda} _{1}\!+\!{\color{blue}\lambda} _{1}^2\!+\!624 N^6\right) \partial_1^3w_1\right)\bigg]\\
   &+\frac{9  {\color{red}\zeta} _{3}^2 {\color{blue}\lambda} _{1}^5}{8192 \pi ^8 N^4} \bigg[20 \left(N^4\!+\!3 N^2\!+\!2\right) \partial_1w_1\!+\!{\color{blue}\lambda} _{1} \left(2 \left(5 N^4\!+\!42 N^2\!+\!73\right) \partial_1^2w_1\!+\!{\color{blue}\lambda} _{1} \left({\color{blue}\lambda} _{1} \left(2 \left(N^2\!+\!9\right) \partial_1^4w_1\!+\!{\color{blue}\lambda} _{1} \partial_1^5w_1\right)\right.\right.\\
   &\left.\left.+\left(N^4\!+\!26 N^2\!+\!93\right) \partial_1^3w_1\right)\right)\bigg]+...
\end{split}\normalsize\raisetag{20pt}\end{equation}
The one-point function in presence of a Wilson loop $\mathcal{A}_{[2]}^{(1,1)}$ has a very interesting property. Indeed, comparing the expansion \eqref{WO2SCQCDfiniteN} with \eqref{WL2N}, 
one can express it in terms of the SCQCD vev of the Wilson loop through a differential operator as follows
\begin{equation}\label{Aderw}
    \mathcal{A}_{[2]}^{(1,1)}={\color{blue}\lambda}_1 \partial_1 w_{1}^{(1)}~,
\end{equation}
with the differential operator being the same found in the pure Gaussian model \eqref{examplegaussWO}. Notice that \eqref{Aderw} is exact for any value of the coupling and for any transcendentality. In the large $N$ limit this result holds with $w_{1}^{(1)}$ replaced by its planar version \eqref{WL2NlargeN}.

Unfortunately, it is not possible to deduce simple formulas like \eqref{Aderw} for higher dimensions $n$.  
The resulting expansions are extremely lengthy and involved, then we present here only few transcendentality orders in the large $N$ limit:
\begin{equation}\label{A2scqcd}\footnotesize\begin{split}
\mathcal{A}_{[3]}^{(1,1)} &=\mathcal{A}_{[3]}^{1}+
\frac{3 {\color{red}\zeta} _{3} {\color{blue}\lambda} _{1}^3 \left(\left({\color{blue}\lambda} _{1}-16\right) \partial_1w_1-8 {\color{blue}\lambda} _{1} \left(7 \partial_1^2w_1+2 {\color{blue}\lambda} _{1} \partial_1^3w_1\right)+2 w_1\right)}{256 \sqrt{2} \pi ^4}\\
&-\frac{5  {\color{red}\zeta} _{5} {\color{blue}\lambda} _{1}^4 \left(\left({\color{blue}\lambda} _{1}-12\right) \partial_1w_1-6 \left(7 {\color{blue}\lambda} _{1}+24\right) \partial_1^2w_1-64 {\color{blue}\lambda} _{1}^2 \left(8 \partial_1^4w_1+{\color{blue}\lambda} _{1} \partial_1^5w_1\right)-12 \left({\color{blue}\lambda} _{1}+68\right) {\color{blue}\lambda} _{1} \partial_1^3w_1-3 w_1\right)}{1024 \sqrt{2} \pi ^6}\\
&-\frac{9 {\color{red}\zeta} _{3}^2 {\color{blue}\lambda} _{1}^5 \left(8 \left({\color{blue}\lambda} _{1}-10\right) \partial_1w_1-{\color{blue}\lambda} _{1} \left(8 {\color{blue}\lambda} _{1} \left(21 \partial_1^3w_1+2 {\color{blue}\lambda} _{1} \partial_1^4w_1\right)-\left({\color{blue}\lambda} _{1}-352\right) \partial_1^2w_1\right)+10 w_1\right)}{32768 \sqrt{2} \pi ^8}+...
\end{split}\normalsize\raisetag{21pt}\end{equation}
\begin{equation}\footnotesize\begin{split}
\mathcal{A}_{[4]}^{(1,1)} &=\mathcal{A}_{[4]}^{1}
-\frac{3 {\color{red}\zeta} _{3} {\color{blue}\lambda} _{1}^4 \left(-3 \partial_1w_1-\left({\color{blue}\lambda} _{1}-36\right) \partial_1^2w_1+4 {\color{blue}\lambda} _{1} \left(11 \partial_1^3w_1+2 {\color{blue}\lambda} _{1} \partial_1^4w_1\right)\right)}{64 \pi ^4}\\
&-\frac{5 {\color{red}\zeta} _{5} {\color{blue}\lambda} _{1}^5 \left(9 \partial_1w_1\!+\!2 \left({\color{blue}\lambda} _{1}\!-\!24\right) \partial_1^2w_1\!-\!\left(52 {\color{blue}\lambda} _{1}\!+\!960\right) \partial_1^3w_1\!-\!8 {\color{blue}\lambda} _{1} \left(\left({\color{blue}\lambda} _{1}\!+\!230\right) \partial_1^4w_1\!+\!8 {\color{blue}\lambda} _{1} \left(11 \partial_1^5w_1\!+\!{\color{blue}\lambda} _{1} \partial_1^6w_1\right)\!\right)\!\right)}{512 \pi ^6}\\
&-\frac{9 {\color{red}\zeta} _{3}^2 {\color{blue}\lambda} _{1}^6 \left(18 \partial_1w_1\!-\!\left(216\!-\!10 {\color{blue}\lambda} _{1}\right) \partial_1^2w_1\!-\!{\color{blue}\lambda} _{1} \left(4 {\color{blue}\lambda} _{1} \left(27 \partial_1^4w_1\!+\!2 {\color{blue}\lambda} _{1} \partial_1^5w_1\right)\!-\!\left({\color{blue}\lambda} _{1}\!-\!344\right) \partial_1^3w_1\right)\!\right)}{8192 \pi ^8}+...
\end{split}\normalsize\raisetag{18pt}\end{equation}
\begin{equation}\footnotesize\begin{split}
\mathcal{A}_{[2,2]}^{(1,1)} &=\mathcal{A}_{[2,2]}^{1}
-\frac{3{\color{red}\zeta} _{3} {\color{blue}\lambda} _{1}^4 }{64 \pi ^4}\left(6 \partial_1^2w_1+{\color{blue}\lambda} _{1} \partial_1^3w_1\right)+\frac{5 {\color{red}\zeta} _{5} {\color{blue}\lambda} _{1}^4 }{256 \pi ^6}\bigg[-6 \partial_1w_1+8 \left({\color{blue}\lambda} _{1}+9\right) \partial_1^2w_1+{\color{blue}\lambda} _{1} \left(\left({\color{blue}\lambda} _{1}+128\right) \partial_1^3w_1\right.\\
&\left.+2 {\color{blue}\lambda} _{1} \left(23 \partial_1^4w_1+2 {\color{blue}\lambda} _{1} \partial_1^5w_1\right)\right)\bigg]
+\frac{9 {\color{red}\zeta} _{3}^2 {\color{blue}\lambda} _{1}^6 \left(58 \partial_1^2w_1+{\color{blue}\lambda} _{1} \left(16 \partial_1^3w_1+{\color{blue}\lambda} _{1} \partial_1^4w_1\right)\right)}{8192 \pi ^8}+...
\end{split}\normalsize\raisetag{21pt}\end{equation}
\begin{equation}\footnotesize\begin{split}
\mathcal{A}_{[5]}^{(1,1)} &=\mathcal{A}_{[5]}^{1}
\!-\!\frac{\!3{\color{red}\zeta} _{3} {\color{blue}\lambda} _{1}^4 \left(\!\left({\color{blue}\lambda} _{1}\!-\!72\right) \!\partial_1w_1\!+\!8 \left(9\! \left(8\!-\!3 {\color{blue}\lambda} _{1}\right) \!\partial_1^2w_1\!+\!32 {\color{blue}\lambda} _{1}^2\! \left(8 \partial_1^4w_1\!+\!{\color{blue}\lambda} _{1} \partial_1^5w_1\right)\!-\!6 \left({\color{blue}\lambda} _{1}\!-\!68\right) {\color{blue}\lambda} _{1} \partial_1^3w_1\right)\!+\!3 w_1\right)}{512 \sqrt{2} \pi ^4}\\
&+\frac{5{\color{red}\zeta} _{5} {\color{blue}\lambda} _{1}^5}{4096 \sqrt{2} \pi ^6} \bigg[11 w_1+2 \left({\color{blue}\lambda} _{1}-104\right) \partial_1w_1+4 \left(\left(120-137 {\color{blue}\lambda} _{1}\right) \partial_1^2w_1-\left(22 {\color{blue}\lambda} _{1}^2-600 {\color{blue}\lambda} _{1}-3840\right) \partial_1^3w_1\right.\\
&\left.+32 {\color{blue}\lambda} _{1} \left(10 \left({\color{blue}\lambda} _{1}+93\right) \partial_1^4w_1
+{\color{blue}\lambda} _{1} \left(\left({\color{blue}\lambda} _{1}+900\right) \partial_1^5w_1+8 {\color{blue}\lambda} _{1} \left(29 \partial_1^6w_1+2 {\color{blue}\lambda} _{1} \partial_1^7w_1\right)\right)\right)\right)\bigg]
+...
\end{split}\normalsize\raisetag{20pt}\end{equation}
\begin{equation}\footnotesize\begin{split}
\mathcal{A}_{[3,2]}^{(1,1)} &=\mathcal{A}_{[3,2]}^{1}
+\frac{3 {\color{red}\zeta} _{3} {\color{blue}\lambda} _{1}^3 \left(4 \left({\color{blue}\lambda} _{1}+8\right) \partial_1w_1+{\color{blue}\lambda} _{1} \left(\left({\color{blue}\lambda} _{1}-32\right) \partial_1^2w_1-8 {\color{blue}\lambda} _{1} \left(13 \partial_1^3w_1+2 {\color{blue}\lambda} _{1} \partial_1^4w_1\right)\right)-4 w_1\right)}{256 \sqrt{2} \pi ^4}\\
&+\frac{5 {\color{red}\zeta} _{5} {\color{blue}\lambda} _{1}^4}{2048 \sqrt{2} \pi ^6} \bigg[21 w_1-12 \left({\color{blue}\lambda} _{1}+18\right) \partial_1w_1-2 \left({\color{blue}\lambda} _{1} \left({\color{blue}\lambda} _{1}+126\right)-288\right) \partial_1^2w_1\\
&+4 {\color{blue}\lambda} _{1} \left(9 \left(5 {\color{blue}\lambda} _{1}+144\right) \partial_1^3w_1+2 {\color{blue}\lambda} _{1} \left(\left(3 {\color{blue}\lambda} _{1}+716\right) \partial_1^4w_1+16 {\color{blue}\lambda} _{1} \left(13 \partial_1^5w_1+{\color{blue}\lambda} _{1} \partial_1^6w_1\right)\right)\right)\bigg]+...
\end{split}\normalsize\end{equation}
\begin{equation}\label{A6scqcd}\footnotesize\begin{split}
\mathcal{A}_{[6]}^{(1,1)} &=\mathcal{A}_{[6]}^{1}  
\!-\!\frac{3 {\color{red}\zeta} _{3} {\color{blue}\lambda} _{1}^5}{256 \pi ^4} 
\bigg[12 \partial_1w_1+32 \left(8 {\color{blue}\lambda} _{1}^2 \left(11 \partial_1^5w_1\!+\!{\color{blue}\lambda} _{1} \partial_1^6w_1\right)\!-\!2 \left({\color{blue}\lambda} _{1}\!-\!115\right) {\color{blue}\lambda} _{1} \partial_1^4w_1\!+\!\left(120\!-\!13 {\color{blue}\lambda} _{1}\right) \partial_1^3w_1\right)\\
&+3 \left({\color{blue}\lambda} _{1}\!-\!128\right) \partial_1^2w_1\!\bigg]
\!+\!\frac{5  {\color{red}\zeta} _{5} {\color{blue}\lambda} _{1}^6 }{2048 \pi ^6}
\bigg[39 \partial_1w_1\!+\!6 \left({\color{blue}\lambda} _{1}\!-\!158\right) \partial_1^2w_1\!-\!4\left(26 {\color{blue}\lambda} _{1}^2\!-\!320 {\color{blue}\lambda} _{1}\!-\!33600\right) \partial_1^4w_1\\
&+4\left(240\!-\!223 {\color{blue}\lambda} _{1}\right) \partial_1^3w_1\!+\!256 {\color{blue}\lambda} _{1} \left(\left({\color{blue}\lambda} _{1}\!+\!1365\right) \partial_1^5w_1\!+\!2 {\color{blue}\lambda} _{1} \left(4 {\color{blue}\lambda} _{1}^2 \partial_1^8w_1\!+\!74 {\color{blue}\lambda} _{1} \partial_1^7w_1\!+\!399 \partial_1^6w_1\right)\right)\bigg]+...
\end{split}\normalsize\raisetag{44pt}\end{equation}
where $\mathcal{A}_{\vec{n}}^1$ are the Gaussian model results \eqref{A6Igauss} and \eqref{examplegaussWO} for $I=1$.

\subsection{Defect correlators in $A_{q-1}$ theories}\label{sec:Aq-1corr}
We study one-point function in presence of Wilson loops in the most general quiver theories $A_{q-1}$ with $q\geq 2$. Unlike the case analysed in section \eqref{sec:WOgauss}, since the $\mathcal{S}_{\text{int}}$ is not zero, correlators of operators belonging to different vector multiplets are non vanishing. Indeed, Wilson loops and local operators can be scattered at will on the quiver generating an enormous amount of possible observables to study. The results displayed in this section shall be considered as the building blocks for twisted and untwisted one-point functions of section \ref{sec:TwistUntw}, which represent the proper observables for holographic perspectives of $A_{q-1}$ theories. 
 
We start by the simplest case, the one-point function of the operator of twist 2 in presence of Wilson loops, which is peculiar since it can be computed exactly through a simple differential operator, as noticed in the previous sections:
  \begin{equation}\label{A2q}
     \mathcal{A}_{[2]}^{(\vec{I},J)}={\color{blue}\lambda}_J\partial_J\,w_{\vec{I}}^{(q)}~,
 \end{equation}
 where $w_{\vec{I}}^{(q)}$ is the vev $\mathcal{W}_{\vec{I}}$ in the $A_{q-1}$ theory. The action of the derivative in \eqref{A2q} drastically changes the behaviour of this observable at large $N$ and also at the orbifold point. Indeed, even if in those limits $w_{\vec{I}}^{(q)}$ is simply equal to its $\mathcal{N}=4$ SYM relative, $\mathcal{A}_{[2]}^{(\vec{I},J)}$ has a non trivial expansion in transcendentality. For instance, if we consider $\vec{I}=1$ and $q=4$, we have
 \begin{equation}\begin{split}\label{Aq=4I=1n=2}
     \mathcal{A}_{[2]}^{(1,1)}=&
     {\color{blue}\lambda} _{1} \partial_1w_1-\frac{3  {\color{red}\zeta} _{3} {\color{blue}\lambda} _{1}^3 \partial_1w_1}{64 \pi ^4}+\frac{5  {\color{red}\zeta} _{5} \left(3 {\color{blue}\lambda} _{1}^4 \partial_1w_1+12 {\color{blue}\lambda} _{1}^4 \partial_1^2w_1+8 {\color{blue}\lambda} _{1}^5 \partial_1^3w_1\right)}{512 \pi ^6}+...\\
     \mathcal{A}_{[2]}^{(1,2)}=&
     \frac{3 {\color{red}\zeta} _{3} {\color{blue}\lambda} _{1}^3 \partial_1w_1}{128 \pi ^4}-\frac{5  {\color{red}\zeta} _{5} {\color{blue}\lambda} _{1}^4 \left(3 \partial_1w_1+12 \partial_1^2w_1+8 {\color{blue}\lambda} _{1} \partial_1^3w_1\right)}{1024 \pi ^6}-\frac{9 {\color{red}\zeta} _{3}^2 {\color{blue}\lambda} _{1}^5 \partial_1w_1}{4096 \pi ^8}+...\\
     \mathcal{A}_{[2]}^{(1,3)}=&
     \frac{9 {\color{red}\zeta} _{3}^2 {\color{blue}\lambda} _{1}^5 \partial_1w_1}{8192 \pi ^8}-\frac{15 {\color{red}\zeta} _{3} {\color{red}\zeta} _{5} \left(7 {\color{blue}\lambda} _{1}^6 \partial_1w_1+12 {\color{blue}\lambda} _{1}^6 \partial_1^2w_1+8 {\color{blue}\lambda} _{1}^7 \partial_1^3w_1\right)}{65536 \pi ^{10}}+...
 \end{split}\end{equation}
 where we consider the cases $J=1,2,3$ equivalent to distances $d=0,1,2$ computed with \eqref{distance}.
 
The one-point coefficients with higher dimensions $n$ can be computed with the method developed in the previous sections. For reasons of space, we present only results at large $N$. 
We first consider the operator and the Wilson loop belonging to the same node that, since the theories $A_{q-1}$ are invariant under cyclic reparametrisations of the node labels, we conventionally choose to be $I=1$. These observables are denoted as $\mathcal{A}_{\vec n}^{(1,1)}$.
The first deviation from the Gaussian model results \eqref{A6Igauss} and \eqref{examplegaussWO} (identified as $\mathcal{A}_{\vec n}^{1}$) is given by
\begingroup
\allowdisplaybreaks
 \begin{equation}\footnotesize\begin{split}\label{exampleWOq1}
\mathcal{A}_{[3]}^{(1,1)}&=
\mathcal{A}_{[3]}^{1}
+\frac{3{\color{red}\zeta} _{3} {\color{blue}\lambda} _{1}^2 \left(2 {\color{blue}\lambda} _{1}-{\color{blue}\lambda} _{2}-{\color{blue}\lambda} _{q}\right) \left(\left({\color{blue}\lambda} _{1}-16\right) \partial_1w_1-8 {\color{blue}\lambda} _{1} \left(7 \partial_1^2w_1+2 {\color{blue}\lambda} _{1} \partial_1^3w_1\right)+2 w_1\right)}{512 \sqrt{2} \pi ^4}+...\\
\mathcal{A}_{[4]}^{(1,1)}&=
\mathcal{A}_{[4]}^{1}
+\frac{3 {\color{red}\zeta} _{3} {\color{blue}\lambda} _{1}^3 \left(2 {\color{blue}\lambda} _{1}-{\color{blue}\lambda} _{2}-{\color{blue}\lambda} _{q}\right) \left(3 \partial_1w_1+\left({\color{blue}\lambda} _{1}-36\right) \partial_1^2w_1-4 {\color{blue}\lambda} _{1} \left(11 \partial_1^3w_1+2 {\color{blue}\lambda} _{1} \partial_1^4w_1\right)\right)}{128 \pi ^4}+...\\
\mathcal{A}_{[2,2]}^{(1,1)}&=
\mathcal{A}_{[2,2]}^{1}
-\frac{3 {\color{red}\zeta} _{3} {\color{blue}\lambda} _{1}^3 \left(4 \left(3 {\color{blue}\lambda} _{1}-{\color{blue}\lambda} _{2}-{\color{blue}\lambda} _{q}\right) \partial_1^2w_1+{\color{blue}\lambda} _{1} \left(2 {\color{blue}\lambda} _{1}-{\color{blue}\lambda} _{2}-{\color{blue}\lambda} _{q}\right) \partial_1^3w_1\right)}{128 \pi ^4}+...\\
\mathcal{A}_{[5]}^{(1,1)}&=
\mathcal{A}_{[5]}^{1}
-\frac{3{\color{red}\zeta} _{3} {\color{blue}\lambda} _{1}^3 \left(2 {\color{blue}\lambda} _{1}-{\color{blue}\lambda} _{2}-{\color{blue}\lambda} _{q}\right) }{1024 \sqrt{2} \pi ^4}\bigg[3 w_1+\left({\color{blue}\lambda} _{1}-72\right) \partial_1w_1+72 \left(8-3 {\color{blue}\lambda} _{1}\right) \partial_1^2w_1\\
&\qquad\qquad\qquad\qquad\qquad\qquad\qquad\qquad\qquad+256 {\color{blue}\lambda} _{1}^2 \left(8 \partial_1^4w_1+{\color{blue}\lambda} _{1} \partial_1^5w_1\right)-48 \left({\color{blue}\lambda} _{1}-68\right) {\color{blue}\lambda} _{1} \partial_1^3w_1\bigg]+...\\
\mathcal{A}_{[3,2]}^{(1,1)}&=
\mathcal{A}_{[3,2]}^{1}
+\frac{3{\color{red}\zeta} _{3} {\color{blue}\lambda} _{1}^2 }{512 \sqrt{2} \pi ^4}\bigg[\left(3 \left({\color{blue}\lambda} _{2}+{\color{blue}\lambda} _{q}\right)-8 {\color{blue}\lambda} _{1}\right) w_1\!+\!{\color{blue}\lambda} _{1} \left(\left(2 {\color{blue}\lambda} _{1}^2\!-\!\left({\color{blue}\lambda} _{2}\!+\!{\color{blue}\lambda} _{q}\!+\!64\right) {\color{blue}\lambda} _{1}\!+\!24 \left({\color{blue}\lambda} _{2}\!+\!{\color{blue}\lambda} _{q}\right)\right) \!\partial_1^2w_1\right.\\
&\;\left.+\left({\color{blue}\lambda} _{1}\!+\!8\right) \left(8 {\color{blue}\lambda} _{1}\!-\!3 \left({\color{blue}\lambda} _{2}\!+\!{\color{blue}\lambda} _{q}\right)\right) \partial_1w_1\!+\!8 {\color{blue}\lambda} _{1} \left(\left(11 \left({\color{blue}\lambda} _{2}\!+\!{\color{blue}\lambda} _{q}\right)\!-\!26 {\color{blue}\lambda} _{1}\right) \partial_1^3w_1\!-\!2 {\color{blue}\lambda} _{1} \left(2 {\color{blue}\lambda} _{1}\!-\!{\color{blue}\lambda} _{2}\!-\!{\color{blue}\lambda} _{q}\right) \!\partial_1^4w_1\right)\right)\bigg]\!+\!...\\
\mathcal{A}_{[6]}^{(1,1)}&=
\mathcal{A}_{[6]}^{1}
-\frac{3{\color{red}\zeta} _{3} {\color{blue}\lambda} _{1}^4\left(2 {\color{blue}\lambda} _{1}-{\color{blue}\lambda} _{2}-{\color{blue}\lambda} _{q}\right)}{512 \pi ^4}  \bigg[12 \partial_1w_1+3 \left({\color{blue}\lambda} _{1}-128\right) \partial_1^2w_1+32 \left(8 {\color{blue}\lambda} _{1}^2 \left(11 \partial_1^5w_1+{\color{blue}\lambda} _{1} \partial_1^6w_1\right)\right.\\
&\qquad\qquad\qquad\qquad\qquad\qquad\qquad\qquad\qquad\qquad\left.-2 \left({\color{blue}\lambda} _{1}-115\right) {\color{blue}\lambda} _{1} \partial_1^4w_1+\left(120-13 {\color{blue}\lambda} _{1}\right) \partial_1^3w_1\right)\bigg]+...
 \end{split}\normalsize\end{equation}
 \endgroup
In section \ref{Sec4.3} we showed that Wilson loops vevs drastically simplify at the orbifold point and are reduced to multiple copies of the $\mathcal{N}=4$ SYM vevs, see \eqref{wqvanishes}, \eqref{wvecIqvanishes}, \eqref{w1dLargeNOrbifold}, \eqref{w[12...q]largeN}. Correlators with local operators are more subtle. Indeed, even if many cancellations occur, they still deviates from the pure Gaussian model results. This is evident from \eqref{exampleWOq1} for observables containing multitrace operators, while the single trace cases are more peculiar. Indeed, at the orbifold point we obtain 
 \begin{equation}\footnotesize\begin{split}
\mathcal{A}_{[3]}^{(1,1)}\bigg|_{{\color{blue}\lambda}_I = {\color{blue}\lambda}}&=
\mathcal{A}_{[3]}^{1}
-\frac{5 {\color{red}\zeta} _{5} {\color{blue}\lambda}^4 \left(8 \left(\partial_{\color{blue}\lambda} w_1+2 {\color{blue}\lambda} \partial_{\color{blue}\lambda}^2w_1\right)-w_1\right)}{2048 \sqrt{2} \pi ^6}\!+\!...\\
\mathcal{A}_{[4]}^{(1,1)}\bigg|_{{\color{blue}\lambda}_I = {\color{blue}\lambda}}&=
\mathcal{A}_{[4]}^{1}
-\frac{35 {\color{red}\zeta} _{7} {\color{blue}\lambda} ^6\left(- \partial_{\color{blue}\lambda} w_1+12  \partial_{\color{blue}\lambda}^2w_1+8 {\color{blue}\lambda} \partial_{\color{blue}\lambda}^3w_1\right)}{16384 \pi ^8}\!+\!...\\
\mathcal{A}_{[5]}^{(1,1)}\bigg|_{{\color{blue}\lambda}_I = {\color{blue}\lambda}}&=
\mathcal{A}_{[5]}^{1}
-\frac{63 {\color{red}\zeta} _{9} {\color{blue}\lambda}^7 \left(8 \left(-\!3 \partial_{\color{blue}\lambda} w_1\!-\!6 \left({\color{blue}\lambda} \!-\!4\right) \partial_{\color{blue}\lambda}^2w_1\!+\!32 {\color{blue}\lambda} \left(3 \partial_{\color{blue}\lambda}^3w_1\!+\!{\color{blue}\lambda}  \partial_{\color{blue}\lambda}^4w_1\right)\right)\!+\!w_1\right)}{1048576 \sqrt{2} \pi ^{10}}\!+\!...\\
\mathcal{A}_{[6]}^{(1,1)}\bigg|_{{\color{blue}\lambda}_I = {\color{blue}\lambda}}&=
\mathcal{A}_{[6]}^{1}
-\frac{231 {\color{red}\zeta} _{11} {\color{blue}\lambda} ^9\! \left(3 \partial_{\color{blue}\lambda} w_1\!-\!32\! \left(3 \partial_{\color{blue}\lambda}^2w_1\!+\!2 \left({\color{blue}\lambda} \!-\!15\right) \!\partial_{\color{blue}\lambda}^3w_1\!-\!8 {\color{blue}\lambda}  \!\left(5 \partial_{\color{blue}\lambda}^4w_1\!+\!{\color{blue}\lambda}  \partial_{\color{blue}\lambda}^5w_1\right)\!\right)\!\right)}{8388608 \pi ^{12}}\!+\!...
 \end{split}\normalsize\end{equation}
 where, analyzing the patterns we can guess the following general expansion
 \begin{equation}\label{orbifoldAq11}
   \mathcal{A}_{[n]}^{(1,1)}\bigg|_{{\color{blue}\lambda}_I = {\color{blue}\lambda}}=
\mathcal{A}_{[n]}^{1} 
-\frac{{\color{blue}\lambda}^n}{2^{n-1}(8\pi^2)^n}\binom{2n}{n}{\color{red}\zeta}_{2n-1}\mathcal{A}_{[n]}^{1} +...
 \end{equation}
 where dots stand for higher transcendentality terms. It's also interesting to notice that for a fixed value of $n$, not only the terms proportional to ${\color{red}\zeta}_{2k-1}$ with $k=2,...,n-1$ disappear, but also all the terms that contain a power of them, even at orders higher than ${\color{red}\zeta}_{2n-1}$. Then, we can conclude that these $\cN=2$ observables approach more and more the corresponding ones in $\mathcal{N}=4$ as $n$ grows.
 
 Increasing the distance $d$ \eqref{distance} between the node 1 and the node in which the single trace operator belongs, there are no longer contributions from the Gaussian model to start with. Indeed, the first non-trivial contributions arise due to the interaction of the operators with the bi-fundamental matter.
 For instance for $d=1$, at large $N$ we have
 \begin{equation}
   \mathcal{A}_{[n]}^{(1,2)}({\color{blue}\lambda}_1,{\color{blue}\lambda}_2)=
\frac{{\color{blue}\lambda}_1{\color{blue}\lambda}_2^{n-1}}{2^{n}(8\pi^2)^n}\binom{2n}{n}{\color{red}\zeta}_{2n-1}\mathcal{A}_{[n]}^{1} +...
 \end{equation}
 that, at the orbifold point turns out to be very similar to the first deviation from the Gaussian model for $d=0$ \eqref{orbifoldAq11}.
 Analyzing this term at different values of $d$ it is possible to recognize an interesting pattern and guess the behaviour for any value of $q\geq 2$ a s a function of $d$ and $n$ as follows
 \begin{equation}\label{LOwOq}
   \mathcal{A}_{[n]}^{(1,d+1)}\!=\!
\begin{cases}
      \frac{{\color{red}\zeta}_{2n-1}^d}{2^{n+d-1}(8\pi^2)^{n d}}\binom{2n}{n}^d{\color{blue}\lambda}_1{\color{blue}\lambda}_{d+1}^{n-1}\!\!\left[{\displaystyle\prod_{i=2}^d{\color{blue}\lambda}^n_i}+\!\!\!{\displaystyle\prod_{j=d+2}^q\!\!{\color{blue}\lambda}^n_j}\right]\mathcal{A}_{[n]}^{1}, & \!\!\!\text{if}\ d=q/2 \\
      \frac{{\color{red}\zeta}_{2n-1}^d}{2^{n+d-1}(8\pi^2)^{n d}}\binom{2n}{n}^d{\color{blue}\lambda}_1{\color{blue}\lambda}_{d+1}^{n-1}\!\!\left[{\displaystyle\prod_{i=2}^d{\color{blue}\lambda}^n_i}\right]\mathcal{A}_{[n]}^{1}, & \!\!\!\text{otherwise}
    \end{cases}+\!...
 \end{equation}
where dots represents higher orders in transcendentality. The rational for this general result varying with the distance $d$ follows the the same line as \eqref{w[12]}, and a direct explanation of \eqref{LOwOq} can be visualized in section \ref{FT:Aq}.

\subsection{Twisted and untwisted operators in presence of Wilson loops}\label{sec:TwistUntw}

All the results collected in the previous sections can be used as building blocks to compute one-point functions of twisted and untwisted operators \eqref{ut} in presence of Wilson loops.
Indeed, the observables $\mathcal{U}$ and $\mathcal{T}$ \eqref{TUdef} can be written in terms of $\mathcal{A}$ as in \eqref{UTinA} and then we identify the latter as the vev on the matrix model \eqref{Adef} computed above. 
The twisted and untwisted operators can be considered the most natural local operators one can build in the circular quiver theories due to the fact that they enjoy good transformation properties under the orbifold action of $\mathbb{Z}_q$.
In this section we consider some examples...

In section \ref{sec:Aq-1corr}, we identified a special observable that can be computed for any $\vec{I}$, $J$ and $q$.
This is the one-point functions in presence of Wilson loops for operators of twist 2, that takes the simple and compact form given in \eqref{A2q}. Starting from this result, we can compute the exact value for $\mathcal{U}$ and $\mathcal{T}$ for $n=2$ that is given by
\begin{equation}\label{UT2}
    \cU_{[2]}^{(\vec{I})}=  \sum_{J=1}^q{\color{blue}\lambda}_J\partial_J\,w_{\vec{I}}^{(q)}\,,\quad\qquad
   \cT_{[2]}^{(\vec{I},J)}= \left({\color{blue}\lambda}_J\partial_J\,w_{\vec{I}}^{(q)}-{\color{blue}\lambda}_{J+1}\partial_{J+1}\,w_{\vec{I}}^{(q)}\right)\,.
\end{equation}
Both the observables have a non trivial expansion in transcendentality, but if we consider the large $N$ limit in the theory at the orbifold point, we can notice some interesting features. Let's take for example the observables \eqref{UT2} for $q=4$ and $\vec{I}=1$. Their value at the orbifold point and in the 't Hooft coupling is written in terms of the expansions \eqref{Aq=4I=1n=2}. Then, considering that $\mathcal{A}_{[2]}^{(1,4)}=\mathcal{A}_{[2]}^{(1,4)}\bigg|_{{\color{blue}\lambda}_{2}\leftrightarrow {\color{blue}\lambda}_{4}}$ for the cyclic symmetry of the quiver, $\cU_{[2]}^{(1)}$ that corresponds to their sum is simply given by
\begin{equation}
    \cU_{[2]}^{(1)}\bigg|_{{\color{blue}\lambda}_I = {\color{blue}\lambda}}={\color{blue}\lambda}_1\partial_1w_1~.
\end{equation}
On the other hand, the twisted correlators are given by the following non-trivial expansions
\begin{equation}\footnotesize\begin{split}
 &\cT_{[2]}^{(1,1)}\bigg|_{{\color{blue}\lambda}_I = {\color{blue}\lambda}}\!\!\!\!=\!{\color{blue}\lambda} \partial_{\color{blue}\lambda} w_1\!-\!\frac{9   \left({\color{red}\zeta}_3 {\color{blue}\lambda} ^3 \partial_{\color{blue}\lambda} w_1\right)}{128 \pi ^4}\!+\!\frac{15   {\color{red}\zeta}_5 {\color{blue}\lambda} ^4 \left(3 \partial_{\color{blue}\lambda} w_1\!+\!12 \partial_{\color{blue}\lambda} ^2w_1\!+\!8 {\color{blue}\lambda} \partial_{\color{blue}\lambda} ^3w_1\right)}{1024 \pi ^6}\!+\!\frac{45   {\color{red}\zeta}_3^2 {\color{blue}\lambda} ^5 \partial_{\color{blue}\lambda} w_1}{8192 \pi ^8}\!-\!\frac{21   {\color{red}\zeta}_7 {\color{blue}\lambda} ^5}{32768 \pi ^8} \bigg[41 \partial_{\color{blue}\lambda} w_1\\
 &+32 \left(9 \partial_{\color{blue}\lambda} ^2w_1+6 \left({\color{blue}\lambda} +5\right) \partial_{\color{blue}\lambda} ^3w_1+8 {\color{blue}\lambda} \left(5 \partial_{\color{blue}\lambda} ^4w_1+{\color{blue}\lambda} \partial_{\color{blue}\lambda} ^5w_1\right)\right)\bigg]-\frac{75   {\color{red}\zeta}_3 {\color{red}\zeta}_5 {\color{blue}\lambda} ^6 \left(7 \partial_{\color{blue}\lambda} w_1+12 \partial_{\color{blue}\lambda} ^2w_1+8 {\color{blue}\lambda} \partial_{\color{blue}\lambda} ^3w_1\right)}{65536 \pi ^{10}}\\
 &+\frac{9   {\color{red}\zeta}_9 {\color{blue}\lambda} ^6}{65536 \pi ^{10}} \bigg[123 \partial_{\color{blue}\lambda} w_1+1020 \partial_{\color{blue}\lambda} ^2w_1+8 \left(\left(85 {\color{blue}\lambda} +930\right) \partial_{\color{blue}\lambda} ^3w_1+40 \left(31 {\color{blue}\lambda} +63\right) \partial_{\color{blue}\lambda} ^4w_1\right.\\
 &\qquad\qquad\qquad\qquad\qquad\qquad\qquad\qquad\left.+8 {\color{blue}\lambda} \left(\left(31 {\color{blue}\lambda} +630\right) \partial_{\color{blue}\lambda} ^5w_1+12 {\color{blue}\lambda} \left(21 \partial_{\color{blue}\lambda} ^6w_1+2 {\color{blue}\lambda} \partial_{\color{blue}\lambda} ^7w_1\right)\right)\right)\bigg]+...
\end{split}\normalsize\raisetag{45pt}\end{equation}
\begin{equation}\footnotesize\begin{split}
 &\cT_{[2]}^{(1,2)}\bigg|_{{\color{blue}\lambda}_I = {\color{blue}\lambda}}\!\!\!\!=\!
 \frac{3   {\color{red}\zeta}_3 {\color{blue}\lambda} ^3 \partial_{\color{blue}\lambda} w_1}{128 \pi ^4}-\frac{5   {\color{red}\zeta}_5 {\color{blue}\lambda} ^4 \left(3 \partial_{\color{blue}\lambda} w_1+12 \partial_{\color{blue}\lambda} ^2w_1+8 {\color{blue}\lambda} \partial_{\color{blue}\lambda} ^3w_1\right)}{1024 \pi ^6}-\frac{27   {\color{red}\zeta}_3^2 {\color{blue}\lambda} ^5 \partial_{\color{blue}\lambda} w_1}{8192 \pi ^8}+\frac{7   {\color{red}\zeta}_7 {\color{blue}\lambda} ^5}{32768 \pi ^8} \bigg[41 \partial_{\color{blue}\lambda} w_1\\
 &+288 \partial_{\color{blue}\lambda} ^2w_1+32 \left(6 \left({\color{blue}\lambda} +5\right) \partial_{\color{blue}\lambda} ^3w_1+8 {\color{blue}\lambda} \left(5 \partial_{\color{blue}\lambda} ^4w_1+{\color{blue}\lambda} \partial_{\color{blue}\lambda} ^5w_1\right)\right)\bigg]+\frac{45   {\color{red}\zeta}_3 {\color{red}\zeta}_5 {\color{blue}\lambda} ^6 \left(7 \partial_{\color{blue}\lambda} w_1+12 \partial_{\color{blue}\lambda} ^2w_1+8 {\color{blue}\lambda} \partial_{\color{blue}\lambda} ^3w_1\right)}{65536 \pi ^{10}}\\
 &-\frac{3   {\color{red}\zeta}_9 {\color{blue}\lambda} ^6}{65536 \pi ^{10}} \bigg[123 \partial_{\color{blue}\lambda} w_1+1020 \partial_{\color{blue}\lambda} ^2w_1+8 \left(\left(85 {\color{blue}\lambda} +930\right) \partial_{\color{blue}\lambda} ^3w_1+40 \left(31 {\color{blue}\lambda} +63\right) \partial_{\color{blue}\lambda} ^4w_1\right.\\
 &\qquad\qquad\qquad\qquad\qquad\qquad\qquad\qquad\left.+8 {\color{blue}\lambda} \left(\left(31 {\color{blue}\lambda} +630\right) \partial_{\color{blue}\lambda} ^5w_1+12 {\color{blue}\lambda} \left(21 \partial_{\color{blue}\lambda} ^6w_1+2 {\color{blue}\lambda} \partial_{\color{blue}\lambda} ^7w_1\right)\right)\right)\bigg]+...
\end{split}\normalsize\raisetag{45pt}\end{equation}
where the remaining two twisted observables are $\cT_{[2]}^{(1,3)}=-\cT_{[2]}^{(1,2)}$ and $\cT_{[2]}^{(1,4)}=-\cT_{[2]}^{(1,1)}$. Notice that for $q>2$ there are different classes of twisted operator, since only few of them share at least a node with the Wilson loop. 

The same behaviour can be verified also considering operators with $n\geq2$ and for many different values of $q$. What we conclude is that in the theory at the orbifold point, the untwisted sector does not perceive the presence of the $\mathbb{Z}_q$ orbifold in the large $N$ limit, confirming the same mechanism of the two-point functions of chiral/anti-chiral operators \cite{Galvagno:2020cgq,Pini:2017ouj}. Then we have 
\begin{equation}\label{UntwlikeN4}
    \cU_{\vec{n}}^{(1)}\bigg|_{{\color{blue}\lambda}_I = {\color{blue}\lambda}}=\mathcal{A}_{\vec{n}}^{1}~,
\end{equation}
namely the untwisted one-point functions correspond to $\cN=4$ results, for any values of $n$ and $q$\footnote{This result can be generalized when a generic number of Wilson loops are inserted and reads $\cU_{\vec{n}}^{(\vec{I})}\big|_{{\color{blue}\lambda}_I = {\color{blue}\lambda}}=\sum_{J=1}^q\mathcal{A}_{\vec{n}}^{(\vec{I},J)}\big|_{\text{Gauss}}$. It is understood that the elements of the sum in $J$ are different from zero only if $\vec{I}$ contains $J$.}.
 
On the contrary, the twisted sector corresponds to the combination of local operator that are sensitive to the orbifold action measuring the discrepancy with respect to the $\cN=4$ theory. For twisted operators which share at least a vector multiplet with the Wilson loop we find the following expansions for the $q>2$ case: 
 \begin{equation}\label{TwistedOrbifold}\footnotesize\begin{split}
\mathcal{T}_{[3]}^{(1,1)}\bigg|_{{\color{blue}\lambda}_I = {\color{blue}\lambda}}&=
\mathcal{A}_{[3]}^{1}
-\frac{15 {\color{red}\zeta} _{5} {\color{blue}\lambda}^4 \left(8 \left(\partial_{\color{blue}\lambda} w_1+2 {\color{blue}\lambda} \partial_{\color{blue}\lambda}^2w_1\right)-w_1\right)}{4096 \sqrt{2} \pi ^6}\!+\!...\\
\mathcal{T}_{[4]}^{(1,1)}\bigg|_{{\color{blue}\lambda}_I = {\color{blue}\lambda}}&=
\mathcal{A}_{[4]}^{1}
-\frac{105 {\color{red}\zeta} _{7} {\color{blue}\lambda} ^6\left(- \partial_{\color{blue}\lambda} w_1+12  \partial_{\color{blue}\lambda}^2w_1+8 {\color{blue}\lambda} \partial_{\color{blue}\lambda}^3w_1\right)}{32768 \pi ^8}\!+\!...\\
\mathcal{T}_{[5]}^{(1,1)}\bigg|_{{\color{blue}\lambda}_I = {\color{blue}\lambda}}&=
\mathcal{A}_{[5]}^{1}
-\frac{189 {\color{red}\zeta} _{9} {\color{blue}\lambda}^7 \left(8 \left(-\!3 \partial_{\color{blue}\lambda} w_1\!-\!6 \left({\color{blue}\lambda} \!-\!4\right) \partial_{\color{blue}\lambda}^2w_1\!+\!32 {\color{blue}\lambda} \left(3 \partial_{\color{blue}\lambda}^3w_1\!+\!{\color{blue}\lambda}  \partial_{\color{blue}\lambda}^4w_1\right)\right)\!+\!w_1\right)}{2097152 \sqrt{2} \pi ^{10}}\!+\!...\\
\mathcal{T}_{[6]}^{(1,1)}\bigg|_{{\color{blue}\lambda}_I = {\color{blue}\lambda}}&=
\mathcal{A}_{[6]}^{1}
-\frac{693 {\color{red}\zeta} _{11} {\color{blue}\lambda} ^9\! \left(3 \partial_{\color{blue}\lambda} w_1\!-\!32\! \left(3 \partial_{\color{blue}\lambda}^2w_1\!+\!2 \left({\color{blue}\lambda} \!-\!15\right) \!\partial_{\color{blue}\lambda}^3w_1\!-\!8 {\color{blue}\lambda}  \!\left(5 \partial_{\color{blue}\lambda}^4w_1\!+\!{\color{blue}\lambda}  \partial_{\color{blue}\lambda}^5w_1\right)\!\right)\!\right)}{16777216 \pi ^{12}}\!+\!...
 \end{split}\normalsize\end{equation}

They possess a non trivial expansion in transcendentality even at the orbifold point, showing the following pattern:
\begin{equation}\label{TwistedGeneral}
   \mathcal{T}_{[n]}^{(1,1)}\bigg|_{{\color{blue}\lambda}_I = {\color{blue}\lambda}}=
\mathcal{A}_{[n]}^{1} \parenth{1 -\frac{{\color{blue}\lambda}^n}{(8\pi^2)^n}\frac{3}{2^n}\binom{2n}{n}{\color{red}\zeta}_{2n-1}} +...
 \end{equation}
For completeness we also report the results for the special case of the symmetric quiver $q=2$, which follow a similar shape, but with slightly different coefficients, due to the presence of a unique twisted operator $\cT_{\vec n}$ (we only report the leading transcendentality deviation for brevity):
 \begin{equation}\label{TwistedOrbifoldq=2}\footnotesize\begin{split}
 \mathcal{T}_{[2]}\bigg|_{{\color{blue}\lambda}_I = {\color{blue}\lambda}}&=
\mathcal{A}_{[2]}^{1}
-\frac{3 {\color{red}\zeta} _{3} {\color{blue}\lambda}^3 \partial_{\color{blue}\lambda} w_1}{32 \pi ^4}\!+\!...\\
\mathcal{T}_{[3]}\bigg|_{{\color{blue}\lambda}_I = {\color{blue}\lambda}}&=
\mathcal{A}_{[3]}^{1}
-\frac{5 {\color{red}\zeta} _{5} {\color{blue}\lambda}^4 \left(8 \left(\partial_{\color{blue}\lambda} w_1+2 {\color{blue}\lambda} \partial_{\color{blue}\lambda}^2w_1\right)-w_1\right)}{1024 \sqrt{2} \pi ^6}\!+\!...\\
\mathcal{T}_{[4]}\bigg|_{{\color{blue}\lambda}_I = {\color{blue}\lambda}}&=
\mathcal{A}_{[4]}^{1}
-\frac{35 {\color{red}\zeta} _{7} {\color{blue}\lambda} ^6\left(- \partial_{\color{blue}\lambda} w_1+12  \partial_{\color{blue}\lambda}^2w_1+8 {\color{blue}\lambda} \partial_{\color{blue}\lambda}^3w_1\right)}{38192 \pi ^8}\!+\!...\\
\mathcal{T}_{[5]}\bigg|_{{\color{blue}\lambda}_I = {\color{blue}\lambda}}&=
\mathcal{A}_{[5]}^{1}
-\frac{63 {\color{red}\zeta} _{9} {\color{blue}\lambda}^7 \left(8 \left(-\!3 \partial_{\color{blue}\lambda} w_1\!-\!6 \left({\color{blue}\lambda} \!-\!4\right) \partial_{\color{blue}\lambda}^2w_1\!+\!32 {\color{blue}\lambda} \left(3 \partial_{\color{blue}\lambda}^3w_1\!+\!{\color{blue}\lambda}  \partial_{\color{blue}\lambda}^4w_1\right)\right)\!+\!w_1\right)}{524288 \sqrt{2} \pi ^{10}}\!+\!...\\
\mathcal{T}_{[6]}\bigg|_{{\color{blue}\lambda}_I = {\color{blue}\lambda}}&=
\mathcal{A}_{[6]}^{1}
-\frac{231 {\color{red}\zeta} _{11} {\color{blue}\lambda} ^9\! \left(3 \partial_{\color{blue}\lambda} w_1\!-\!32\! \left(3 \partial_{\color{blue}\lambda}^2w_1\!+\!2 \left({\color{blue}\lambda} \!-\!15\right) \!\partial_{\color{blue}\lambda}^3w_1\!-\!8 {\color{blue}\lambda}  \!\left(5 \partial_{\color{blue}\lambda}^4w_1\!+\!{\color{blue}\lambda}  \partial_{\color{blue}\lambda}^5w_1\right)\!\right)\!\right)}{4194304 \pi ^{12}}\!+\!...
 \end{split}\normalsize\end{equation}

As a final remark, we mention the possibility to combine the observables defined in section \ref{Sec:2} in all the possible ways. Since \eqref{ut} and \eqref{WuWt} are defined as linear combinations of the operators $ O_{\vec n}^{(J)}$ and $W_I$ it is understood that any correlators involving them can be written in terms of the results obtained in the previous sections. We leave to the reader the possibility to compute them using the set of results we included in the notebook \texttt{WLcorrelators.nb} attached to this manuscript.
However we want to point out some interesting features of this correlators in the theory $A_{q-1}$ at the orbifold point and in the 't Hooft limit.
Since the twisted Wilson loops are vanishing in this limit \eqref{wtorb}, we conclude:
\begin{equation}
    \vev{\mathcal{W}^t_I \;U_{\vec{n}}}_q=
    \vev{\mathcal{W}^t_I \;T^{(J)}_{\vec{n}}}_q=
    \vev{\mathcal{W}^t_I :\mathcal{O}_{\vec{n}}^{(J)}:}_q=0~,\qquad\quad {\color{blue}\lambda}_I={\color{blue}\lambda} \;\&\;N\rightarrow \infty~.
\end{equation}
On the other hand, the untwisted Wilson loop reduces to \eqref{wuorb}, then the one-point function of a local operator takes the following form:
\begin{equation}
    \vev{\mathcal{W}^u :\mathcal{O}_{\vec{n}}^{(J)}:}_q=\mathcal{A}_{\vec{n}}^{(1,J)}~,\qquad\quad {\color{blue}\lambda}_I={\color{blue}\lambda} \;\&\;N\rightarrow \infty~,
\end{equation}
while the one of the twisted and untwisted operators are given by
\begin{equation}\begin{split}
    \vev{\mathcal{W}^u U_{\vec{n}}}_q&=\mathcal{U}_{\vec{n}}^{(1)}=\mathcal{A}_{\vec{n}}^{(1)}~,\\
    &\qquad\qquad\qquad\qquad\qquad\qquad\qquad {\color{blue}\lambda}_I={\color{blue}\lambda}~, \;\&\;N\rightarrow \infty\\
    \vev{\mathcal{W}^u T_{\vec{n}}^{(I)}}_q&=\mathcal{T}^{(1,I)}_{\vec{n}}=\mathcal{A}_{\vec{n}}^{(1,I)}-\mathcal{A}_{\vec{n}}^{(1,I+1)}~,
\end{split}\end{equation}
where we used \eqref{UTinA}.

\section{Diagrammatic interpretation in the planar limit}\label{Sec:FieldTheory}
In this section we discuss the diagrammatic interpretation of the results shown in the previous sections, with the aim of visualizing the main features and differences of the theories discussed in the present paper ($\cN=4$, SCQCD and $A_{q-1}$ quivers) also at the level of traditional Feynman diagrams. Moreover, comparing the matrix model results and their Feynman graph representation allows to identify the solution of very involved high loop Feynman integrals.

\subsection{$\cN=4$ SYM: contribution from rainbow diagrams}
We first review the behavior of the maximally supersymmetric case, where starting from Feynman diagrams computations many exact results were obtained for the circular Wilson loop vev \cite{Erickson:2000af,Drukker:2000rr} and for the one-point functions with chiral operators. 

\begin{figure}[!t]
\begin{minipage}[t]{.5\textwidth}
        \centering
        \includegraphics[width=.7\textwidth]{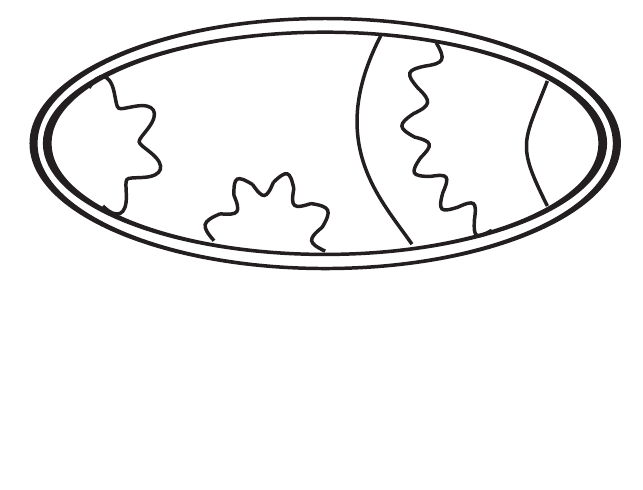}
        \subcaption{Wilson loop vev}\label{Fig:WL}
    \end{minipage}
    \begin{minipage}[t]{.5\textwidth}
        \centering
        \includegraphics[width=.7\textwidth]{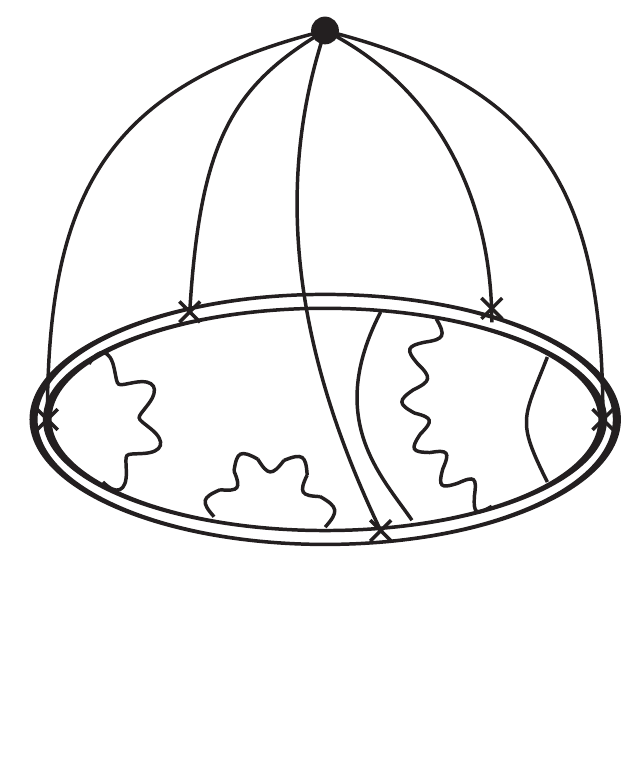}
        \subcaption{One-point function with chiral operators}\label{Fig:WLO}
    \end{minipage}  

    \caption{Typical diagrams contributing to $\vev{W}$ and $\vev{W O_{\vec{n}}}$ in the $\cN=4$ theory. The double line represents the Wilson loop, the wavy line is the gauge field, while the straight line stands for the scalar field in the Wilson loop expression.}
\end{figure}

The crucial point for $\vev{W}$ is that all the diagrams which contain internal vertices cancel, and the whole contribution for each order in perturbation theory is given by rainbow diagrams, see Figure \ref{Fig:WL}, i.e. planar combinations of the tree level propagators of the gauge and the scalar fields appearing in the Wilson loop:
\begin{align}\label{gaugescalarprop}
    \vev{\varphi^a(x)\bar \varphi^b(0)}_{\mathrm{tree}} = \frac{\delta^{ab}}{4\pi^2x^2}~,\hspace{1cm} \vev{A^a_{\mu}(x)\bar A^b_{\nu}(0)}_{\mathrm{tree}} = \frac{\delta^{ab}\delta_{\mu\nu}}{4\pi^2x^2}~.
\end{align}
The combination of the gauge and scalar propagators, represented as a unique wavy/straight line, is independent of distances, and the Wilson loop vev has only a coupling dependence:
\begin{align}\label{N4WLvev}
    \vev{W} &= 1+ \parbox[c]{.13\textwidth}{\includegraphics[width = .13\textwidth]{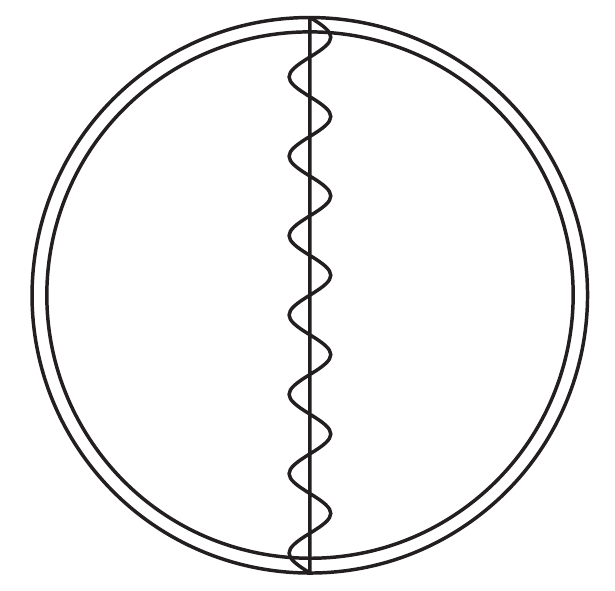}}+\parbox[c]{.13\textwidth}{\includegraphics[width = .13\textwidth]{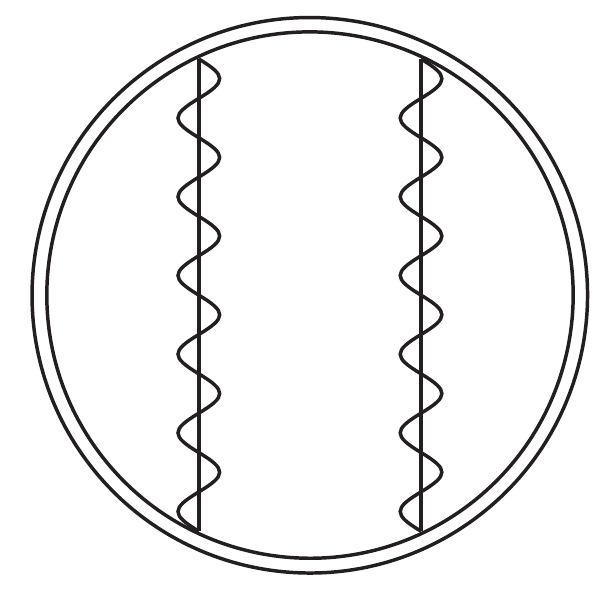}}+\parbox[c]{.13\textwidth}{\includegraphics[width = .13\textwidth]{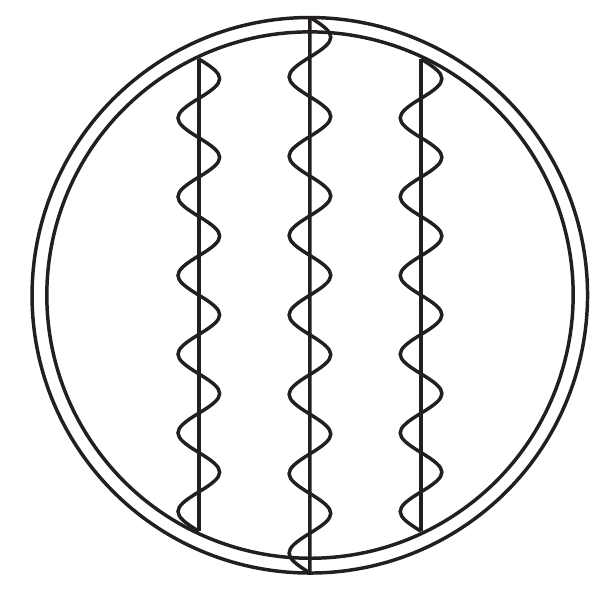}}+\dots\notag \\
    &=1+\frac{{\color{blue}\lambda}}{8}+\frac{{\color{blue}\lambda}^2}{192}+\frac{{\color{blue}\lambda}^3}{9216}+\dots = \frac{2}{\sqrt{{\color{blue}\lambda}}} I_1(\sqrt{{\color{blue}\lambda}})~.
\end{align}

This pure combinatorial problem can be immediately resummed using a matrix model. Such matrix model has then been derived in a rigorous way using supersymmetric localization \cite{Pestun:2007rz}, and leads to the well-known exact result of $\cN=4$.

As for the one-point function with chiral operators, the situation is similar. Due to residual conformal symmetry the spacetime dependence of $\vev{W O_{\vec{n}}(x)}$ is completely fixed to the average distance between the local operator and the Wilson loop \cite{Billo:2018oog}, and the tree level is given by connecting $O_{\vec{n}}$ to $W$ with $n$ scalar propagators. Considering perturbative corrections, the diagrams with internal vertices correcting the Wilson loop cancel as before, and there are no perturbative corrections to the scalar legs belonging to the chiral operator (see \cite{Semenoff:2001xp,Pestun:2002mr,Giombi:2012ep,Bonini:2014vta}). Hence the only diagrams contributing to $\vev{W O_{\vec{n}}}$ are the internal planar corrections to the Wilson loop as before, as depicted in Figure \ref{Fig:WLO}. This fact explains the exact results \eqref{WLexactLambda} for any values of the coupling ${\color{blue}\lambda}$.

\subsection{Perturbative SCQCD}\label{FT:SCQCD}
Moving to the $\cN=2$ case, the situation is more elaborate. The present perturbative analysis makes use of $\cN=1$ superspace formalism, which has been developed in a series of papers \cite{Billo:2017glv,Billo:2018oog,Billo:2019job,Billo:2019fbi, Galvagno:2020cgq}.
\paragraph{Wilson loop vev}
The perturbative analysis of the Wilson loop vev in SCQCD has been originally explored in \cite{Andree:2010na} and fully developed in \cite{Billo:2019fbi} also for other classes of $\cN=2$ theories with matter content in symmetric/anti-symmetric representation of the gauge group. The presence of a non-trivial matter content ($2N$ fundamental hypermultiplets in the present case) generates a wide number of non-trivial perturbative corrections and there is no possibility of obtaining exact results for any ${\color{blue}\lambda}$ as in $\cN=4$ case. However it is possible to identify the first corrections to each transcendentality order ${\color{red}\zeta}_{2n-1}$, namely the $n$-loops correction to the gluon/scalar propagator of eq. \eqref{gaugescalarprop} (see section 4.3 of \cite{Billo:2019fbi} for a detailed explanation).
\begin{align}
      \parbox[c]{.2\textwidth}{\includegraphics[width = .2\textwidth]{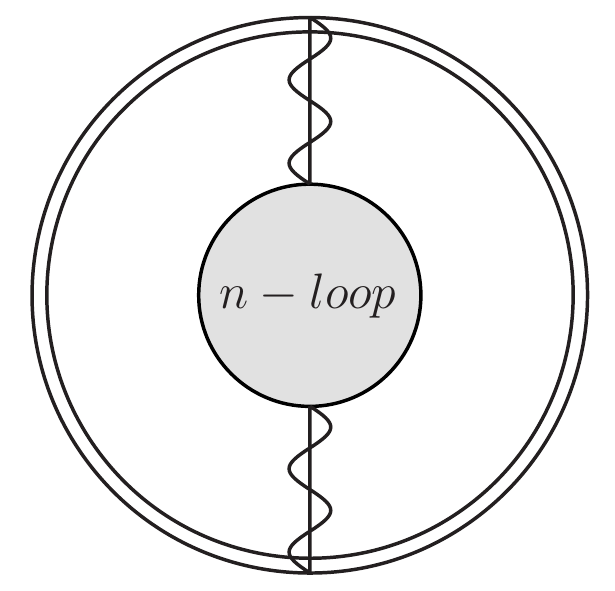}}\propto {\color{blue}\lambda}^{n}{\color{red}\zeta}_{2n-1}~,
\end{align}

For example the three-loops correction to the Wilson loop vev is given by the insertion of the following two-loops correction of the scalar propagator (explicitly computed in \cite{Billo:2017glv}):
\begin{align}\label{scalarprop2L}
    \parbox[c]{.2\textwidth}{\includegraphics[width = .2\textwidth]{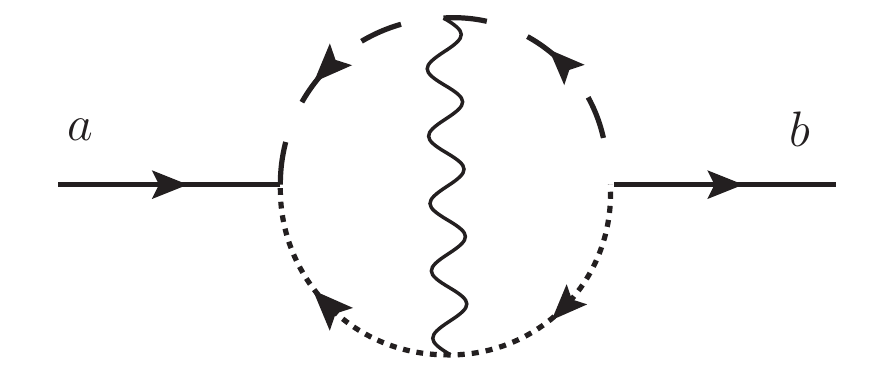}} = \frac{-3{\color{red}\zeta}_3{\color{blue}\lambda}^2}{(8\pi^2)^2N^2}(N^2+1) \frac{\delta^{ab}}{4\pi^2x^2} ~,
\end{align}

 to be summed to the correction to the gauge propagator (which differ simply by its Lorentz coupling, as for the tree level \eqref{gaugescalarprop}). Together they reproduce the first ${\color{red}\zeta}(3)$ term of the matrix model expansion.

\paragraph{One-point function}
Considering the one-point function of chiral operators, the situation is pretty similar: the first correction to ${\color{red}\zeta}_3$ transcendentality term has been achieved in \cite{Billo:2018oog}, and is given by the insertion of the following subdiagrams.
\begin{align}\label{WLO2loop}
    \parbox[c]{.2\textwidth}{\includegraphics[width = .2\textwidth]{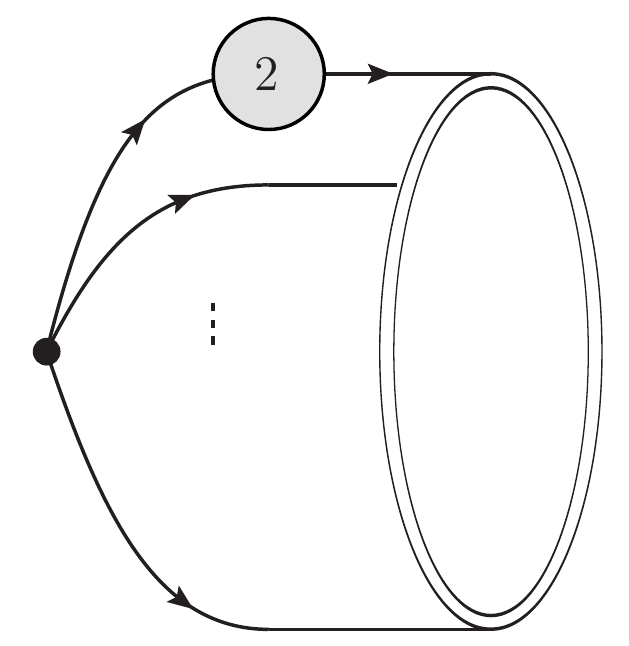}}~,\hspace{1.5cm} \parbox[c]{.2\textwidth}{\includegraphics[width = .2\textwidth]{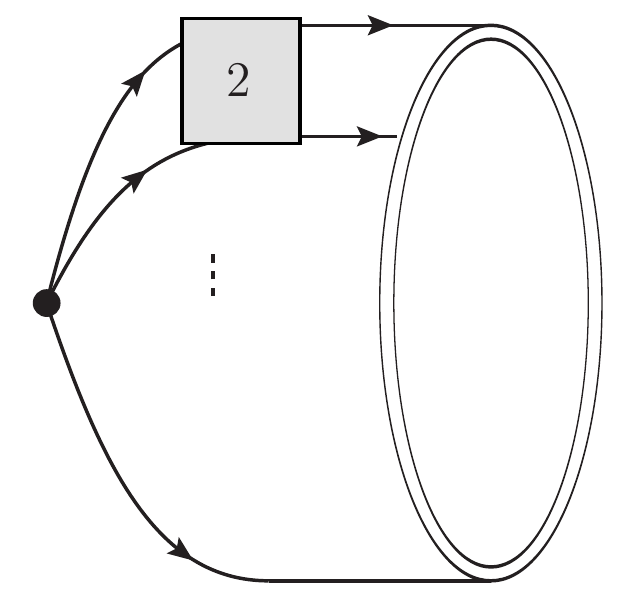}}~.
\end{align}

The two-loop corrected propagator is given in \eqref{scalarprop2L}, while the second diagram corresponds to the following 4-legs diagram:
\begin{align}\label{zeta3box}
    \parbox[c]{.2\textwidth}{\includegraphics[width = .2\textwidth]{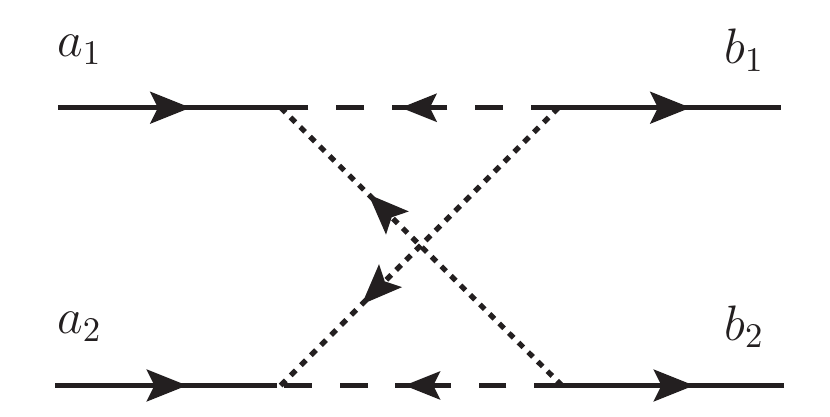}} = \frac{-6{\color{red}\zeta}_3{\color{blue}\lambda}^2}{(16\pi^2)^2N^2}\parenth{\delta^{a_1b_1}\delta^{a_2b_2}+\delta^{a_1b_2}\delta^{a_2b_1}+\delta^{a_1a_2}\delta^{b_1b_2}} \frac{1}{(4\pi^2x^2)^2}~.
\end{align}
The combination of all the possible diagrams displayed in \eqref{WLO2loop} yields the 2-loops correction to the one-point coefficient $\mathcal{A}_n^{(1)}$ for the first term of ${\color{red}\zeta}_3$ transcendentality, for the results displayed in equations \eqref{A2scqcd}-\eqref{A6scqcd}.

\paragraph{Exponentiation of the ${\color{red}\zeta}_3$ term}
We can extend the previous diagrammatic analysis for SCQCD, in particular we justify at the level of Feynman diagrams the exponentiation of the ${\color{red}\zeta}_3$ term observed from matrix model calculations in section \ref{sec:WLSCQCD}.
The two possible subdiagrams responsible for ${\color{red}\zeta}_3$ corrections are the two-loops propagator \eqref{scalarprop2L} and the 4-legs diagram \eqref{zeta3box}. However, looking at their color factors it is clear that, when these diagrams are inserted in a single trace, the 4-legs diagram is subleading as $N^{-2}$ with respect to the two-loops propagator at each perturbative order.
Hence, we can derive at the diagrammatical level the full ${\color{red}\zeta}_3$ transcendentality: at each perturbative order, the ${\color{red}\zeta}_3$ contribution is given by a single correction of a gauge/scalar propagator applied to the original $\cN=4$ expansion, depicted in equation \eqref{N4WLvev}. The structure is the following:
\begin{align}\label{N2WLvev}
    \vev{W}\Big|_{{\color{red}\zeta}_3} &=  \parbox[c]{.13\textwidth}{\includegraphics[width = .13\textwidth]{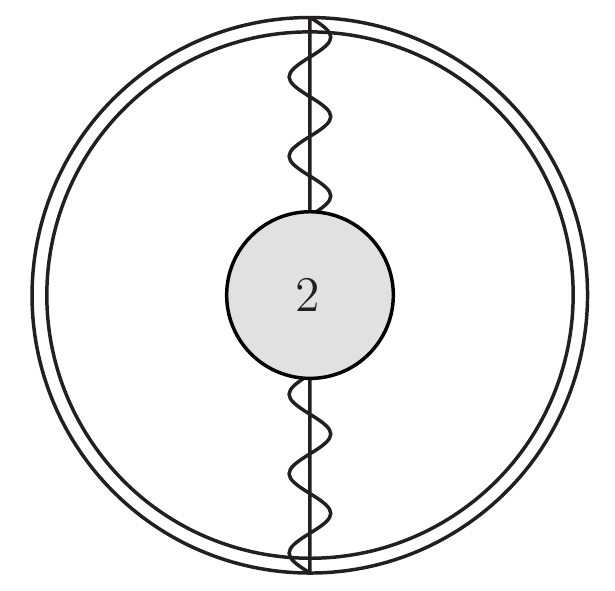}}+2~\parbox[c]{.13\textwidth}{\includegraphics[width = .13\textwidth]{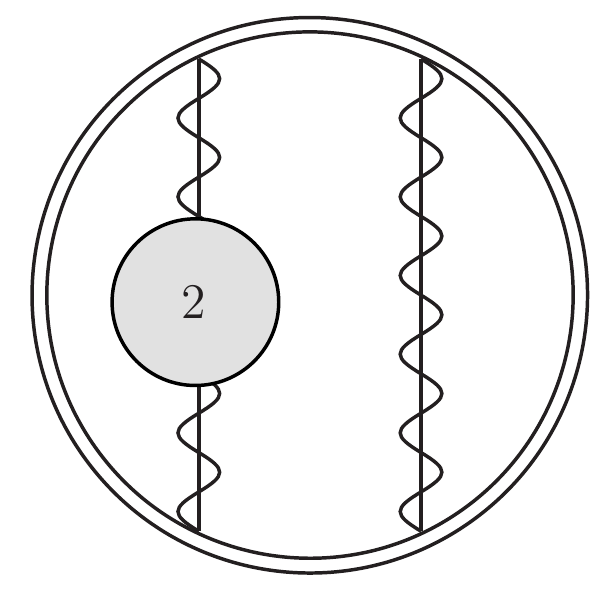}}+3~\parbox[c]{.13\textwidth}{\includegraphics[width = .13\textwidth]{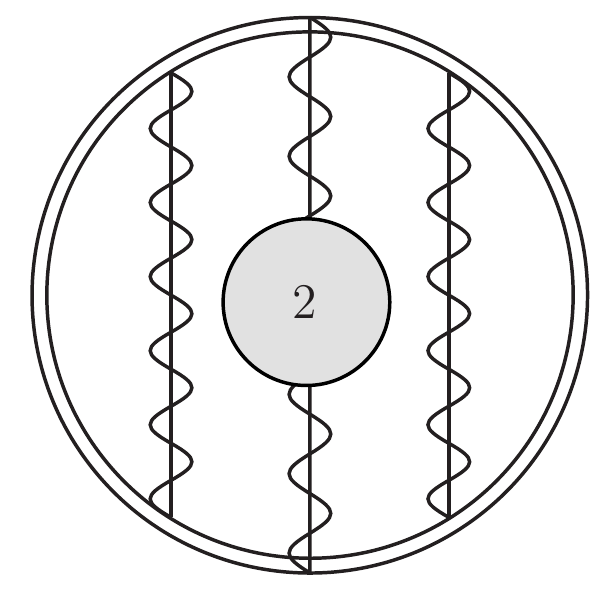}}+\dots\notag \\
    &=\frac{-3{\color{red}\zeta}_3{\color{blue}\lambda}^2}{(8\pi^2)^2} \left(\frac{{\color{blue}\lambda}}{8}+\frac{{\color{blue}\lambda}^2}{96}+\frac{{\color{blue}\lambda}^3}{3072}+\dots\right) = \frac{-3{\color{red}\zeta}_3{\color{blue}\lambda}^2}{(8\pi^2)^2}~ {\color{blue}\lambda}\partial_{\color{blue}\lambda} w~.
\end{align}
where $w$ is the Wilson loop vev in $\cN=4$ SYM.
Such mechanism can be generalized for higher powers of ${\color{red}\zeta}_3$ transcendentality. The general ${\color{red}\zeta}_3^k$ term is given by $k$ insertions of a two-loops bubble inside the $\cN=4$ result. This purely combinatorial problem gives rise to the exponentiation described in equation \eqref{expresum1}.

From this sketchy summary it is evident that the perturbative analysis of $\cN=2$ SCQCD using Feynman diagrams allows the direct computation of a rather limited number of contributions, especially if compared with $\cN=4$ case, where a Feynman diagram analysis lead to exact results in ${\color{blue}\lambda}$. Only transcendentality ${\color{red}\zeta}_3$ is really treatable\footnote{The first contribution to transcendentality ${\color{red}\zeta}_5$ at the diagrammatical level has been achieved in \cite{Billo:2019fbi}.}, while it is hard to extend this analysis further.

\subsection{$A_{q-1}$ theories: cancellations at the orbifold point}\label{FT:Aq}
 We shall see how perturbative computations using Feynman diagrams can be pushed very far in perturbation theory for $A_{q-1}$ quivers, proving again their role as interpolating theory between SCQCD and $\cN=4$.
\paragraph{Wilson loop vev}
It is interesting to understand the cancellation properties for $\vev{W_I}_q$ at the orbifold point, seen in section \ref{Sec4.3}, at the level of Feynman diagrams.

The set of Feynman diagrams contributing to $w_I^{(q)}$ is the same as the SCQCD case, provided that we include the contributions from all the nodes. This additional contributions evaluated at the orbifold point are responsible for the cancellations discussed above. We consider again the two-loop correction to the scalar propagator as an example\footnote{As in \eqref{N4WLvev}, the same happens for the gauge field propagator, with the only addition of the $\delta_{\mu\nu}$ Lorentz factor.}. 
Considering the Lagrangian and the Feynman rules represented in Appendix \ref{App:FieldTheory}, the two-loops propagator in a $A_{q-1}$ theory is given by three contributions:
\begin{align}\label{Aqscalarprop}
    \parbox[c]{.2\textwidth}{\includegraphics[width = .2\textwidth]{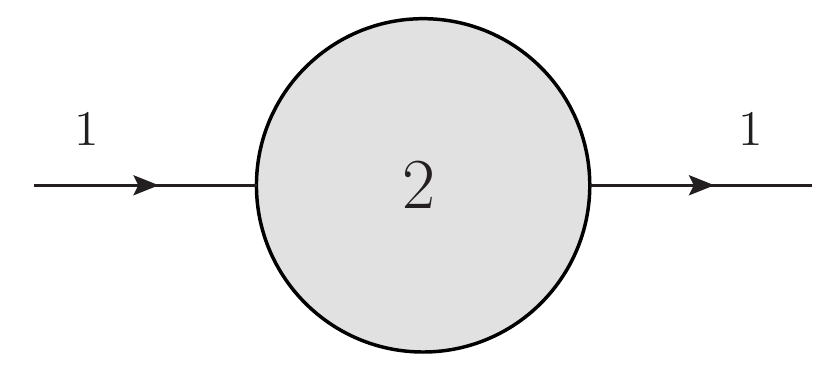}} &= \parbox[c]{.2\textwidth}{\includegraphics[width = .2\textwidth]{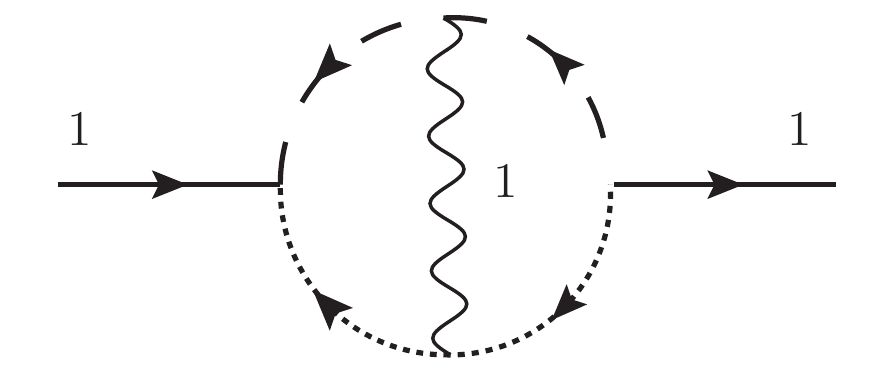}}+\parbox[c]{.2\textwidth}{\includegraphics[width = .2\textwidth]{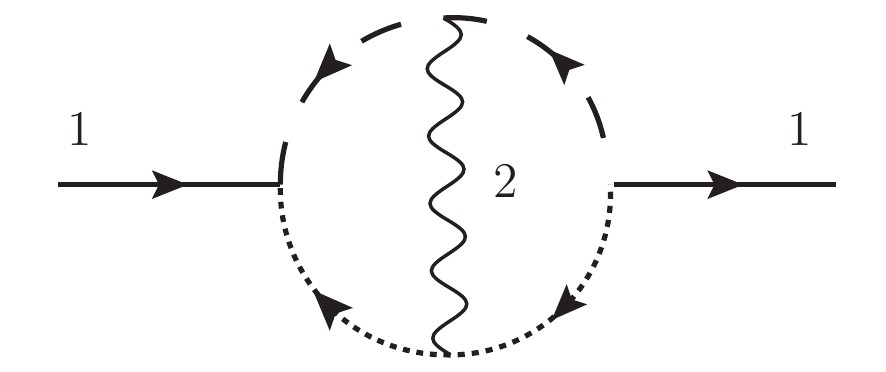}}+\parbox[c]{.2\textwidth}{\includegraphics[width = .2\textwidth]{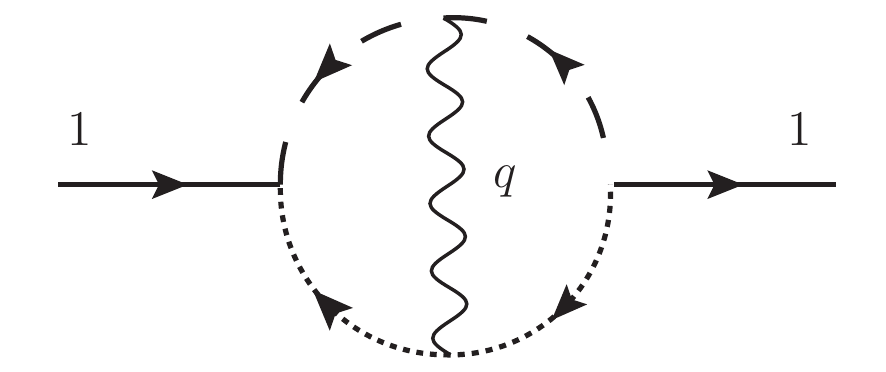}}\notag \\
  &=\frac{-3{\color{red}\zeta}_3}{(8\pi^2)^2}\frac{{\color{blue}\lambda}_1}{N^2}\parenth{{\color{blue}\lambda}_1(N^2+1)-\frac{1}{2}({\color{blue}\lambda}_2+{\color{blue}\lambda}_q)(N^2-1)} \frac{\delta^{ab}}{4\pi^2x^2}  ~.
\end{align}
the spacetime integral is the same as \eqref{scalarprop2L}, whereas the color factor gets the influence of the neighboring nodes, as computed in \eqref{A9}. From the result \eqref{Aqscalarprop} it is clear that in the planar limit and at the orbifold point the two-loops correction to the scalar propagator exactly cancels.
Analogous cancellations among neighboring nodes happen also for all higher contributions in perturbation theory involving hypermultiplets. The only residual diagrams must not involve matter fields and therefore return precisely the $\cN=4$ contribution, represented in Figure \ref{Fig:WL}. This mechanism nicely explains the result \eqref{wqvanishes} for quiver theories at the orbifold point.

\paragraph{One-point functions: operator and Wilson loop in the same node.} It is now interesting to see which diagrams are preserved at the orbifold point and provide non-trivial contributions for the one-point coefficient. We start by the case of $W_I$ and $O_n^{(J)}$ in the same node, namely $J=I$ , in order to have a direct comparison with $\cN=4$ and SCQCD. The massive cancellations of diagrams involving matter hypermultiplets seen for the Wilson loop vev happen for this observable as well, however it is possible to identify some diagrams that are preserved by the planar limit and the orbifold point. The first deviations from the $\cN=4$ results for $\vev{W_I O_n^{(J)}(x)}$ are due to the following diagram, made of a hypermultiplet loop with $2n$ adjoint legs:
\begin{align}\label{Davydichev11}
\parbox[c]{.15\textwidth}{\includegraphics[width = .15\textwidth]{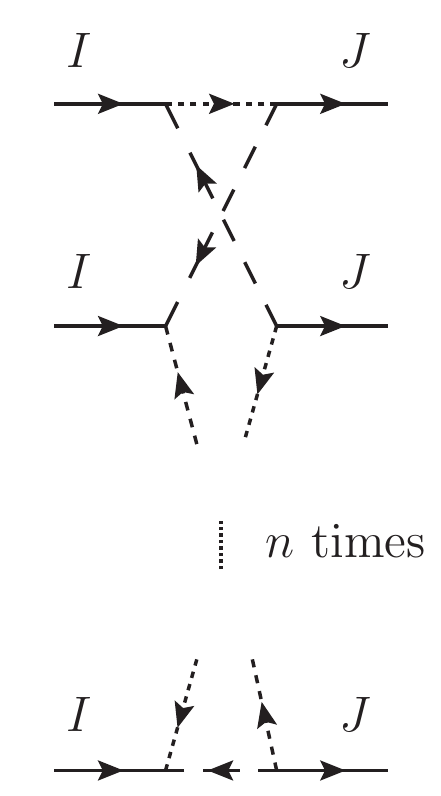}} \quad
=~ \Big(\frac{-1}{16\pi^2 N}\Big)^{n} \,{\color{blue}\lambda}_I^{n/2}{\color{blue}\lambda}_J^{n/2}\,\binom{2n}{n} \,\frac{{\color{red}{\color{red}\zeta}}_{2n-1}}{n} \,\times \,
\Big(\frac{1}{4\pi^2 x^2}\Big)^n~.
\end{align}
The labels on the scalar fields indicate the node the adjoint scalar field belong to. The result \eqref{Davydichev11} is fully derived in Appendix D of \cite{Galvagno:2020cgq} with the help of the uniqueness relations\footnote{For a recent review of the star-triangle relation together with their Mathematica implementation see \cite{Preti:2018vog,Preti:2019rcq}.} (see also \cite{Billo:2017glv} and \cite{Beccaria:2020hgy}).
Inserting this subdiagram with $I=J=1$ into $\vev{W_1 O^{(1)}_n(x)}$, it is possible to explain the full correction (for any values of ${\color{blue}\lambda}_I$) proportional to the first transcendentality term ${\color{red}\zeta}_{2n-1}$: we correct the full $\cN=4$ result, which is the sum of all rainbow diagrams as depicted in Figure \ref{Fig:WLO}, with the subdiagram \eqref{Davydichev11}, obtaining the following combination:
\begin{align}\label{AnDiag}
\mathcal{A}_n^{(1,1)} = \sum_{\mathrm{rainbow}} \parbox[c]{.18\textwidth}{\includegraphics[width = .18\textwidth]{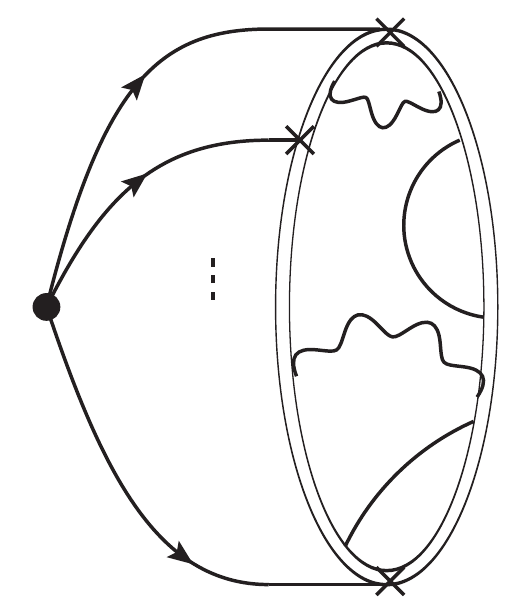}} +\!\! \sum_{\mathrm{rainbow}} \parbox[c]{.25\textwidth}{\includegraphics[width = .25\textwidth]{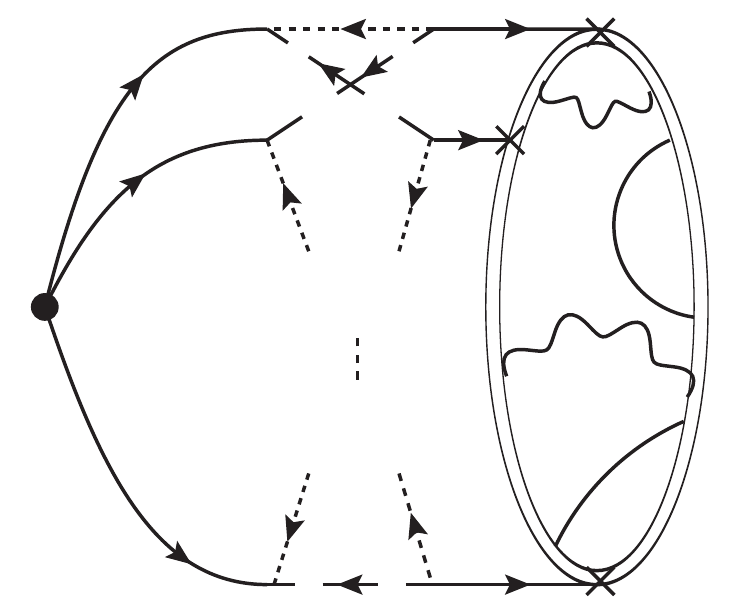}}\!+\!\mathrm{higher~ transc}~.
\end{align}
The general result displayed in \eqref{orbifoldAq11} is perfectly reproduced at the first trascendentality deviation, for any values of ${\color{blue}\lambda}_I$. We see how also for the one-point functions the perturbative expansions are extremely more accessible with respect to the SCQCD case, and such result also at a diagrammatic level lead to some resummation, as in $\cN=4$ case.

A similar analysis allows to compute the first transcendentality term for the one-point functions with operators in different nodes with respect to the Wilson loop. We consider the generic distance $d$ case $\vev{W_1 O_n^{(d+1)}}$. The subdiagram \eqref{Davydichev11} can be seen as a building block which allows to connect neighboring nodes
\begin{align}
\label{MultiDavydichevd1}
\parbox[c]{.35\textwidth}{\includegraphics[width = .35\textwidth]{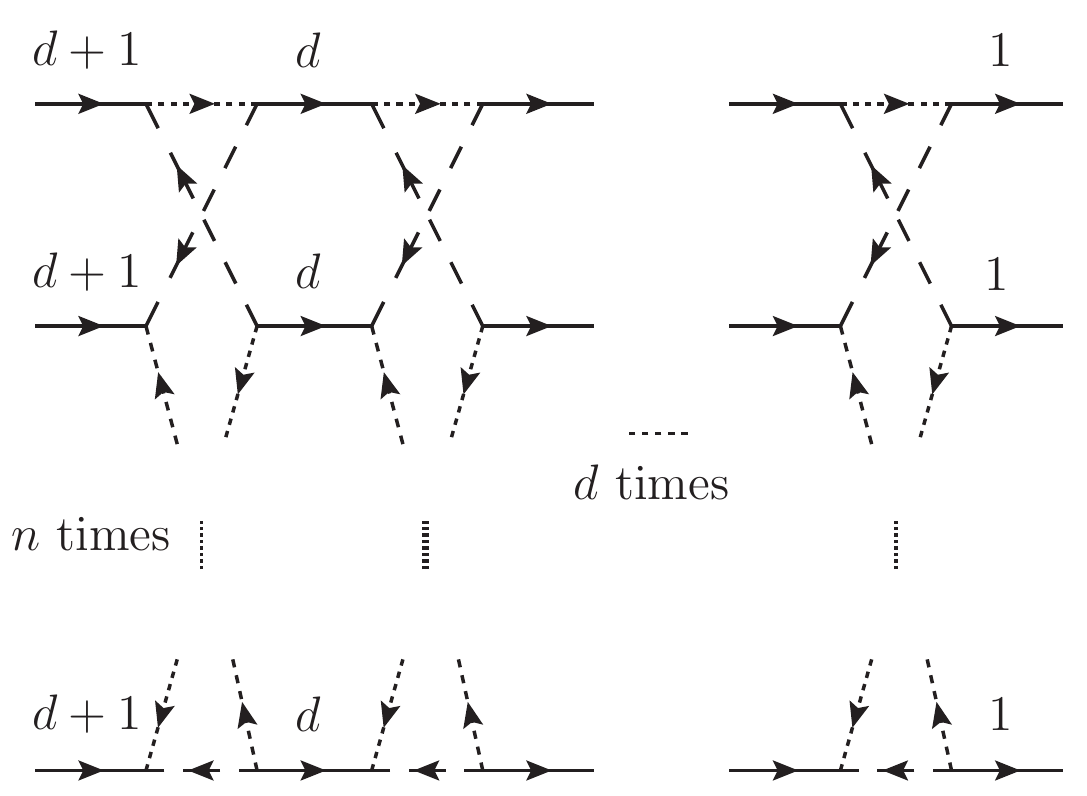}} \!= n^{d-1}   \left[\parenth{\frac{-1}{16\pi^2}}^{n}\binom{2n}{n}\frac{{\color{red}\zeta}_{2n-1}}{n}\right]^d\prod_{I=1}^{d+1}{\color{blue}\lambda}_I^{n/2}\Big(\frac{1}{4\pi^2 x^2}\Big)^n ~,
\end{align}
Therefore the first transcendentality correction to $\vev{W_1 O_n^{(d+1)}}$ is given by the following diagram.
\begin{align}\label{AnDiagd1}
\mathcal{A}_n^{(1,d+1)} = \sum_{\mathrm{rainbow}} \parbox[c]{.4\textwidth}{\includegraphics[width = .4\textwidth]{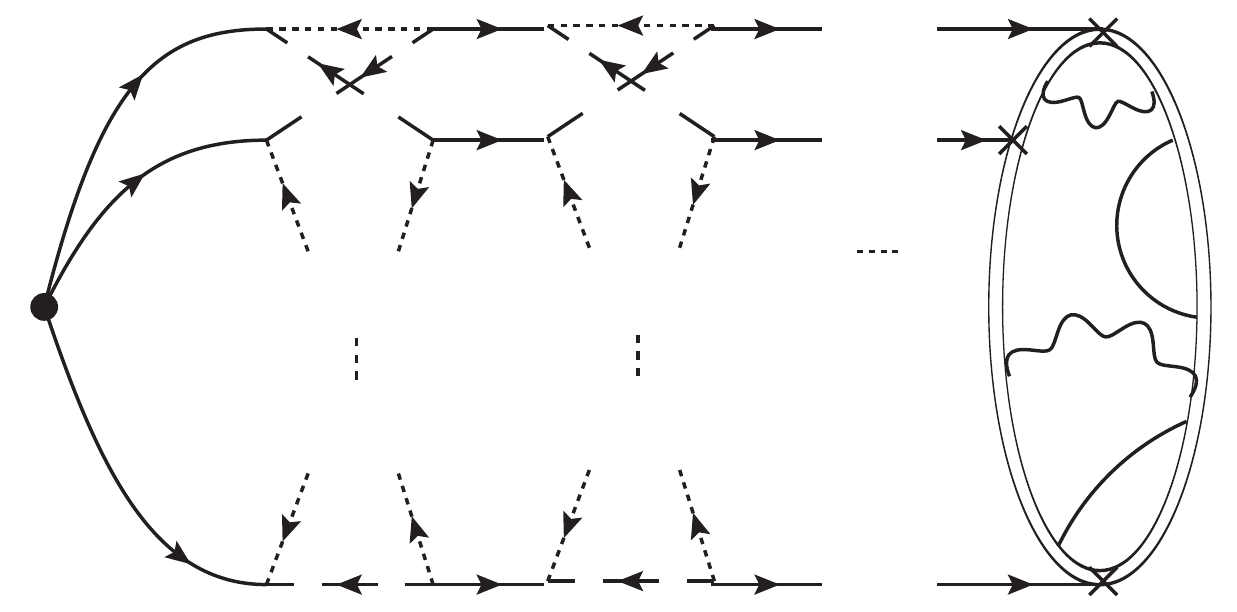}} +\mathrm{higher~ transcendentalities}~,
\end{align}
which is the consistent diagrammatic interpretation of 
\eqref{LOwOq}.

The Feynman diagram in \eqref{MultiDavydichevd1} have the same structure the fermionic wheels appearing in the integrable fishnet theories studied in \cite{Caetano:2016ydc,Kazakov:2018gcy,Pittelli:2019ceq,Levkovich-Maslyuk:2020rlp,PittelliPreti2021}. In particular, a suitable scaling limit of the $\gamma$-deformation of the $\mathcal{N}=2$ $A_{q-1}$ theories produces a conformal non-supersymmetric integrable field theory that generates diagrams given only by combinations of the one in \eqref{MultiDavydichevd1}. Those Yukawa vertices are coming from the vector multiplet part of the Lagrangian, while the one we are analyzing in this section are related to the hypermultiplets. However, since the $\gamma$-deformation acts also on the hypers even if it is subleading respect to the vector multiplet part, one can in principle modify the scaling limit in order to select only the hypermultiplet contributions.
It could be nice to explore this possibility in the future and compare it with our general results.

These diagrams are also responsible for the results for the untwisted and twisted operators, see \eqref{UTinA}: for the untwisted combination of operators the transcendentality terms involving hypermultiplets mutually cancel and return the pure $\cN=4$ results. For the twisted combinations, instead, we get non-trivial corrections which confirm the results shown in section \ref{sec:TwistUntw}.

\subsubsection{Correlators of coincident Wilson loops}\label{FT:WLcoincident}
A further proof that \eqref{Davydichev11} represents a crucial building block for diagrammatic expansions of $\cN=2$ theories comes from correlators of multiple Wilson loops, belonging to different nodes. Indeed it is possible to explain at a diagrammatical level the results of section \ref{Sec4.3}.

\begin{figure}[!t]
\begin{center}
\includegraphics[scale=0.55]{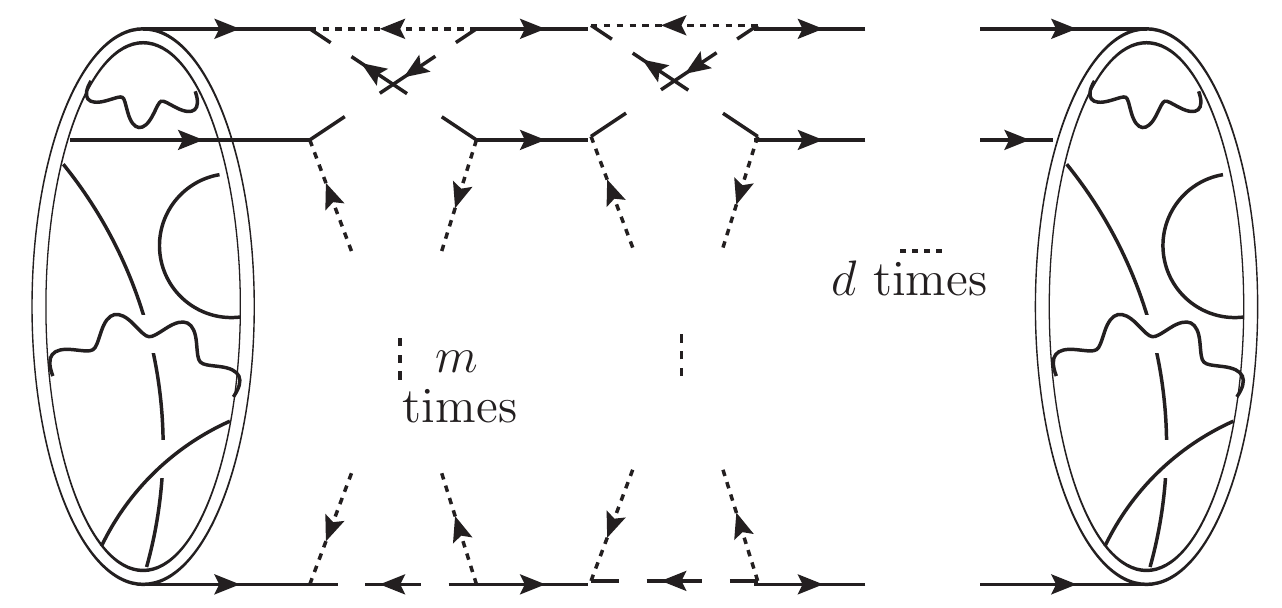}
\caption{Diagrams contributing with transcendentality ${\color{red}\zeta}_{2m-1}^d$ to the connected part of $w_{[1,d+1]}^{(q)}$ at finite $N$ and generic values of the couplings.}
\label{Fig:WlWl}
\end{center}
\end{figure}

Considering the correlator of two Wilson loops belonging to two nodes at a distance $d$, whose matrix model result is shown in \eqref{w[12]}, the connected part of $w_{[1,d+1]}^{(q)}$ is reproduced by the multiple insertion of \eqref{Davydichev11}, following the pattern depicted in Figure \ref{Fig:WlWl}.

Notice that diagrams like those in Figure \ref{Fig:WlWl} explain all the transcendentality terms ${\color{red}\zeta}_{2m-1}^d$ in the expansion of $w_{[1,d+1]}^{(q)}$, and in particular the ${\color{red}\zeta}_{3}^d$ term, explicitly written in \eqref{w[12]}. Besides, this diagrammatic explanation holds for any values of the couplings and at finite N. As explained in section \ref{FT:SCQCD}, the $2m$-legs building blocks \eqref{Davydichev11} are subleading in $N$ when inserted in the Wilson loop expression. This explains the factorization \eqref{w1dLargeN} in the large-$N$ limit.

\begin{figure}[!t]
\begin{center}
\includegraphics[scale=0.55]{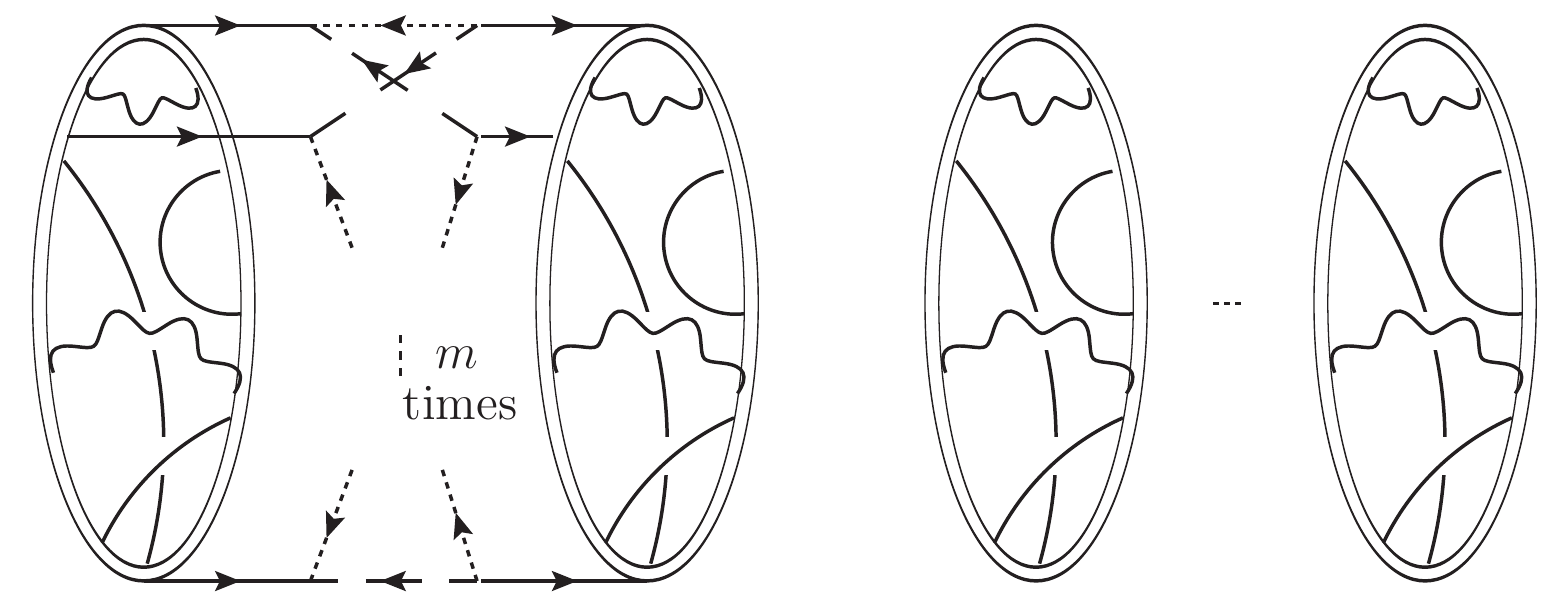}
\caption{Diagrams contributing with transcendentality ${\color{red}\zeta}_{2m-1}$ to the connected part of $w_{[1,2,\dots, q]}^{(q)}$ at finite $N$ and generic values of the couplings.}
\label{Fig:Wlmultiple}
\end{center}
\end{figure}

The second interesting example of correlator of coincident Wilson loops which can be explained at a diagrammatical level is the observable $w_{[1,2,\dots, q]}^{(q)}$. In this case we have a Wilson loop for each node, therefore the first nontrivial connected contribution at transcendentality ${\color{red}\zeta}_{2m-1}$ is given by the insertion of a $2m$-legs building block between two neighboring Wilson loops, as displayed in Figure \ref{Fig:Wlmultiple}.

In particular the first transcendentality term ${\color{red}\zeta}_3$ is explicitly shown in equation \eqref{w[12...q]} and is captured by a diagram analogous to Figure \ref{Fig:Wlmultiple} with $m=2$. Again, these connected contributions are all subleading in $N$ and therefore the factorization properties \eqref{w[12...q]largeN} hold in the planar limit and at the orbifold point.

\vskip 1.5cm
\noindent {\large {\bf Acknowledgments}}
\vskip 0.2cm
We thank M. Beccaria, M. Bill\`o, M. Frau, A. Lerda, A. Pini and K. Zarembo for many useful discussions and suggestions.

\noindent
The work of F.G. was supported by a grant from the Swiss National Science Foundation, as well as via the NCCR SwissMAP.
The work of M.P. was supported by European Research Council
 (ERC) under the European Union’s Horizon 2020 research and innovation programme (grant agreement No. 865075) EXACTC.
\vskip 1cm

\begin{appendix}
\section{Correlators of chiral operators with multiple coincident Wilson loops}\label{App:OmultipleW}
For completeness we include the most general case of our analysis, namely the insertion of a generic chiral operator $O_{\vec n}^{(I)}$ inside a correlator of multiple coincident Wilson loops, corresponding to the observable $\cA_{\vec n}^{(\vec I,J)}$. We discuss separately the pure Gaussian case case and the $\cN=2$ single node (SCQCD) and multiple nodes ($A_{q-1}$) cases.

\paragraph{Pure Gaussian matrix model}
As shown in section \ref{sec:WN4}, correlators involving several coincident Wilson loops must be separated in two cases depending if the loop operators belong or not to the same vector multiplet. To clarify this point when it is applied to defect correlators, at first we consider two coincident Wilson loops $\mathcal{W}_{[I,J]}$, then we generalize for any number of them.
When $I\neq J$, due to the factorization of the Wilson loops shown in \eqref{factN4IneqJ}, we simply obtain
\begin{equation}
    \vev{\mathcal{W}_{[I,J]} :\mathcal{O}^{(I)}_{\vec{n}}:}_0\equiv
     \mathcal{A}_{\vec{n}}^{([I,J],I)}=
     \mathcal{A}_{\vec{n}}^{I}({\color{blue}\lambda}_I,N)w_J({\color{blue}\lambda}_J,N)
\end{equation}
where $\mathcal{A}_{\vec{n}}^{I}$ are the correlators computed above and $w_J$ the Wilson loop vev of $\cN=4$ SYM \eqref{exactWL}. On the other hand, if $J=I$ only a specific class of operators admits exact results in terms of derivatives of known functions. Indeed, following the discussion of section \ref{sec:WLSCQCD}, when more than one Wilson loop belong to a node of the quiver, only terms proportional to the $t$-function \eqref{t222222} can be written exactly. In the present case, this means that we can consider only operators with $\vec{n}=[2,2,...,2]$ obtaining
\begin{equation}\label{WO2222gauss}
    \vev{\mathcal{W}_{[I,I]} :\mathcal{O}^{(I)}_{[\underbrace{2,2,...,2}_{\text{$m$-times}}]}:}_0\equiv
     \mathcal{A}_{[2,2,...,2]}^{([I,I],I)}=
     \mathcal{A}_{[2,2,...,2]}^{I}({\color{blue}\lambda}_I,N)
     \qquad w_I\rightarrow w_{[I,I]}
\end{equation}
where we have to substitute the $\cN=4$ vev of the Wilson loop with \eqref{wIIexactN=4}. However, all the remaining cases can be computed perturbatively cutting the sums at an enough high order and solving all the $t$-functions with the recursion \eqref{recursion}. For instance
\begin{equation}\small\begin{split}\label{A3III}
  \mathcal{A}_{[3]}^{([I,I],I)}
 \! =&\frac{(N^4\!-\!5N^2+4){\color{blue}\lambda}_I^2}{16\sqrt{2}N^4}\!+\!
  \frac{(3N^6\!-\!7N^4\!-\!28N^2+32){\color{blue}\lambda}_I^3}{256\sqrt{2}N^6}
  \!+\!\frac{(N^2\!-\!1)(N^2\!-\!4)(7N^4\!+\!55N^2\!-\!180){\color{blue}\lambda}_I^4}{7680\sqrt{2}N^8}\\
  &+\frac{(N^2\!-\!1)(N^2\!-\!4)(3N^6+50N^4-276N^2+480){\color{blue}\lambda}_I^5}{73728\sqrt{2}N^{10}}+O({\color{blue}\lambda}_I^6)
\end{split}\normalsize\raisetag{20pt}\end{equation}

The generalization to $\mathcal{A}_{\vec{n}}^{(\vec{I},J)}$ is pretty straightforward. Indeed, given the factorization of the Wilson loops on the quiver, this is equal to the right-hand side of \eqref{factorizedwvecIN4} substituting the $w$ that contains the index $J$ of the operator with the corresponding $\mathcal{A}$. Moreover, analyzing the pattern for the lowest multitrace operators in \eqref{examplegaussWO} together with the
formula \eqref{WO2222gauss},
we can conclude that
\begin{equation}
    \mathcal{A}_{[\underbrace{2,2,...,2}_{\text{$m$-times}}]}^{(\vec{I},J)}={\color{blue}\lambda}_J^m\partial_J^m w_{\vec{I}}
\end{equation}

 \paragraph{SCQCD}
 In SCQCD all the Wilson loops belong to the same vector multiplet and then $\vec{I}$ is constrained to be a vector of only ones. Given the experience gathered from the previous sections, we know that for this observable it is possible to compute exactly only the terms in the transcendentality expansion that are proportional to pure powers of ${\color{red}\zeta}_3$ and only if we consider the operators $\mathcal{O}^{(1)}_{[2,2,...,2]}$ (see for instance \eqref{WO2222gauss}).
In general, it is sufficient to select the ${\color{red}\zeta}_3^m$ terms in the observables with a single Wilson loop and substitute $w_1$ as follows
\begin{equation}
    \mathcal{A}_{[2,2,...,2]}^{([1,1,...,1],1)}=
    \mathcal{A}_{[2,2,...,2]}^{(1,1)}\qquad w_1\rightarrow w_{[1,1,...,1]}
\end{equation}
As an example we have
\begin{equation}\footnotesize\begin{split}
  \mathcal{A}_{[2,2]}^{([1,1],1)}\bigg|_{{\color{red}\zeta}_3^m}&={\color{blue}\lambda} _{1}^2 \partial_1^2w_{[1,1]}-\frac{3{\color{red}\zeta} _{3} {\color{blue}\lambda} _{1}^4 }{64 \left(\pi ^4 N^2\right)}\bigg[6 \left(N^2+3\right) \partial_1^2w_{[1,1]}+{\color{blue}\lambda} _{1} \left(\left(N^2+9\right) \partial_1^3w_{[1,1]}+{\color{blue}\lambda} _{1} \partial_1^4w_{[1,1]}\right)\!\bigg]\\
  &+\frac{9 {\color{red}\zeta} _{3}^2 {\color{blue}\lambda} _{1}^6}{8192 \pi ^8 N^4} \bigg[\left(58 N^4\!+\!408 N^2\!+\!710\right) \partial_1^2w_{[1,1]}\!+\!{\color{blue}\lambda} _{1} \left(4 \left(4 N^4\!+\!57 N^2\!+\!173\right) \partial_1^3w_{[1,1]}\right.\\
  &\left.+\!{\color{blue}\lambda} _{1} \left({\color{blue}\lambda} _{1} \left(2 \left(N^2\!+\!13\right) \partial_1^5w_{[1,1]}\!+\!{\color{blue}\lambda} _{1} \partial_1^6w_{[1,1]}\right)\!+\!\left(N^4\!+\!40 N^2\!+\!219\right) \partial_1^4w_{[1,1]}\right)\right)\bigg]+...\\
  \mathcal{A}_{[2,2,2]}^{([1,1],1)}\bigg|_{{\color{red}\zeta}_3^m}&={\color{blue}\lambda} _{1}^3 \partial_1^3w_{[1,1]}-\frac{3 {\color{red}\zeta} _{3} {\color{blue}\lambda} _{1}^5}{64 \left(\pi ^4 N^2\right)} \bigg[9 \left(N^2+5\right) \partial_1^3w_{[1,1]}+{\color{blue}\lambda} _{1} \left(\left(N^2+13\right) \partial_1^4w_{[1,1]}+{\color{blue}\lambda} _{1} \partial_1^5w_{[1,1]}\right)\!\bigg]\\
  &+\!\frac{9 {\color{red}\zeta} _{3}^2 {\color{blue}\lambda} _{1}^7}{8192 \pi ^8 N^4} \bigg[6 \left(19 N^4\!+\!210 N^2\!+\!587\right) \partial_1^3w_{[1,1]}\!+\!{\color{blue}\lambda} _{1} \left(2 \left(11 N^4\!+\!222 N^2\!+\!967\right) \partial_1^4w_{[1,1]}\right.\\
  &\left.+{\color{blue}\lambda} _{1} \left({\color{blue}\lambda} _{1} \left(2 \left(N^2+17\right) \partial_1^6w_{[1,1]}+{\color{blue}\lambda} _{1} \partial_1^7w_{[1,1]}\right)+\left(N^4+54 N^2+401\right) \partial_1^5w_{[1,1]}\right)\right)\bigg]+...
\end{split}\normalsize\end{equation}
In the large $N$ limit the substitution factorizes as \eqref{factorizedwvecIN4}.

Also in this case the correlator involving the operator $:\mathcal{O}_2^{(1)}:$ is pretty special. Indeed, similarly to the case analyzed in \eqref{Aderw}, when the one point function of the shortest operator is computed in presence of any number of coincident Wilson loops, it can be computed exactly in terms of the SCQCD multiple Wilson loop vev as follows
\begin{equation}\label{Aderwmultiple}
    \mathcal{A}_{[2]}^{([1,1,...,1],1)}={\color{blue}\lambda}_1 \partial_1 w_{1,1,...,1}^{(1)}
\end{equation}
All the remaining cases, can be easily studied in perturbation theory. For instance 
\begin{equation}\footnotesize\begin{split}
    &\mathcal{A}_{[3]}^{([1,1],1)}=
    \mathcal{A}_{[3]}^{([1,1],1)}\bigg|_{\text{Gauss}} \!\!-\frac{9 {\color{blue}\lambda} _{1}^4 \left(N^2\!+\!3\right) 
    \left(N^4\!-\!5 N^2\!+\!4\right)
    {\color{red}\zeta} _{3}}{1024 \sqrt{2} \pi ^4 N^6}
    \!-\!\frac{{\color{blue}\lambda} _{1}^5\left(N^4\!-\!5 N^2\!+\!4\right) }{16384 \sqrt{2} \pi ^6 N^8} 
    \bigg[12 \pi ^2 \left(N^2\!+\!4\right) \left(3 N^2\!+\!8\right) {\color{red}\zeta} _{3}\\
    &-\!5 \left(22 N^4\!\!+\!31 N^2\!-\!53\right)\! {\color{red}\zeta} _{5}\!\bigg]
    \!+\!\frac{{\color{blue}\lambda} _{1}^6\!\left(N^4\!\!-\!5 N^2\!+\!4\right) }{262144 \sqrt{2} \pi ^8 N^{10}} 
    \!\bigg[108 N^2 \!\left(3 N^4\!\!+\!20 N^2\!\!+\!37\right) \!{\color{red}\zeta} _{3}^2\!-\!8 \pi ^4\! \!\left(N^2\!+\!5\right)\! \left(7 N^4\!+\!55 N^2\!-\!180\right) \!{\color{red}\zeta} _{3}\\
    &+10 \pi ^2 \left(43 N^6+279 N^4+202 N^2-464\right) {\color{red}\zeta} _{5}-105 \left(11 N^6+12 N^4-16 N^2+29\right) {\color{red}\zeta} _{7}\bigg]+O\left({\color{blue}\lambda} _{1}^7\right)
\end{split}\normalsize\raisetag{18pt}\end{equation}
where $\mathcal{A}_{[3]}^{([1,1],1)}\bigg|_{\text{Gauss}}$
is the Gaussian model result given by \eqref{A3III} for $I=1$.

\paragraph{$A_{q-1}$ theories}
We extend the analysis for a theory with $q$ nodes. Since the possible choices of the positions of the operator and Wilson loops on the quiver are endless, we only show here few examples at large $N$ and we leave the majority of the extended results to be consulted on the attached notebook \texttt{WLcorrelators.nb}. Since we have showed the general result involving the shortest operator $:\mathcal{O}_{[2]}^{(J)}:$ in \eqref{A2q}, from now on we will consider only operators with dimension $n\geq 3$. In the previous section we learned that operators of this kind have to be computed using standard perturbation theory, namely truncating the sums up to  a high enough cut-off and computing the $t$-functions with the recursion relation.
However, there are some interesting exceptions that we can still study exactly using the method described in the previous sections. For instance, the ${\color{red}\zeta}_3^m$ terms when the operator is a multitrace  with dimension $\vec{n}=[2,2,...,2]$ lying on the same node of the coincident Wilson loops. As an example, for $q=2$ we have
\begin{equation}\footnotesize\begin{split}
  &\mathcal{A}_{[2,2]}^{([1,1],1)}\bigg|_{{\color{red}\zeta}_3^m}=
  {\color{blue}\lambda} _{1}^2 \partial_1^2w_{[1,1]}-\frac{3 {\color{red}\zeta} _{3} {\color{blue}\lambda} _{1}^3 \left(\left(6 {\color{blue}\lambda} _{1}-4 {\color{blue}\lambda} _{2}\right) \partial_1^2w_{[1,1]}+{\color{blue}\lambda} _{1} \left({\color{blue}\lambda} _{1}-{\color{blue}\lambda} _{2}\right) \partial_1^3w_{[1,1]}\right)}{64 \pi ^4}\\
  &+\frac{\!9 {\color{red}\zeta} _{3}^2 {\color{blue}\lambda} _{1}^3 \left(8\!\left(\tfrac{29}{4} {\color{blue}\lambda} _{1}^3\!-\!9 {\color{blue}\lambda} _{2} {\color{blue}\lambda} _{1}^2\!+\!4 {\color{blue}\lambda} _{2}^2 {\color{blue}\lambda} _{1}\!-\! {\color{blue}\lambda} _{2}^3\right) \!\partial_1^2w_{[1,1]}\!+\!{\color{blue}\lambda} _{1} \!\!\left({\color{blue}\lambda} _{1}\!-\!{\color{blue}\lambda} _{2}\right)\!\! \left(\!\left({\color{blue}\lambda} _{1}\!-\!{\color{blue}\lambda} _{2}\right)\! {\color{blue}\lambda} _{1}^2 \partial_1^4w_{[1,1]}\!+\!2 \!\left(8 {\color{blue}\lambda} _{1}^2\!-\!5 {\color{blue}\lambda} _{2} {\color{blue}\lambda} _{1}\!+\!{\color{blue}\lambda} _{2}^2\right)\! \partial_1^3w_{[1,1]}\right)\!\right)}{8192 \pi ^8}
\end{split}\normalsize\raisetag{46pt}\end{equation}
where $w_{[1,1]}$ is the vev of two coincident Wilson loops in $\cN=4$ \eqref{wIIexactN=4}. 
The remaining parts of the expansion with different transcendentality can be systematically computed in perturbation theory. 

The only cases in which we can compute each term of the  transcendentality expansion exactly for $n\geq 3$ and multiple coincident Wilson loops are when the latter appear at most once for any node of the quiver. Following the examples given in section \eqref{Sec4.3}, for instance we can consider the configuration in which the one-point function of an operator belonging to the node 1 is computed in presence of $q$ Wilson loops, one for each node.
At large $N$ we have
\begin{equation}\small\begin{split}
    \mathcal{A}_{[n]}^{([1,2,...,q],1)}=
    \mathcal{A}_{[n]}^{(1,1)}\prod_{i=2}^q& w_{i}^{{(q)}}+
    \frac{{\color{red}\zeta}_{2n-1}}{8^{n+1}\pi^{2n}}
    \binom{2n}{n}  {\color{blue}\lambda}_1^{n-1}w_1\times\\
    &\times\bigg[\left(\prod_{i=3}^q w_{i}\right){\color{blue}\lambda}_2
    \mathcal{A}_{[n]}^{(2,2)}\bigg|_{\text{Gauss}}+\left(\prod_{i=2}^{q-1}w_{i}\right){\color{blue}\lambda}_q 
    \mathcal{A}_{[n]}^{(q,q)}\bigg|_{\text{Gauss}}\bigg]+...
\end{split}\end{equation}
The first term of the right-hand side corresponds to $ \mathcal{A}_{[n]}^{(1,1)}w_{[1,2,...,q]}^{{(q)}}$ that in the 't Hooft limit factorizes according to \eqref{w[12...q]}. All the other terms represent the deviation from this factorization and they start at order ${\color{red}\zeta}_{2n-1}$ in transcendentality. The theory at the orbifold point doesn't present any peculiar behavior for these observables a part of a drastic simplification of their transcendentality expansions. However, in the next section we identify a special class of operators whose one-point functions in presence of Wilson loops present some interesting properties at the orbifold point.

\section{Action, Feynman rules and color factors}\label{App:FieldTheory}
We write the action of $A_{q-1}$ theories using the $\cN=1$ superspace formalism.\\
The $\cN=2$ vector field in the node $I$ is decomposed into a $\cN=1$ vector multiplet $V_I$ and a $\cN=1$ chiral multiplet $\Phi_I$, both transforming in the adjoint representation of $SU(N)_I$. The $\cN=2$ matter hypermultiplet is formed by two $\cN=1$ chiral multiplets $\big(Q, \widetilde{Q} \big)$, transforming in the bifundamental $\big(\square, \bar{\square}\big)  ~\mathrm{of~SU}(N)_I\times \mathrm{SU}(N)_J$.

We write the superspace action for a generic quiver theory $A_{q-1}$:
\begin{align}\label{Stot}
S_{q-1} &= \sum_{I=1}^q \Bigg[ \frac{1}{8g_I^2} \left(\int d^4x\,d^2\theta\, \tr(W_I^\a W^I_\a)\!+\!\mathrm{h.c.}\!\right) +\!2\!\int d^4x\,d^4\theta\, \tr\!\left( e^{-2g_IV_I}\Phi_I^\dagger e^{2g_IV_I}\Phi_I\right) \notag \\
&+ \!\int d^4x\,d^4\theta\,\bigg( \tr \left( Q^\dagger e^{2g_IV_I} Q e^{-2g_{I+1}V_{I+1}}\right) + \tr \left(\widetilde{Q} e^{-2g_IV_I}\widetilde{Q}^\dagger e^{2g_{I+1}V_{I+1}} \right) \bigg) \notag \\
&+ \left(\ii \sqrt{2}g_I\!\int\!d^4x\,d^2\theta\,\widetilde{Q} \Phi_I Q +\mathrm{h.c.}\right)+ \left(\ii \sqrt{2}g_{I+1}\!\int\!d^4x\,d^2\theta\,\widetilde{Q} \Phi_{I+1} Q +\mathrm{h.c.}\right) \Bigg]
\end{align}
where $g_I$ are all the Yang-Mills couplings and $W^I_\a$ is the super field strength of $V_I$ defined as:
\begin{align}
W^I_\a=-\frac{1}{4}\bar{D}^2\left(e^{-2g_IV_I}D_\a e^{2g_IV_I}\right)~.
\end{align}
Notice that, since we are considering necklace quivers, the node $I=q+1$ is identified with $I=1$.

The Feynman rules for the action \eqref{Stot} are fully derived in \cite{Galvagno:2020cgq}, here we only emphasise the elements directly needed to follow section \ref{Sec:FieldTheory}. All the superfields are expanded in terms of the generators of the gauge group, according to their representation (adjoint or bifundamental):
\begin{equation}
V_I = V_I^a (T_a)^u_{~v}~,~~~~~\Phi_I = \Phi_I^a (T_a)^u_{~v}~,~~~~~ Q=Q^A(B_A)^u_{~\hat{v}}~,~~~~~\tilde{Q} = \tilde{Q}_A(B^A)^{\hat{u}}_{~v}~.
\end{equation}
The indices $a,b,\dots$ are adjoint indices, $A,B,\dots$ (anti-)bifundamental indices, $u,v,\hat u,\hat v$ (anti-)fundamental indices.\\
The matrices $T_a,B_A$ obey the following relations:
\begin{align}\label{matrixRelations}
\left[T_a,T_b\right] = \mathrm{i} f_{abc}T^c~,~~~~~ (T_a)^u_{~v}(T^a)^w_{~z} = \delta^u_z\delta^w_v-\frac{1}{N} \delta^u_v\delta^w_z~,~~~~~(B_A)^u_{~\hat{v}}(B^A)^{\hat{u}}_{~v} = \delta^u_v \delta^{\hat u}_{\hat v}~.
\end{align}

The explicit expressions of the propagators and the vertices can be derived from \eqref{Stot} and are fully presented in Appendix B of \cite{Galvagno:2020cgq}. However in the present paper we only need to compute color factors, hence we report the expressions for the color part of the vertices that are needed in the main text:

\begingroup
\allowdisplaybreaks
\begin{align}\label{Somevertices}
V_I\,Q^\dagger Q:\parbox[c]{.2\textwidth}{\includegraphics[width = .2\textwidth]{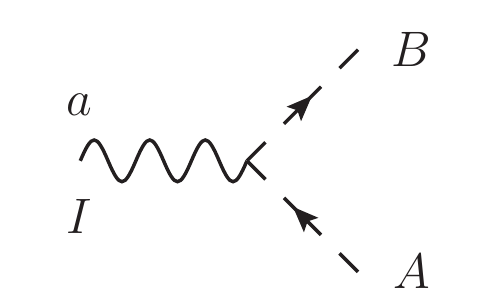}} &= 2 g_I \tr \parenth{T^a B^A B_B}~,  \notag \\
V_I\,\widetilde Q\widetilde Q^\dagger: ~~\parbox[c]{.2\textwidth}{\includegraphics[width = .2\textwidth]{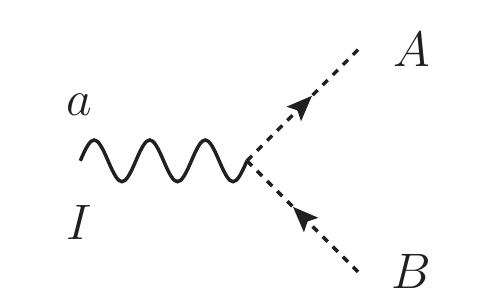}} &= -2 g_I \tr \left(T^a B_A B^B\right)~, \notag \\
\Phi_I\,Q\widetilde Q: ~~\parbox[c]{.2\textwidth}{\includegraphics[width = .2\textwidth]{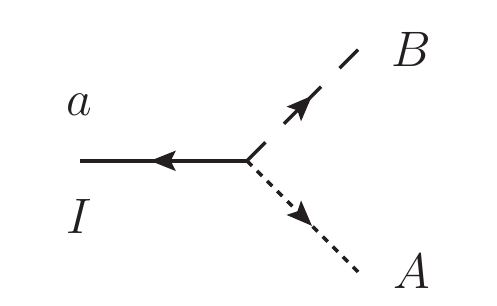}} &= \sqrt{2}\, \ii\, g_I  \tr \left(T^a B^A B_B\right) ~, \notag \\
\Phi_I^\dagger\,Q^\dagger\widetilde Q^\dagger: ~~\parbox[c]{.2\textwidth}{\includegraphics[width = .2\textwidth]{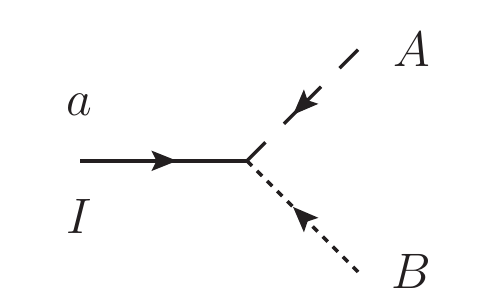}} &= -\sqrt{2}\,\ii\, g_I  \tr \left(T^a B^A B_B\right)~.
\end{align}
\endgroup

\paragraph{Color factor of the two-loops scalar propagator.}
These vertices are sufficient to compute some interesting color factors of $A_{q-1}$ theories, in particular for the two-loops correction to the scalar propagator belonging to $I$-th node. The generic expression for the color factor reads:
\begin{align}\label{2loopscolor}
&\parbox[c]{.25\textwidth}{\includegraphics[width = .25\textwidth]{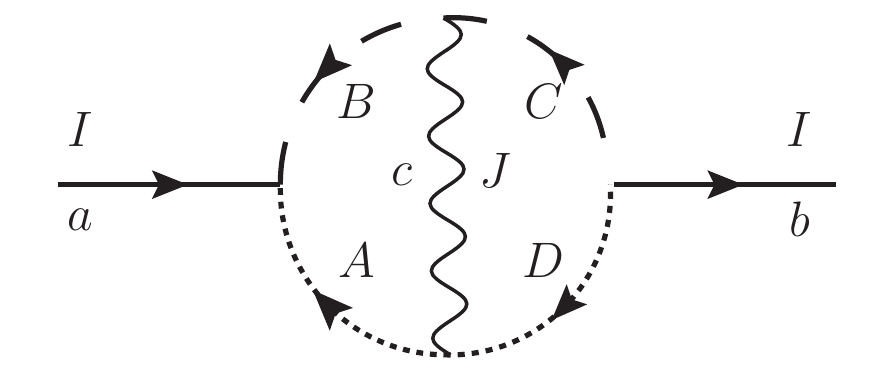}} = C_{IJ}^{ab}({\color{blue}\lambda}_I,{\color{blue}\lambda}_J,N)~,
\end{align}
and we need to distinguish two possible cases.
\begin{itemize}
\item
$I=J$. In this case the result corresponds to the SCQCD case, derived in several papers \cite{Billo:2017glv,Billo:2018oog,Billo:2019fbi} and reads: 
\begin{equation}
C_{II}^{ab}({\color{blue}\lambda}_I,N) = \frac{4{\color{blue}\lambda}_I^2}{N^2} (N^2+1)\delta^{ab}~.
\end{equation}
\item
$I\neq J$. This case corresponds to the contributions coming from neighboring nodes. Here we need the Feynman rules \eqref{Somevertices} as well as the matrix relations \eqref{matrixRelations}. We find:
\begin{equation}
C_{IJ}^{ab}({\color{blue}\lambda}_I,{\color{blue}\lambda}_J,N) = -\frac{2{\color{blue}\lambda}_I{\color{blue}\lambda}_J}{N^2} (N^2-1)\delta^{ab}~.
\end{equation}
\end{itemize}
 
The computation of the color part of the two-loops propagator of the adjoint scalar field in node 1 returns the following combination:
\begin{equation}\label{A9}
C_{11}^{ab}+C_{12}^{ab}+C_{1q}^{ab} = \frac{4{\color{blue}\lambda}_1}{N^2}\parenth{{\color{blue}\lambda}_1(N^2+1)-\frac{1}{2}({\color{blue}\lambda}_2+{\color{blue}\lambda}_q)(N^2-1)}\delta^{ab}~,
\end{equation}
which gives rise to the cancellations typical of $A_{q-1}$ theories discussed in section \ref{FT:Aq}.

\end{appendix}

\bibliographystyle{nb}
\bibliography{biblio}

\end{document}